\documentclass[pra,twocolumn,superscriptaddress,amssymb]{revtex4}
\usepackage[resetlabels]{multibib}
\newcites{appendix}{Appendix literature}

\usepackage[colorlinks=true,linkcolor=magenta,citecolor=magenta,urlcolor=magenta,linktocpage=true]{hyperref}
\usepackage{natbib}
\usepackage[utf8]{inputenc}
\usepackage[english]{babel}
\usepackage[T1]{fontenc}
\usepackage{amsmath}
\usepackage{cleveref}
\usepackage{svg}
\usepackage{tabularx}
\usepackage{array, makecell}

\usepackage{listings}
\usepackage{xcolor}
\makeatletter
\let\cat@comma@active\@empty
\makeatother

\usepackage{listings}
\usepackage{xcolor}
\definecolor{codegreen}{rgb}{0,0.6,0}
\definecolor{codegray}{rgb}{0.5,0.5,0.5}
\definecolor{codepurple}{rgb}{0.58,0,0.82}
\definecolor{backcolour}{rgb}{0.95,0.95,0.92}
\lstdefinestyle{mystyle}{
    backgroundcolor=\color{backcolour},   
    commentstyle=\color{codegreen},
    keywordstyle=\color{magenta},
    numberstyle=\tiny\color{codegray},
    stringstyle=\color{codepurple},
    basicstyle=\ttfamily\footnotesize,
    breakatwhitespace=false,         
    breaklines=true,                 
    captionpos=b,                    
    keepspaces=true,                 
    numbers=none,                    
    numbersep=5pt,                  
    showspaces=false,                
    showstringspaces=false,
    showtabs=false,                  
    tabsize=2
}
\usepackage{psfrag,graphicx}
\usepackage{dcolumn}
\usepackage{bm}
\usepackage{amsfonts,amssymb,amsmath}        
\usepackage{slashed}
\usepackage[utf8]{inputenc}
\usepackage{graphicx}
\usepackage[caption=false]{subfig}
\usepackage{cancel}
\usepackage{xcolor}
\usepackage{amsmath}
\setcitestyle{numbers,square}

\newcommand{\be}{\begin{equation}}
\newcommand{\ee}{\end{equation}}
\newcommand{\bq}{\begin{eqnarray}}
\newcommand{\eq}{\end{eqnarray}}

\newcommand{\ket}[1]{\left |#1 \right\rangle}
\newcommand{\bra}[1]{\left \langle #1 \right |}
\newcommand{\ketbra}[2]{\left|#1\right\rangle\left\langle#2\right|}

\newcommand{\hhat}[1]{\hat{\hat{#1}}}  

\makeatletter

\renewenvironment{widetext@grid}{%
  \par\ignorespaces
  \setbox\widetext@top\vbox{%
   \vskip15\p@
   \hb@xt@\hsize{%
    \leaders\hrule\hfil
    \vrule\@height6\p@
   }%
   \vskip6\p@
  }%
  \setbox\widetext@bot\hb@xt@\hsize{%
    \vrule\@depth6\p@
    \leaders\hrule\hfil
  }%
  \onecolumngrid
  \let\set@footnotewidth\set@footnotewidth@ii
}{%
  \par
  \twocolumngrid\global\@ignoretrue
  \@endpetrue
}%

\begin{document}

\title{Kondo QED: The Kondo effect and photon trapping in a two-impurity Anderson model ultra-strongly coupled to light}
\author{Po-Chen Kuo}
\affiliation{Department of Physics, National Cheng Kung University, 701 Tainan, Taiwan}
\affiliation{Center for Quantum Frontiers of Research and Technology, NCKU, 70101 Tainan, Taiwan}
\affiliation{Theoretical Quantum Physics Laboratory, Cluster for Pioneering Research, RIKEN, Wakoshi, Saitama 351-0198, Japan}
\author{Neill Lambert}
\email{nwlambert@gmail.com}
\affiliation{Theoretical Quantum Physics Laboratory, Cluster for Pioneering Research, RIKEN, Wakoshi, Saitama 351-0198, Japan}
\author{Mauro Cirio}
\email{cirio.mauro@gmail.com}
\affiliation{Graduate School of China Academy of Engineering Physics, Haidian District, Beijing, 100193, China}
\author{Yi-Te Huang}
\affiliation{Department of Physics, National Cheng Kung University, 701 Tainan, Taiwan}
\affiliation{Center for Quantum Frontiers of Research and Technology, NCKU, 70101 Tainan, Taiwan}
\author{Franco Nori}
\affiliation{Theoretical Quantum Physics Laboratory, Cluster for Pioneering Research, RIKEN, Wakoshi, Saitama 351-0198, Japan}
\affiliation{Center for Quantum Computing, RIKEN, Wakoshi, Saitama 351-0198, Japan}
\affiliation{Physics Department, The University of Michigan, Ann Arbor, Michigan 48109-1040, USA.}
\author{Yueh-Nan Chen}
\email{yuehnan@mail.ncku.edu.tw}
\affiliation{Department of Physics, National Cheng Kung University, 701 Tainan, Taiwan}
\affiliation{Center for Quantum Frontiers of Research and Technology, NCKU, 70101 Tainan, Taiwan}

\date{\today} 
\begin{abstract}
The Kondo effect is one of the most studied examples of strongly correlated quantum many-body physics. Another type of strongly correlated physics that has only recently been explored in detail (and become experimentally accessible) is that of ultrastrong coupling between light and matter. Here, we study a system which we denote as "Kondo QED") that combines both phenomena, consisting of a two-impurity Anderson model ultra-strongly coupled to a single-mode cavity. While presented as an abstract model, it is relevant for a range of future hybrid cavity-QED systems. Using the hierarchical equations of motion approach we show that the ultrastrong coupling of cavity photons to the electronic states (impurity) noticeably suppresses the electronic Kondo resonance due to the destruction of many-body correlations of the Kondo cloud. We observe this \emph{transfer of correlations from the Kondo cloud to the cavity} by computing the entropy and mutual information of the impurity-cavity subsystems. In addition, in the weak lead-coupling limit and at zero-bias, the model exhibits a ground-state photon accumulation effect originating entirely from counter-rotating terms in the impurity-cavity interaction. Interestingly, in the strong lead-coupling limit, this accumulation is ``Kondo-enhanced'' by new transition paths opening when increasing the hybridization to the leads. This suggests a new mechanism for the generation of real photons from virtual states.  We further show that the suppression of the Kondo effect is stable under broadening of the cavity resonance as a consequence of the interaction to an external bosonic continuum. Our findings pave the way for the simultaneous control of both the Kondo QED effect and a photon accumulation effect using the ultrastrong coupling of light and matter.

\end{abstract}
\maketitle
\section{Introduction}
Understanding the properties of strongly correlated open quantum systems remains one of the significant challenges in quantum many-body physics, with applications in quantum computation~\cite{Smith2019}, machine learning~\cite{Carleo2017}, quantum optics~\cite{Thomas2013,PuertasMartnez2019}, and condensed matter physics~\cite{bruus2004}. The Kondo resonance, arising from the strong quantum correlations formed between magnetic impurities and the surrounding electrons, has not only provided a testing ground for fundamental theories, but also for quantitative comparisons between theory and experiments~\cite{Kouwenhoven2001}. 

Quantum dots (or single molecules~\cite{Scott2010,Jeong2001}) are often used as controllable impurities and can be engineered to manifest the Kondo effect~\cite{Hur2015,Kalish2004,Sprinzak2002,Keller2013,Shang2018,Pustilnik2004}. In addition, they are promising for a range of technological applications, like single-electron transistors~\cite{Wingreen2004,Park2002,Natelson2004,Scott2010}. Importantly, for our purpose, it has been demonstrated experimentally that both charge and spin degrees of freedom can be coupled to microwave photons~\cite{Frey2012,Bruhat2016,Halbhuber2020,Woerkom2018}. So far, electronic systems (ESs), like quantum dots or impurity spin, operating in the Kondo regime in concert with electron-photon interactions~\cite{Cottet2015,Rubio2022} have offered a way to non-invasively probe quantum correlations in fermionic many-body systems through the phases and amplitudes of the photonic signals~\cite{Desjardins2017,GuoPing2021}. In these studies the electronic properties are largely unaffected by the cavity photons due to the weak electron-photon interaction. 

At the same time, it has been shown that light-matter coupling can be tuned to be on the same order of magnitude as the bare frequencies of the isolated systems \cite{Anton2019,Solano2019, Stockklauser2017}. In this ultra-strong coupling (USC) regime, virtual processes which simultaneously create or annihilate both light and matter excitations (usually neglected by the rotating wave approximation in the weak and strong coupling regimes \cite{Anton2019,Solano2019}) become important. Interestingly, these processes, enabled by the so-called counter-rotating terms in the Hamiltonian, induce an hybridization between light and matter even in the ground state, which becomes dressed by virtual excitations \cite{Stassi2013}.
This allows for the emergence of counterintuitive phenomena in various fields, such as quantum optics~\cite{Stassi2013,Anton2017,Garziano2016}, transport~\cite{Mauro2016,Cirio2019}, chemistry~\cite{Felipe2016,MartnezMartnez2018,Schfer2022}, and condensed matter~\cite{GarciaVidal2021,Bloch2022}. USC can also be realized in the context of open quantum systems~\cite{Lambert2019}. Additional novel potential applications have been explored in relation to quantum information processing~\cite{Pierre2011,Wendin2017}, quantum memories~\cite{PhysRevA.97.033823}, quantum plasmonics~\cite{Tame2013}, and quantum thermodynamics~\cite{Stella2018,Ivander2022}.

\subsection{Kondo QED: An overview}

In this section, we start with an intuitive overview and explanation of our results. In this work, we provide a qualitative description of how Kondo resonance behaves in the presence of light-matter interactions, which we term Kondo Quantum Electrodynamics (QED). In particular, we are interested in what occurs  when those interactions are allowed to be of order of other system energies (the USC regime). 

We investigate two scenarios: (i) a single-mode cavity, and (ii) a bosonic continuum, both ultra-strongly coupled to a two-impurity electronic system (ES). In both cases the ES is sandwiched between two fermionic environments that are designed to interact solely with the lower energy impurity $|g\rangle$ [see Fig.~\ref{TLC_schematic} (a)]. 

Each impurity can have the following four electronic configurations: a vacant impurity with no electrons, two states where the impurity is occupied by an electron with either spin up or spin down, and a state featuring double occupancy (with both a spin-up and a spin-down electron).
 

When the coupling between the ES and the leads is large, strong correlations are established between the impurity and the electrons in the leads at the Fermi energy, allowing them to form spin-antisymmetric states. The `bare' basis of single- and double-electron occupancy described above are then insufficient to describe this regime. In addition, these correlations cause the emergence of a zero frequency peak in the density of states (DOS) of the electronic system, as shown in Fig.~\ref{TLC_schematic} (b). This peak is a well-known spectroscopic signature of the Kondo effect~\cite{Hewson1993,Ralf2008,Nicolas2009,Yan2012}, and it has been experimentally detected~\cite{STM}.


Because of the many-body nature of the fermionic leads, one must resort to sophisticated numerical methods to describe this parameter regime. One such method, the hierarchical equation of motion (HEOM) approach, allows us to calculate the density of states (DOS) in a numerically exact manner, accounting for strongly non–Markovian bath correlations without resorting to perturbative approximations~\cite{Jin2008,Kato2016,Numericallyexact2020,lambert2020bofinheom,Koyanagi2022}. This means that the zero-frequency Kondo peak, which describes the many-body correlations between electrons in the lead and the electron system (ES), can be accurately captured by the HEOM method.

In order to characterize the impact of the light-matter interactions in the USC limit on the Kondo effect, which will also hybridize with the impurity states, we employ the HEOM method in two ways: first, by explicitly including the light degrees of freedom in the system part of the model (for the cavity limit), and second, by including them in the so-called auxiliary HEOM degrees of freedom (for the continuum limit). From this method we calculate the DOS and find that the hybridization between light and matter (emerging as a consequence of the ultrastrong coupling of the cavity field with the ES) can reduce the electronic Kondo resonance, by suppressing the correlations with the Kondo cloud. 
 
This result can be understood as a competition between ES-cavity and ES-leads hybridizaton. Stronger correlations between the system and the cavity intuitively decrease the electron availability to form delocalized states with the leads, thereby modulating the Kondo effect.
 
Furthermore, in the Ultrastrong Coupling (USC) regime, the  presence of quantum-fluctuations in the light-matter ground state \cite{Anton2019} enables new transition paths which result in a steady-state photon accumulation effect in the cavity. This intriguing effect manifests under both weak and strong matter-lead coupling conditions.


 When the system and the leads interact weakly, this photon accumulation effect is enhanced as the light-matter coupling increases. This can be intuitively understood using an effective master equation valid in the limit of small system-lead-coupling. This clearly demonstrates how, in this weak-coupling limit, increasing light-matter coupling leads to increased virtual photon accumulation due to larger ground-state light-matter hybridization~\cite{Mauro2016}. The trapping (conversion from virtual to real) of those photons occurs due to an inherent light-matter decoupling mechanism which activates when both impurities are occupied.



However when the lead-system interaction enters the non-pertubative Kondo  regime, a different trend is observed. We first see a larger overall photon accumulation magnitude, but with a counter-intuitive suppression of the photon accumulation effect taking place as the light-matter coupling is increased. This suppression can be traced back to the cavity-induced decoupling effect which reduces the importance of higher-order transitions enabled by the combination of system-lead and light-matter hybridization.


In addition, another surprising feature is that despite being triggered by the counter-rotating terms in the light-matter Hamiltonian, the photon-accumulation effect appears for relatively weak light-matter couplings and, as mentioned above, its magnitude can be ``Kondo-enhanced'' in the strong matter-lead coupling regime. This feature may provide a new way to indirectly detect the presence of ground-state virtual photons in the Rabi Hamiltonian~\cite{Childress2004,Hagenm2017,Hagenm2018,Bartolo2018}.

In parallel, we also note that the time required to accumulate a photon in the cavity remains sensitive to the intensity of the light-matter and matter-lead interaction strengths.

The rest of this article is organized as follows. In section II.A, we start by introducing our primary cavity-based model. In Sec. III, we lay out our principal findings, specifically highlighting the suppression of the Kondo peak provoked by the ultrastrong coupling to the single-mode cavity, and the photon accumulation effect which is reciprocally influenced by the Kondo effect. In Sec. IV we consider a continuum bosonic bath rather than a single-mode cavity, and show how it impacts the Kondo resonance. In Section V, we wrap up with a conclusion of our findings, their potential applications, and directions for future research. For those interested in more detail, we have included an Appendix. It covers the step-by-step derivation of the DOS using HEOM, an examination of the convergence of the DOS, a derivation of a modified master equation for degenerate systems, a description of the dressed states of the Kondo QED system, and an explanation of the reabsorption of the diamagnetic term in the light-matter interaction.

\section{The model}
\subsection{Kondo QED: A two-level ES ultrastrongly coupled to cavity photons}
\begin{figure}[]
\includegraphics[width = 0.9\columnwidth]{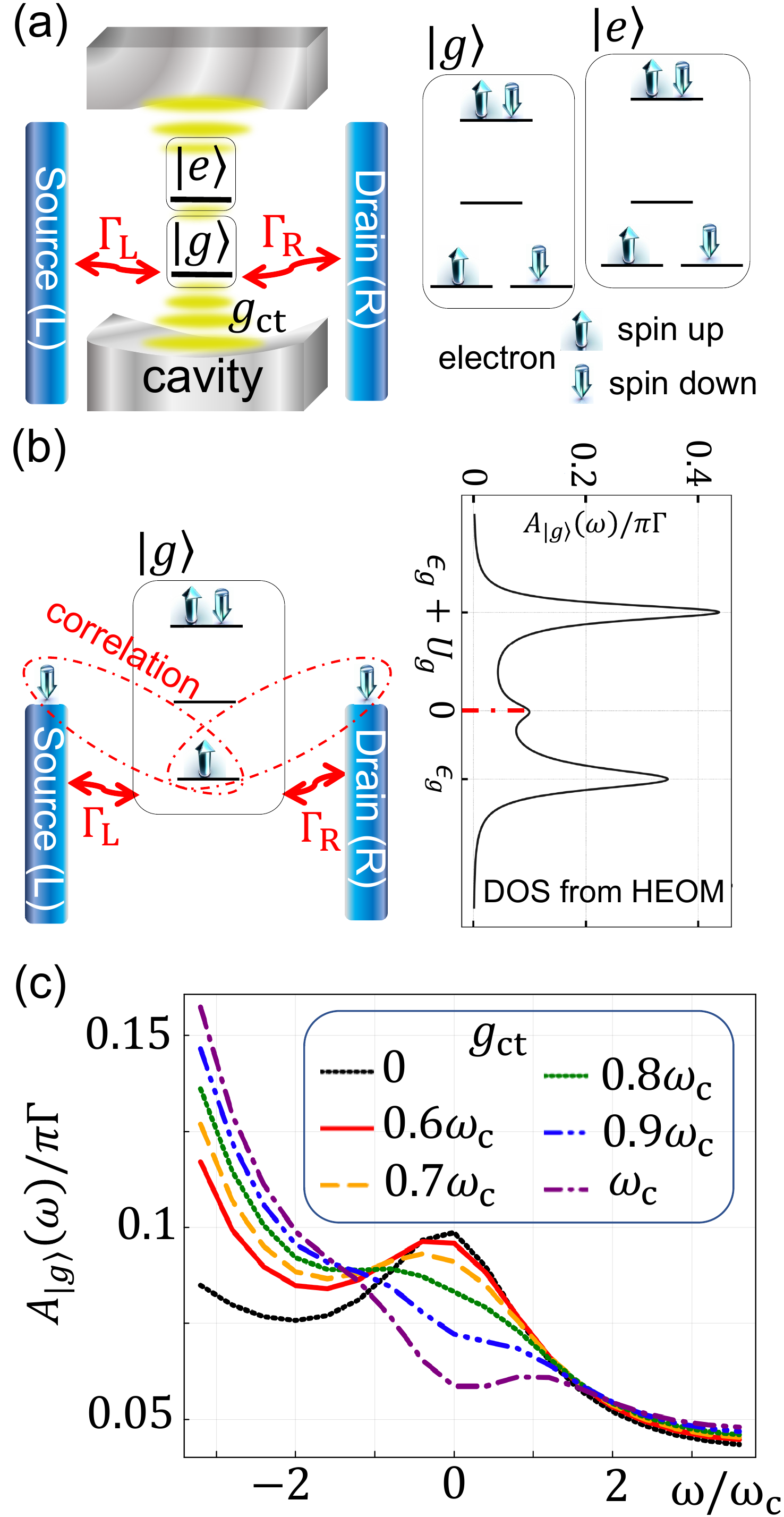}
\caption{(a) An electronic system, described by two states having energy $\epsilon_{g}=-6\omega_{\text{c}}$ and $\epsilon_{\text{e}}=-5\omega_{\text{c}}$, coupled to a high quality factor single-mode cavity with strength $g_{\text{ct}}$ and sandwiched between left and right leads with coupling strengths $\Gamma_{\text{L}}$ and $\Gamma_{\text{R}}$, respectively. 
On the right, we represent the four possible electronic states of each impurity $|g\rangle$ and $|e\rangle$: (bottom) two states where the impurity is occupied by an electron with either spin up or spin down, (middle) an empty state with no electrons, and (top) a double-occupied state with both spin-up and spin-down electrons. (b) Illustration of the Kondo regime, characterized by strong correlations between system electrons in $|g\rangle$ and the leads electrons at the Fermi level. As a consequence, a Kondo peak at the Fermi level (zero frequency) in the DOS appears which requires non-perturbative methods, such as the HEOM approach.
(c) Suppression of the Kondo peak around zero frequency for the equilibrium DOSs $(\mu_{L}=\mu_{R}=0)$ of the ES due to increasing the transverse light-matter coupling $g_{\text{ct}}$ from $0$ (black dotted) to $\omega_{\text{c}}$ (purple dash-dot) at temperature $T=0.5\omega_{\text{c}}$. The Coulomb repulsion between spin-up and spin-down electrons is set to $U_{n}=15\omega_{\text{c}}$. The coupling strengths to both leads are $\Gamma_{\text{L}}=\Gamma_{\text{R}}=\Gamma=\omega_{\text{c}}$ with bandwidth $W_{\text{f}}=10\omega_{\text{c}}$. The truncation of the Fock space dimension, of the HEOM tiers, and of the Pad\'{e} series, are set to $N_{\text{b}}=3$, $N_{\text{c}}=3$, and $u=5$, respectively.}\label{TLC_schematic}
\end{figure}

Here, we consider a two-level ES ultrastrongly coupled to a single-mode cavity and sandwiched between left and right leads as shown in Fig.~\ref{TLC_schematic}(a). Such an ultrastrongly coupled ES-cavity can be practically implemented by various near-future cavity
QED and circuit-QED setups, such as hybrid superconducting circuits~\cite{Xiang2013}, semiconductor quantum wells coupled to a microcavity circuits~\cite{Geiser2012}, molecular excitons coupled to a metal-clad microcavity~\cite{Gubbin2014}, and hybrid solid-state architectures~\cite{Delbecq2011,Toida2013,Mi2017,Desjardins2017}, especially quantum-dot based systems~\cite{Frey2012,Delbecq2013,Viennot2014}. 
We model this setup by the Hamiltonian (with $\hbar=1$ throughout) 
\begin{equation}
\begin{aligned}
H_{\text{T}} = H_{\text{s}} + H_{\text{f}} + H_{\text{ef}},
\end{aligned}
\end{equation}
where the system Hamiltonian 
\begin{equation}
\begin{aligned}
H_{\text{s}} = H_{\text{e}} + H_{\text{c}} + H_{\text{ec}}
\end{aligned}
\end{equation}
describes the ES ($H_e$), the single-mode cavity ($H_{\text{c}}$), and their interactions ($H_{\text{ec}}$). Here, the Hamiltonian of the electronic system is given by
\begin{equation}
\begin{aligned}
H_{\text{e}} &= \sum_{n=g,e}\sum_{\sigma=\uparrow,\downarrow}
          \epsilon_{n}\hat{n}_{n\sigma}
          +U_{n}\hat{n}_{n\uparrow}\hat{n}_{n\downarrow}, 
\end{aligned}
\end{equation}
where $d_{n\sigma}^{\dagger}$ creates an electron at level $n=\text{g},\text{e}$ with energy $\epsilon_{n}$. Here, $\hat{n}_{n\sigma}=d_{n\sigma}^{\dagger}d_{n\sigma}$ is the electronic number operator with spin $\sigma$. The Coulomb repulsion energy $U_{n}$ represents a non-linear effect and requires both spin up $\uparrow$ and spin down $\downarrow$ electrons to occupy the same state. 
The two-level ES is near resonance with the fundamental frequency ($\omega_{\text{c}}$) of the single mode $a$ inside the cavity with Hamiltonian 
\begin{equation}
\begin{aligned}
H_{\text{c}} = \omega_{\text{c}}a^{\dagger}a. 
\end{aligned}
\end{equation}
 In addition, the light-matter coupling between the electronic system and the cavity (c) photons, known as the light-matter coupling, is described by
\begin{equation}
\begin{aligned}
H_{\text{ec}} = \sum_{\sigma=\uparrow,\downarrow}g_{\text{ct}}(d_{g\sigma}^{\dagger}d_{e\sigma} +d_{e\sigma}^{\dagger}d_{g\sigma})(a^{\dagger}+a),
\end{aligned}
\end{equation}
where the coupling constant $g_{\text{ct}}$ can originate from a purely transverse (t) engineered interaction \cite{Neill2003,Beaudoin2016,Neill2018,STqubit2019}.
It is important to note that we assume $g_{\text{ct}}$ and $\omega_{\text{c}}$ have been implicitly renormalized by the $A^2$ term\cite{Anton2019}, as demonstrated in Appendix~\ref{a_diamagnetic}. To study the influence of the light-matter USC on the Kondo effect, we further assume a high-quality microwave cavity so that the dissipation to its bosonic environment can be neglected.
The leads (labeled by $\alpha$) are electronic reservoirs described by the Hamiltonian $\epsilon_{\alpha,k}$,
\begin{equation}
\begin{aligned}
H_{\text{f}} = \sum_{\alpha}\sum_{k}\epsilon_{\alpha,k}c_{\alpha,k}^{\dagger}c_{\alpha,k},
\end{aligned}
\end{equation}
where $c_{\alpha,k}^{\dagger}$ creates a fermion (f) in the state $k$ of the lead $\alpha$. Importantly, the electrons in the leads are assumed to couple to only the lowest level $|g\rangle$ of the ES. Hence, the interactions between the ES and two separate leads can be characterized by the interaction Hamiltonian 
\begin{equation}
\begin{aligned}
H_{\text{ef}} = \sum_{k}\sum_{\alpha=\text{L},\text{R}}\sum_{\sigma=\uparrow,\downarrow}
         g_{\alpha,k}(c_{\alpha,k}^{\dagger}d_{g\sigma}
         +d_{g\sigma}^{\dagger}c_{\alpha,k}).
\end{aligned}
\end{equation}
 The interaction between the electronic system (ES) and the fermionic (f) leads can be fully characterized by the Lorentzian spectral density
\begin{equation}\label{eq:LorentzF}
     J_{\text{f}_{\alpha}}(\omega)=\frac{1}{2\pi}\frac{\Gamma_\alpha W_{\text{f}}^2}{(\omega-\mu_\alpha)^2+W_{\text{f}}^2},
\end{equation}
where $\Gamma_\alpha$ represents the coupling strength between the system and the $\alpha$-lead with bandwidth $W_{\text{f}}$ and chemical potential $\mu_\alpha$.

\section{Results}
\subsection{Suppression of the Kondo peak}
We now compute the DOS of the electronic system. This is a key quantity in describing the Kondo effect \cite{Nicolas2009} and can be engineered to improve electronic device performance.
The DOS of this ultra-strongly coupled ES–cavity system can be calculated as in Eq.~(\ref{sys_DOS}) using a parity-dependent HEOM based on a recent canonical derivation of the influence superoperator~\cite{Mauro2022,huang2023heomjl}, see Appendix~\ref{DOS_deriv} for more details. 

Here, to have a better resolution of the Kondo effect, we restricted our analysis to the DOS of the lowest state $A_{\ket{g}}(\omega)$. We further set a large repulsion energy as $U_{n}=15\omega_{\text{c}}$, to avoid any overlap between the Kondo peak and other resonances when increasing the transverse coupling. By varying the cavity coupling from $g_{\text{ct}}=0.6\omega_{\text{c}}$ to $g_{\text{ct}}=\omega_{\text{c}}$ (deep in the USC regime), the Kondo peak gradually disappears as shown in Fig.~\ref{TLC_schematic}(d). 

We note that, to optimize the memory requirement of the simulation, we truncated the Fock space to three photons. While this is typically insufficient to achieve convergent USC effects, an increase in the truncation only slightly affects the zero-frequency component of the DOS in the Kondo regime. The convergence properties of the whole DOS with respect to the truncation of $N_{\text{c}}$ and $N_{\text{b}}$ are shown in Appendix~\ref{DOS_conv}.
\begin{figure}[]
\includegraphics[width = 0.98\columnwidth]{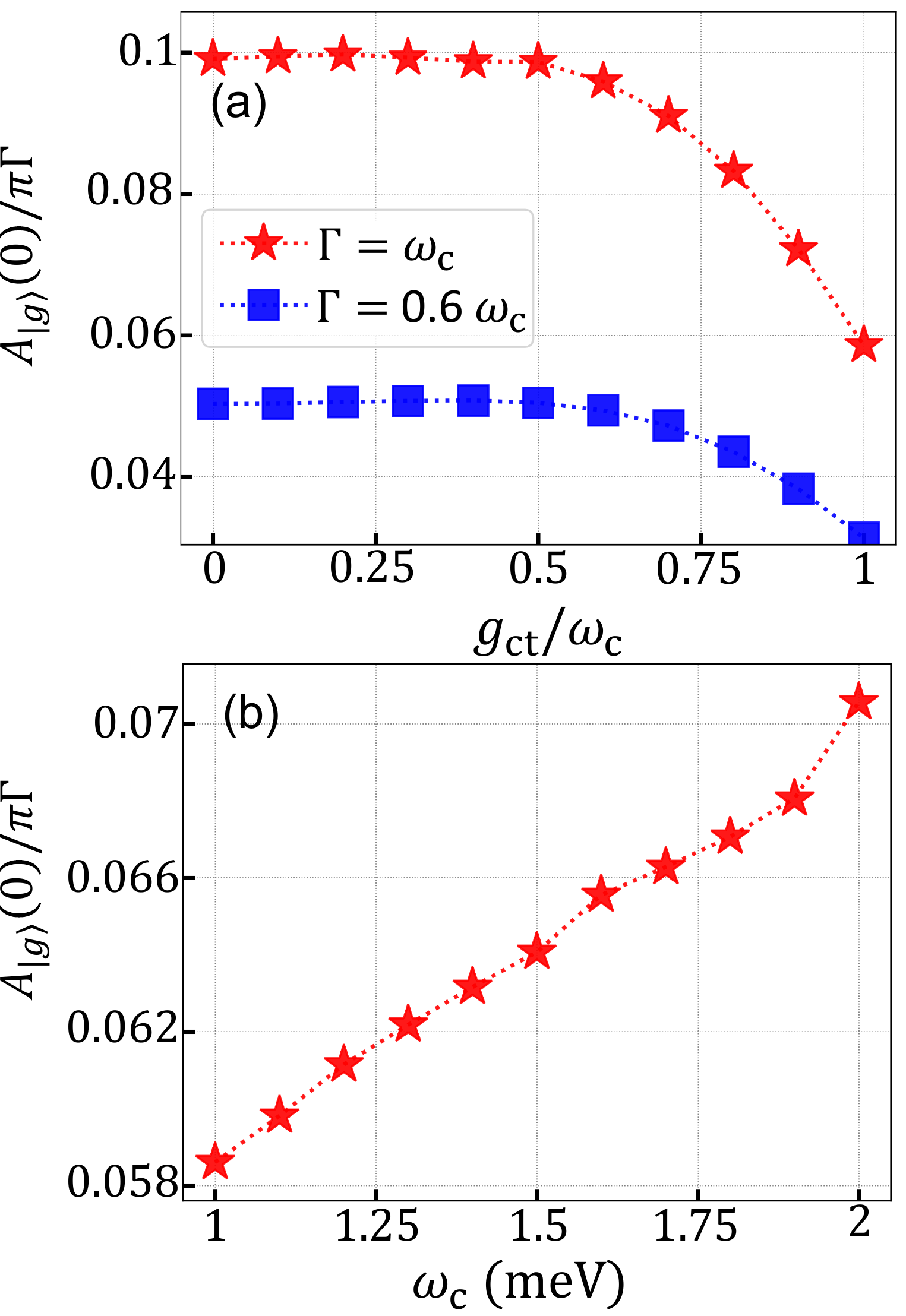}
\caption{(a) The zero-frequency component $A_{|g\rangle}(0)$ of the DOS as a function of the transverse coupling strength $g_{\text{ct}}$, with $\Gamma=0.6\omega_{\text{c}}$ (red star markers) and $\Gamma=\omega_{\text{c}}$ (blue square markers). (b) The zero-frequency component of the DOS, $A_{|g\rangle}(0)$, plotted against the cavity frequency, $\omega_{\text{c}}$. The suppression of the Kondo peak diminishes as $\omega_{\text{c}}$ increases.}\label{Fig1-1}
\end{figure}

Remarkably, in order to observe a noticeable impact on the zero-frequency component of the DOS in the Kondo regime, the strength of the transverse coupling $g_{\text{ct}}$ has to be in the deep-strong coupling regime, i.e., it has to be comparable to the cavity resonant frequency $\omega_{\text{c}}$.
As illustrated in Fig.~\ref{Fig1-1}, when we decrease the lead coupling from $\Gamma_{\alpha}=\omega_{\text{c}}$ to $0.6\omega_{c}$, the suppression of the Kondo peak, $A_{\ket{g}}(0)$, only occurs as $g_{\text{ct}}$ approaches $\omega_{\text{c}}$, not at the reduced value of $\Gamma_{\alpha}$. Similarly, as the cavity resonant frequency $\omega_{\text{c}}$ increases, even up to twice its original value,  we see a reduction in the Kondo suppression effect, correlated with leaving the USC regime into the strong coupling range. Consequently, maintaining the cavity coupling within the deep ultrastrong coupling regime is also a crucial prerequisite for observing a noticeable Kondo suppression.
\begin{figure}[]
\includegraphics[width=\columnwidth]{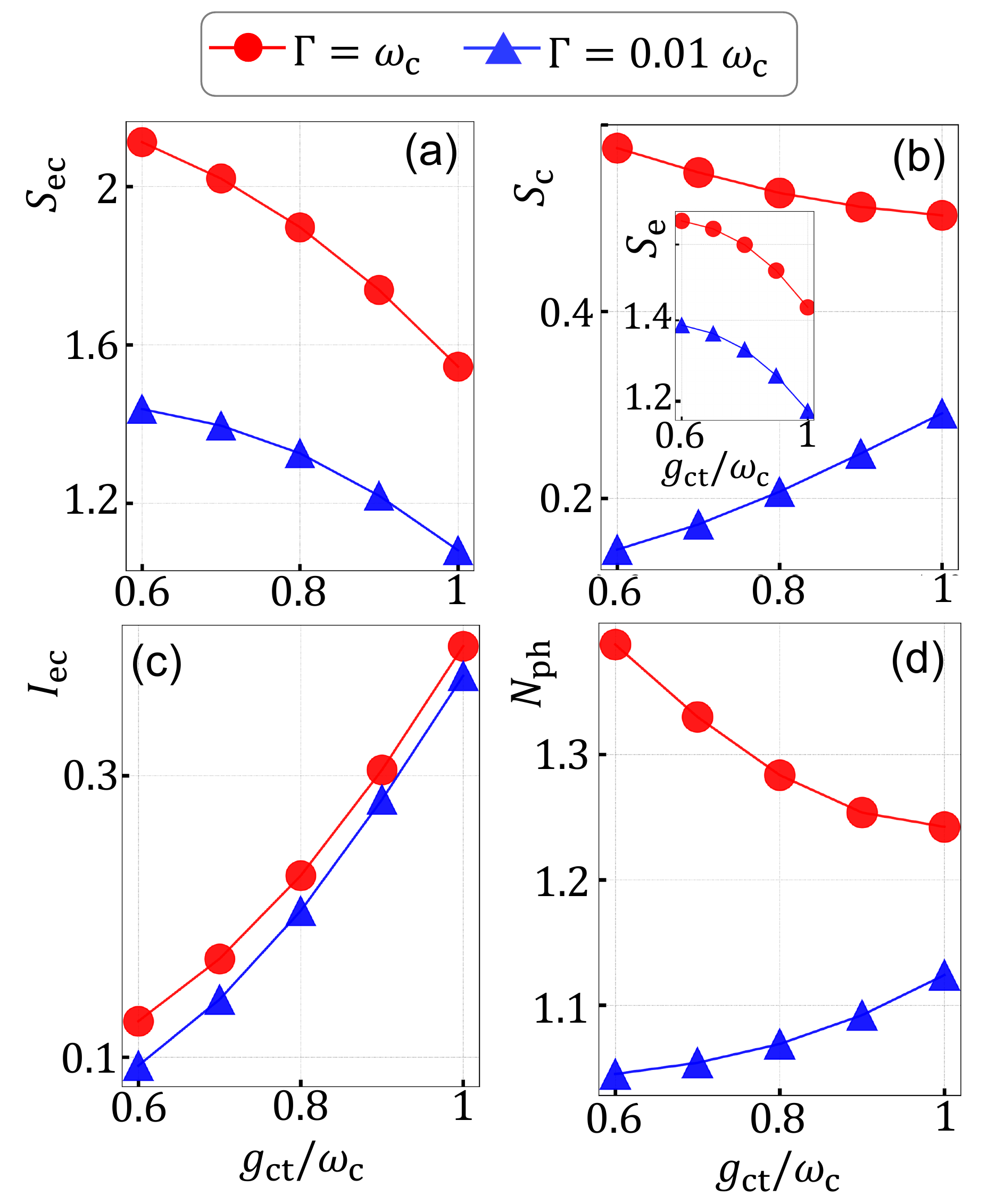}
\caption{Correlations between subsystems. As a function of the light-matter coupling strength $g_{\text{ct}}$, we show (a) the Von Neumann Entropy $(S_{\text{ec}})$ (b) the cavity ($S_{\text{c}}$) and ES ($S_{\text{e}}$) entropy, (c) the mutual information $(I_{\text{ec}})$, and (d) the average photon number $(N_{\text{ph}})$. The scaling of the correlations depends on the coupling strength to the leads (Kondo regime in red and non-Kondo regime in blue).
}\label{S_I}
\end{figure}

Additionally, the suppression of the Kondo effect shows that a potential reduction of the correlations between the system and the leads occurs in the USC light-matter coupling regime. To clarify this, we analyzed the von Neumann entropy of the steady-state reduced density operator $\rho_{\mathrm{ec}}(\infty)=\text{Tr}_{\text{f}}[\rho_{\text{T}}(\infty)]$ by partially tracing over the Hilbert space of the fermionic (f) leads
\begin{equation}
\begin{aligned}
S_{\mathrm{ec}}=-\text{Tr}_{\mathrm{ec}}\left\{\rho_{\mathrm{ec}}(\infty)\mathrm{ln}\left[\rho_{\mathrm{ec}}(\infty)\right]\right\}.
\label{ent_s}
\end{aligned}
\end{equation}%
As shown in Fig.~\ref{S_I}(a), increasing the light-matter coupling $g_{\text{ct}}$ results in reducing the entropy of the system, indicating a decoupling of the ES from the leads.
 
In Fig.~\ref{S_I}(b) we also show the von Neumann entropy of the cavity 
\begin{equation}
\begin{aligned}
S_{\mathrm{c}}=-\text{Tr}_{\mathrm{c}}\left\{\rho_{\mathrm{c}}(\infty)\mathrm{ln}\left[\rho_{\mathrm{c}}(\infty)\right]\right\},
\end{aligned}
\end{equation}
and ES 
\begin{equation}
\begin{aligned}
S_{\mathrm{e}}=-\text{Tr}_{\mathrm{e}}\left\{\rho_{\mathrm{e}}(\infty)\mathrm{ln}\left[\rho_{\mathrm{e}}(\infty)\right]\right\}
\end{aligned}
\end{equation}
alone, where $\rho_{\mathrm{c}}=\text{Tr}_{\mathrm{e,f}}[\rho_{\text{T}}(\infty)]$ and $\rho_{\mathrm{e}}=\text{Tr}_{\mathrm{c,f}}[\rho_{\text{T}}(\infty)]$. 
Combined with the mutual information 
\begin{equation}
\begin{aligned}I_{\mathrm{ec}}=S_{\mathrm{c}}+S_{\mathrm{e}}-S_{\mathrm{ec}},
\end{aligned}
\end{equation}
see Fig.~\ref{S_I}(c), these quantities show that correlations between ES and cavity increase when the $g_{\mathrm{ct}}$ increases. This is consistent with the mentioned suppression of the Kondo peak in the DOS, providing more evidence that the USC light-matter coupling decouples the ES from the leads (akin to the decoupling seen in the tunnel-coupled two-impurity Anderson model~\cite{Spinelli2015}).

\subsection{The photon accumulation}

In analyzing the behavior of the cavity photons in the Kondo regime we see that, surprisingly, the average photon number ($N_{\text{ph}}$) in the cavity is large for small coupling and decreases with increasing light-matter coupling $g_{\text{ct}}$, as shown in Fig. 2(d), which is the opposite of what one might naively expect in USC physics.


\begin{figure*}[]
\includegraphics[width = \textwidth]{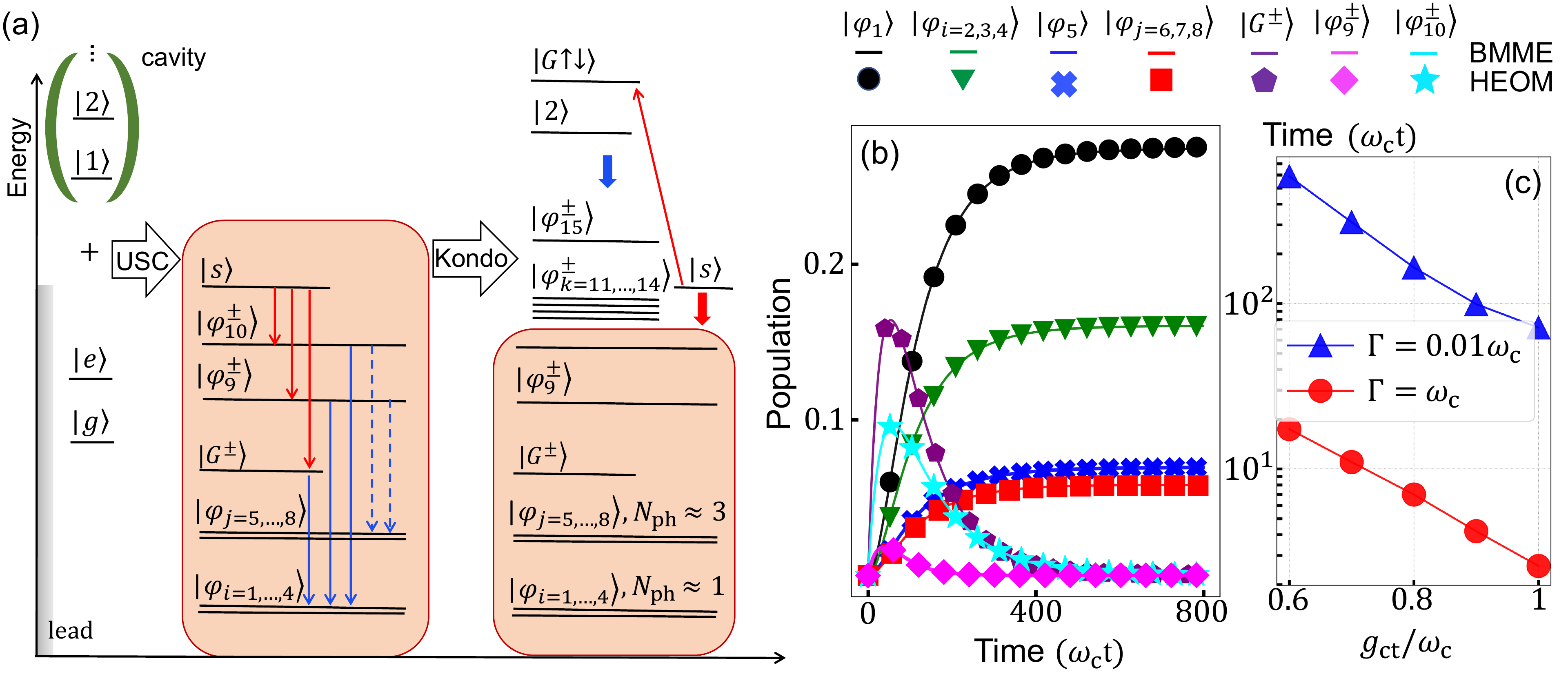}
\caption{(a) On the left, we show an energy diagram for the uncoupled system $(g_{\text{ct}}=0,\Gamma=0)$. The cavity states $|i\rangle$ describe $i$ photons while the ES states ($|e\rangle$ and $|g\rangle$) can be occupied by electrons with arbitrary spin configuration.  As the USC of cavity to ES occurs, the tunneling of an electron from the leads causes a transition from the empty state $|s\rangle$ to the intermediate transient states $|G^{\pm}\rangle$ and $|\varphi_{l=9,10}^{\pm}\rangle$ (red arrow).  Shortly afterwards, another entering electron will participate in this process, further inducing the transition to the stationary dressed states $|\varphi_{i=1,...,4}\rangle$ (blue arrow) and $|\varphi_{j=5,...,8}\rangle$ (blue dashed line arrow). By increasing the ES-leads couplings, there is a transition from the non-Kondo to the Kondo regime where the dressed states with higher energy, i.e., $|G\uparrow\downarrow\rangle$ and $|\varphi_{k=11,...,15}\rangle$, can be excited. (b) The short-time dynamics of different dressed states obtained by solving the HEOM are shown with markers while the solid-lines represent the solutions of the Born-Markov quantum master equation (BMME). The two methods are in good agreement in the non-Kondo regime ($\Gamma=0.01\omega_{\text{c}}$). (c) Evolution time needed to accumulate an expected number of photons equal to one as a function of the light-matter coupling $g_{\text{ct}}$ in the non-Kondo ($\Gamma=0.01\omega_{\text{c}}$) and the Kondo ($\Gamma=\omega_{\text{c}}$) regimes.}\label{state_config_2}
\end{figure*}

To better analyze this effect and its relationship to the Kondo suppression  in more detail, it is useful to consider the weak ES-lead coupling regime. In this non-Kondo regime, shown in Fig.~\ref{state_config_2}, we can employ a Born-Markov quantum master equation (BMME) to describe the influence of the leads on the ES-cavity system. This approach uses a Lindblad dissipator written in terms of a decomposition on the different eigenstates $\ket{\varphi_{i}}$  of $H_{\text{s}}$ with eigenenergies $\epsilon_{\mathbb{I}}$, i.e.,
\begin{equation}
\begin{aligned}
\partial_t\rho_{\text{s}}(t)&=-i[H_{\text{s}},\rho_{\text{s}}(t)]
+\sum_{\alpha\sigma}
\sum_{\epsilon_{k}-\epsilon_{l}=\omega}\sum_{p=\pm}
\gamma_{\alpha,l\rightarrow k}(\omega)\times\\&
\Big\{
p\ketbra{\varphi_{k}}{\varphi_{l}}\rho_{\text{s}}^{p}(t)\ketbra{\varphi_{l}}{\varphi_{k}}
-\frac{1}{2}
\{\ketbra{\varphi_{l}}{\varphi_{l}},\rho_{\text{s}}^{p}(t)\}
\Big\},
\end{aligned}
\end{equation}
where $[\cdot,\cdot]$ ($ \{\cdot,\cdot\}$) denotes the commutator (anticommutator). Note that the density operator in this Born-Markov quantum master equation is allowed to contain both even and odd parity, i.e., 
\begin{equation}
\begin{aligned}
\rho_{\text{s}}(t) = \delta_{p,1}\rho_{\text{s}}^{+}(t)+\delta_{p,-1}\rho_{\text{s}}^{-}(t).
\end{aligned}    
\end{equation}
Here, we assume that there is no coherence in the initial system states (otherwise, the Born-Markov quantum master equation should be modified to take into account degenerate energy levels in the eigenoperator decomposition). The details of the derivations are shown in Appendix~\ref{BMME_dr}. The Born-Markov quantum master equation provides information on the transitions between different system eigenstates which happen at the rates 
\begin{equation}
\begin{aligned}
\gamma_{\alpha,l\rightarrow k}(\omega)= 
2\pi\sum_{\nu=\pm 1}\sum_{u=\uparrow,\downarrow}
\vert\langle \varphi_{k}|d^{\nu}_{gu}|\varphi_{l}\rangle\vert^{2}
J_{\text{f}_{\alpha}}(\omega)n_{\text{f}_{\alpha}}(\omega),
\end{aligned}
\end{equation}
written in terms of the spectral density of leads $J_{\text{f}_{\alpha}}$ and Fermi-Dirac distribution $n_{\text{f}_{\alpha}}={1}/{(e^{\beta\omega}+1)}$, with $\beta=(k_{\text{B}}T)^{-1}$ ($k_{\text{B}}=1$).
Assuming an initially empty electronic system, an electron will rapidly enter the system due to the higher potential of the leads relative to the impurity energies. Due to the USC between cavity and ES, this electron can enter the ground ($|G^{\pm}\rangle$ with $N_{\text{ph}}\approx 0.5$) and higher ($|\varphi_{l=9,10}^{\pm}\rangle$ with $N_{\text{ph}}\approx 1.5$) photon-dressed states as shown in Fig.~\ref{state_config_2}(a).

Importantly, these states contain components in which the ES is excited to the impurity $2$ (state $|e\rangle$) and virtual photons are present in the cavity, due to the counter-rotating terms in the light-matter interaction. These components allow for a non-zero rate to two-electron states, where both impurity $1$ (state $|g\rangle$) and $2$ (state $|e\rangle$) are occupied, and physical photons populate the cavity, through the paths
\begin{equation}
\begin{array}{lllll}
|s\rangle&\xrightarrow{\gamma_{s\rightarrow G}}&|G^{\pm}\rangle&\xrightarrow{\gamma_{G\rightarrow i}}&
|\varphi_{i=1,...,4}\rangle,\\
|s\rangle&\xrightarrow{\gamma_{s\rightarrow l}}&
|\varphi_{l=9,10}^{\pm}\rangle&\xrightarrow{\gamma_{l\rightarrow j}}&
|\varphi_{j=1,...,8}\rangle,
\end{array}
\end{equation}
where $|s\rangle$ represents the state empty of photons and electrons. Here, the stationary dressed states can be approximately
written as 
\begin{equation}
\begin{aligned}
|\varphi_{i=1,...,4}\rangle &\approx 
f_{i_1}\ket{\uparrow,\downarrow,1}+f_{i_2}\ket{\uparrow,\uparrow,1}\\&
+f_{i_3}\ket{\downarrow,\downarrow,1}+f_{i_4}\ket{\downarrow,\uparrow,1}
\end{aligned}
\end{equation}
and 
\begin{equation}
\begin{aligned}
|\varphi_{j=5,...,8}\rangle &\approx f_{j_1}\ket{\uparrow,\downarrow,3}+f_{j_2}\ket{\uparrow,\uparrow,3}\\&+f_{j_3}\ket{\downarrow,\downarrow,3}+f_{j_4}\ket{\downarrow,\uparrow,3}.
\end{aligned}
\end{equation}

Note that $\ket{\uparrow(\downarrow),\uparrow(\downarrow),n_{\text{c}}}$ represents the uncoupled eigenstate containing the electrons with spin up $\uparrow$ (spin down $\downarrow$) configuration singly occupying the lower $|g\rangle$ and higher $|e\rangle$ energy levels. Here, $f_{i(j)_k}$ is the corresponding amplitude of each uncoupled eigenstate. In this sense, $|\varphi_{i=1,...,4}\rangle$ and $|\varphi_{j=5,...,8}\rangle$ mainly contain around 1 and 3 photons, but, because of the double electron occupation, they are uncoupled from the cavity, see Appendix~\ref{composition_dr} for more details. 

As shown in Fig.~\ref{S_I}(d), in the long time limit of the non-Kondo regime, photons accumulate in the cavity, even for arbitrarily small light-matter coupling in a time which depends on $g_{\text{ct}}$. A larger $g_{\text{ct}}$ enhances this ground-state photon accumulation rate as shown in Fig.~\ref{state_config_2}(c). Importantly, this photon is not virtual, and will eventually decay into the electromagnetic environment, allowing for a potential observation of this effect.

\begin{figure*}[]
\includegraphics[width = 0.95\textwidth]{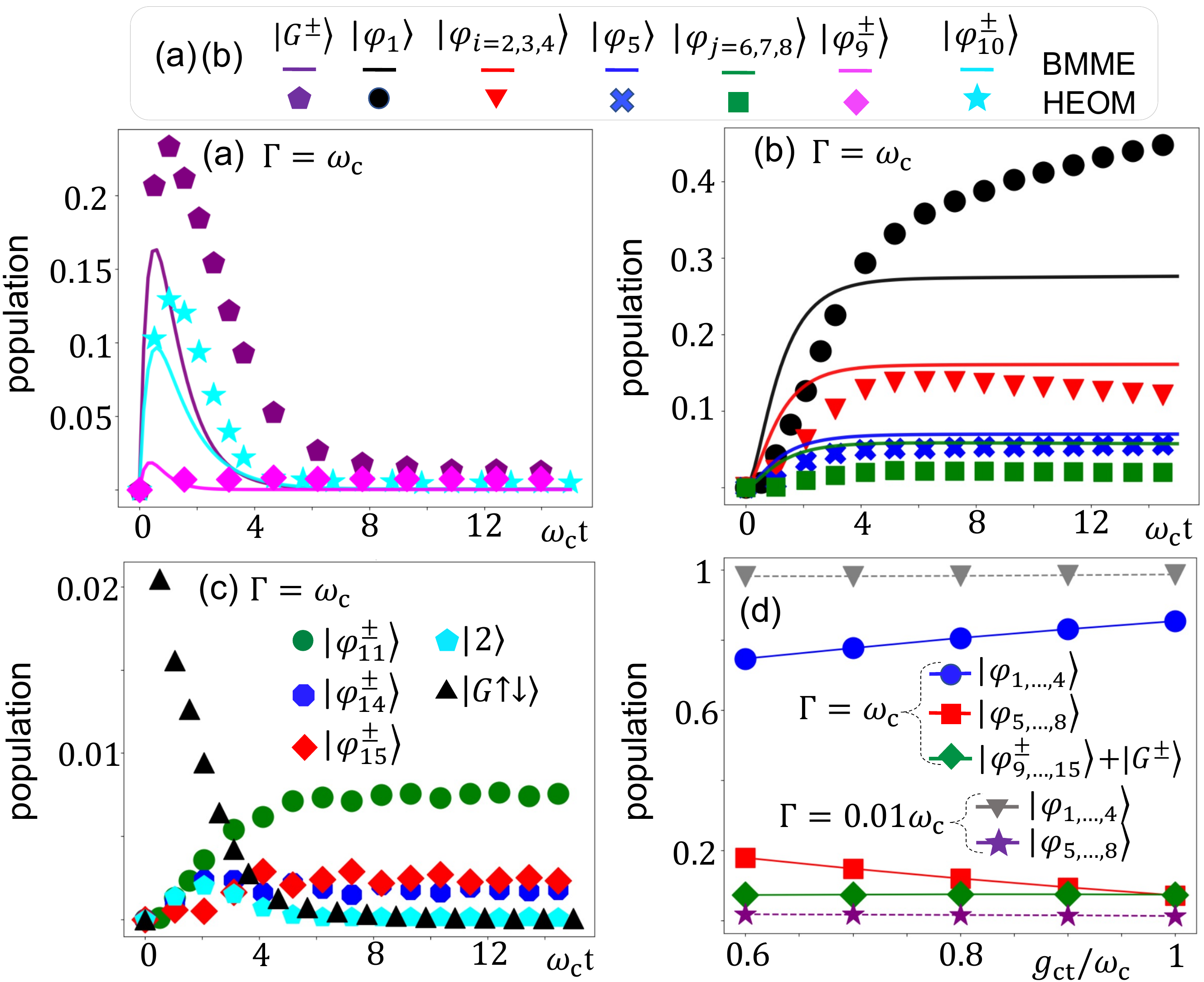}
\caption{Short-time quantum dynamics showing the failure of the BMME (solid curves) in comparison with the HEOM (markers) in the Kondo regime. Due to the Born-Markov approximation, the BMME can be seen to underestimate the populations of $|G^{\pm}\rangle$ and $|\varphi_{l=9,10}^{\pm}\rangle$ in (a) but to overestimate the populations of $|\varphi_{i=1,...,4}\rangle$  and $|\varphi_{j=5,...,8}\rangle$ in (b). (c) shows that the populations of dressed states with higher energy can only be obtained by solving a non-perturbative method such as the HEOM.
(d) shows the populations of stationary dressed states as a function of $g_{\text{ct}}$ in both the non-Kondo and Kondo regime. 
}\label{Kondo_state_config}
\end{figure*}

As we increase the ES-lead coupling to reach the Kondo regime, this strong ES-lead coupling can allow for higher-order transitions to transient states, such as $|G\uparrow\downarrow\rangle$ with double occupation in the lower state and $|2\rangle$ with no electrons but two photons, leading to the dressed states $|\varphi_{11}\rangle$ and $|\varphi_{k=13,14,15}\rangle$ as illustrated in Fig.~\ref{state_config_2}(a). Meanwhile, compared to the non-Kondo regime, such strong ES-lead coupling in the Kondo regime can drastically enhance the ground-state photon accumulation rate, as displayed in Fig.~\ref{state_config_2}(c). 
As expected, in the Kondo regime, the dynamics of the corresponding low energy states cannot be described by the Born-Markov master equation, as shown in Fig.~\ref{Kondo_state_config}(a) and Fig.~\ref{Kondo_state_config}(b). This master equation also fails to model the non-perturbative effects causing transitions to higher excited states such as $|G\uparrow\downarrow\rangle$, $|2\rangle$, and $|\varphi_{k=11,14,15}\rangle$, which did not play a role in the non-Kondo regime, as seen in  Fig.~\ref{state_config_2}(a) and Fig.~\ref{Kondo_state_config}(c).


In the long-time limit, shown in Fig.~\ref{Kondo_state_config}(d), single photon dressed states ($|\varphi_{i=1,...,4}\rangle$) dominate the steady-state occupation in the non-Kondo regime. The photon occupation of these states increases as we increase $g_{\text{ct}}$ (due to an increase in the expected photon number of intermediate transient states) giving rise to the increase of $N_{\text{ph}}$, see Fig.~\ref{S_I}(d). However, in the Kondo-regime, at weaker light-matter couplings, we see larger average populations because of access to new transition paths involving three-photon dressed states ($|\varphi_{j=5,...,8}\rangle$) and even higher energy dressed states ($|\varphi^{\pm}_{k=9,...,15}\rangle$), resulting in Kondo-enhanced dressed photon accumulation. At the same time, these pathways are suppressed as we increase the light-matter coupling, thereby isolating the ES-cavity system from the leads (also causing the suppression of the Kondo peak) and reducing the photon accumulation effect, as shown in Fig.~\ref{S_I} (d).

\subsection{Two-impurity Anderson model ultrastrongly coupled to a bosonic continuum}
\begin{figure}[]
\includegraphics[width =0.9\columnwidth]{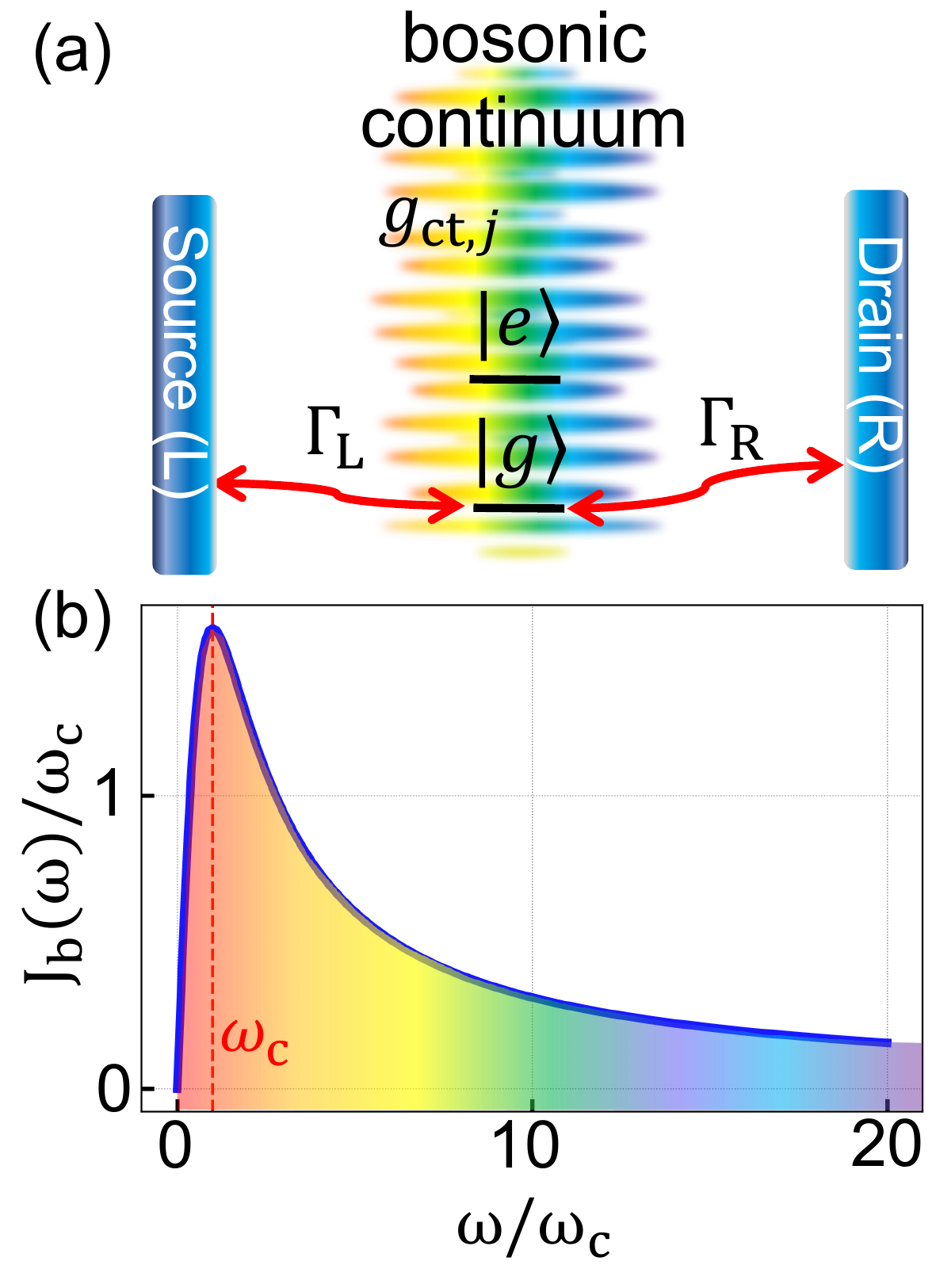}
\caption{(a) An electronic system, characterized by two states with energies $\epsilon_{g}=-6\omega_{\text{c}}$ and $\epsilon_{\text{e}}=-5\omega_{\text{c}}$, is coupled to the $j$th mode in a bosonic continuum with strength $g_{\text{ct},j}$. Simultaneously, this electronic system is situated between left and right leads, with coupling strengths $\Gamma_{\text{L}}$ and $\Gamma_{\text{R}}$, respectively. (b) The spectral density of the bosonic continuum is characterized by the Drude-Lorentz model with its peak at $\omega = \omega_{\text{c}}$.}\label{Fig5}
\end{figure}
\begin{figure}[]
\includegraphics[width = \columnwidth]{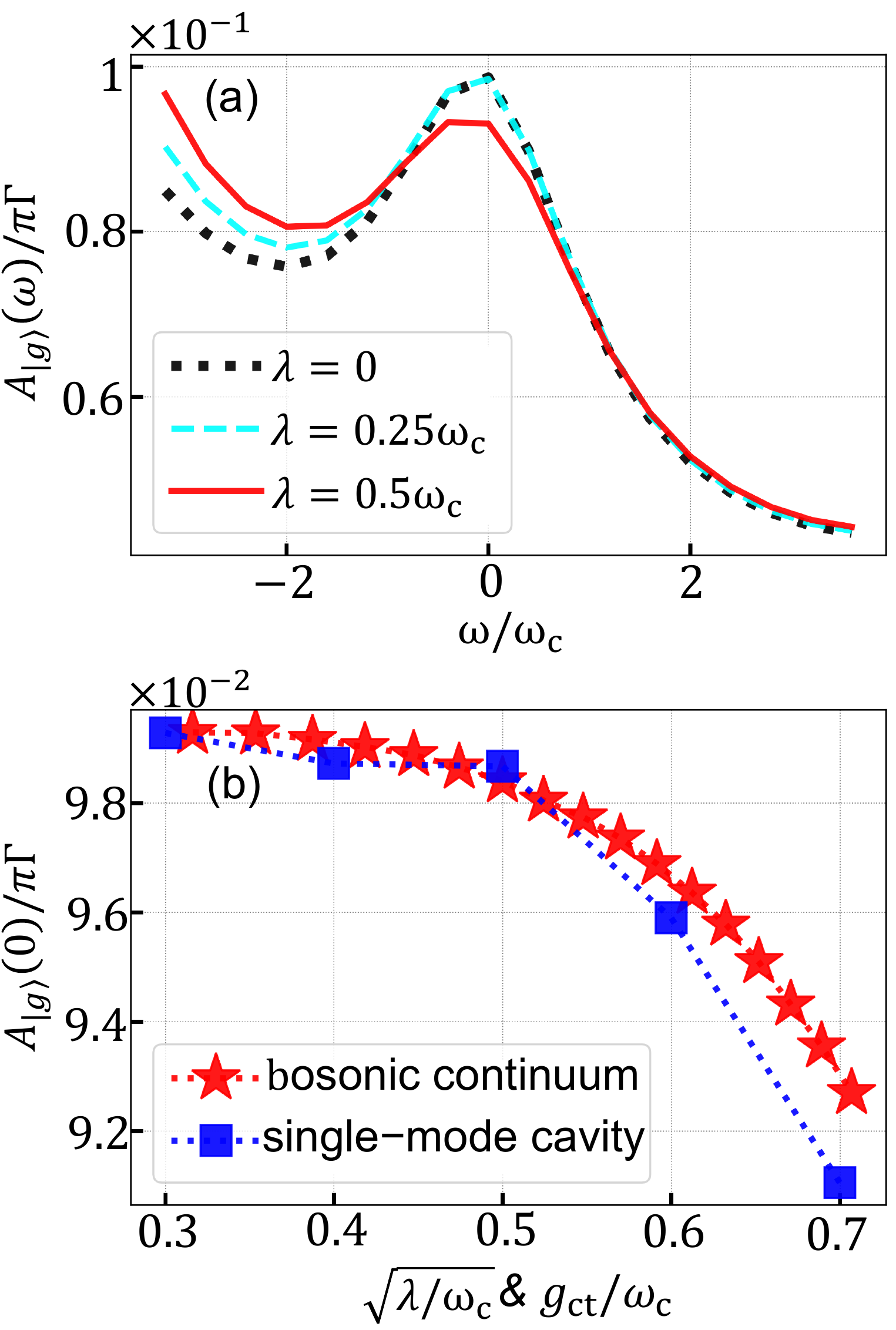}
\caption{(a) The DOS of the ES ultrastrongly coupled to a bosonic continuum with $\lambda=0$, $\lambda=0.25\omega_{\text{c}}$, and $\lambda=0.5\omega_{\text{c}}$ is represented by black dotted, light blue dashed, and red solid curves, respectively. At $\lambda=0.5\omega_{\text{c}}$, the decrease in the zero-frequency DOS peak, $A_{|g\rangle}(0)$, indicates the suppression of the Kondo effect. (b) The height of the zero-frequency DOS peak, $A_{|g\rangle}(0)$, is plotted as a function of the coupling strength. Blue-square and red-star markers represent the coupling of the ES to a single-mode with coupling strength $g_{\text{ct}}$ and to a bosonic continuum with rescaled coupling strength $\sqrt{\lambda\omega_c}$, respectively. The Kondo effect is suppressed as the effective coupling to the bosonic environment is increased.}\label{Fig6}
\end{figure}
In this section, we generalize the previous analysis based on the resonant interaction to single-mode cavities to allow for a coupling to a continuum of environmental modes. This setting is usually introduced to investigate a richer domain of phenomena such as hybridization with the bath~\cite{Lambert2019}, unraveling of multiple excitation bound states~\cite{Cirac2016}, harvesting single photons from the vacuum~\cite{Zueco2019}, developing robust long-distance entanglement protocols~\cite{Jorge2020}, and also to promote advancements in the design of quantum computing and sensing devices~\cite{Segal2023}.
Thus, in addition to single-mode cavities, we also explore systems comprised of a two-level ES ultrastrongly coupled to a bosonic continuum and positioned between two leads, as illustrated in Fig~\ref{Fig5}(a). 

The Hamiltonian describing the bosonic continuum (b) fields $b_j$ with energy $\omega_{\text{b},j}$ is denoted as 
\begin{equation}
\begin{aligned}
H_{\text{b}} = \sum_{j}\omega_{\text{b},j}b^{\dagger}_{j}b_{j}\label{H_bcon}.
\end{aligned}
\end{equation}
The interaction between the ES and the bosonic continuum can be characterized by 
\begin{equation}
\begin{aligned}
H_{\text{eb}} = \sum_{j}\sum_{\sigma=\uparrow,\downarrow}g_{\text{bt},j}(d_{g\sigma}^{\dagger}d_{e\sigma} +d_{e\sigma}^{\dagger}d_{g\sigma})(b^{\dagger}_{j}+b_{j}),\label{H_eb}
\end{aligned}
\end{equation}
Assuming the coupling is of the transverse type (t), as referred to in Eq.\ref{H_cb}, we recast the interaction Hamiltonian into the correlation function given by Eq.~\ref{Cb}. Subsequently, we characterize the spectral density of the bosonic continuum using the Drude-Lorentz model,
\begin{equation}\label{eq:Lorentz}
     J_{\text{b}}(\omega)=\frac{2\lambda W_{\text{b}}\omega}{\omega^2+W_{\text{b}}^2},
\end{equation}
where $\lambda$ represents the coupling strength between the electronic subsystem (ES) and the bosonic continuum, which functions as a bosonic reservoir with a bandwidth of $W_\text{b}$. Assume that $W_\text{b}=\omega_{\text{c}}$ as depicted in Fig.~\ref{Fig5}(b), the bosonic continuum can still have a high amount of energy in the particular mode, which can simulate the single-mode cavity in the previous example but with broadening energy distribution contributed by other modes. This assumption offers an advantage in comparing the effects of the coupled cavity between its distinctive single and multimode characteristics. 

To examine the influence of the ultrastrongly coupled bosonic continuum on the Kondo effect, we determine the DOS employing the hybrid HEOM approach, which encompasses both fermionic and bosonic hierarchy (refer to Eq.~(\ref{HEOM})). 

For the numerical implementation, we truncate the bosonic hierarchy to $N_{\text{b}}=4$ and set the number of exponents in the bosonic correlation to $m_{\text{max}}=10$ (see Eq.\ref{C_b_pade}). The convergence properties of the entire DOS concerning the truncation of $N_{\text{b}}$ and $m_{\text{\text{max}}}$ are further shown in Appendix~\ref{DOS_conv}. Upon implementing a coupling strength of $\lambda = 0.25 \omega_{\text{c}}$ for the bosonic continuum on ES, there is virtually no impact on the Kondo peak in the DOS, as demonstrated in Fig.\ref{Fig6}(a). Moreover, when we increase $\lambda$ to $0.5 \omega_{\text{c}}$, a noticeable suppression of the Kondo effect becomes apparent. 

To compare these results with the single-mode cavity case depicted in Fig.~\ref{TLC_schematic}(b), it is important to introduce a renormalization of the parameter $\lambda$ to account for the broad nature of the system-bath coupling described by Eq.~(\ref{eq:Lorentz}). This could be done by mapping, in specific parameter regimes, the overdamped spectral density in Eq.~(\ref{eq:Lorentz}) to its underdamped version \cite{IlesSmith2014,IlesSmith2016}. 

Alternatively, it is also possible to follow a more intuitive route and consider that the continuum can be approximately replaced by a single effective ancillary mode whose coupling strength to the system scales as the residue $\sqrt{\lambda\omega_c}$ of the spectral density $J_\text{b}(\omega)$ at $i\omega_c$, see Eq.~(E18) in \cite{LuoSi}. Using this scaling, we observe a remarkable similarity in the Kondo peak suppression (as a function of the effective, normalized coupling strengths to both the bath $g_\text{ct}/\omega_c$ and $\sqrt{\lambda/\omega_c}$) between the cases of a single-mode and a continuum bath. These results suggest that, even in the presence of broadening, the resonant interaction to an environmental bosonic mode continues to play a dominant role in suppressing the Kondo correlation between electrons in the ES and the leads.



\section{Conclusion}
In summary, we have shown that the ultrastrong light-matter interaction can suppress the Kondo screening and simultaneously allow for a Kondo-enhanced steady-state photon-trapping effect (via counter-rotating terms and virtual transitions), which is reduced due to lead isolation as the light-matter coupling is increased.

For the latter, our results indicate that, in the Kondo regime, an  increase of the ultrastrong coupling to the cavity does not necessarily imply stronger light-matter hybridization.
While  increasing the lead coupling allows more electrons to participate in the light-matter hybridization, simultaneously increasing the ultrastrong coupling to the cavity generates competition between cavity-induced delocalization across the electronic system (ES) and ES-lead coupling-induced delocalization of electrons between the ES and leads.
This competition gives rise to the counterintuitive decoupling effect between the system and the leads which, in turn, further reduces the photon accumulation effect and the suppression of the Kondo correlation. Therefore, our work offers a comprehensive perspective to gain both qualitative and quantitative understanding of the complex interplay between light-matter coupling and Kondo physics at the nanoscale.


We note that any physical cavity inevitably interacts with its own electromagnetic environment resulting in photon loss \cite{DeLiberato2017}. At the same time, the choice of different gauges needs to be done very carefully~\cite{Bernardis2018,DiStefano2019}. More general conditions, including non-zero photonic decay, will be considered in future work. Furthermore, the USC Kondo-photon interaction considered in this work can also be combined with different impurity configurations, e.g., coupling of both impurities to leads, additional spin-orbit coupling~\cite{Luke2022}, arbitrary mixing longitudinal and transverse couplings of the cavity photons~\cite{Neill2018}, or RKKY interaction~\cite{Spinelli2015} to generalize the possible physics observable in this interesting regime, and allow for other potential ways to tailor the competition between the formation of electron-photon dressed states and many-body entangled states in the Kondo effect.
\section{Acknowledgements}
We acknowledge Stephen Hughes and Xiao Zheng for helpful suggestions and discussions.  N.L.~acknowledges partial support from JST PRESTO through Grant No.~JPMJPR18GC, and the Information Systems Division, RIKEN, for the use of their facilities. M.C. acknowledges support from NSFC (Grants No.~12050410264 and No.~11935012) and NSAF (Grant No.~U1930403). F.N. is supported in part by: Nippon Telegraph and Telephone Corporation (NTT) Research, the Japan Science and Technology Agency (JST) [via the Quantum Leap Flagship Program (Q-LEAP), and the Moonshot R\&D Grant Number JPMJMS2061], and the Asian Office of Aerospace Research and Development (AOARD) (via Grant No. FA2386-20-1-4069). F.N. and N.L.~acknowledge the Foundational Questions Institute Fund (FQXi) via Grant No.~FQXi-IAF19-06. YNC acknowledges the support of the National Science and Technology Council, Taiwan (MOST Grants No. 111-2123-M-006-001).



\setcounter{equation}{0}
\setcounter{section}{0}
\setcounter{table}{0}
\appendix 
\begin{widetext}
\section{Canonical derivation of the density of states}\label{DOS_deriv}
The density of states (DOS) can be written in a compact form as \cite{Hewson1993,Yan2012}
\begin{equation}
A_{n\sigma}(\omega) = \frac{i}{2\pi}\int^{\infty}_{-\infty}dt e^{i\omega t}
                              \left[G^{\text{R}}_{n\sigma}(t)-G^{\text{A}}_{n\sigma}(t)\right],
                              \label{sys_DOS}
\end{equation}
in terms of the retarded $G^{\text{R}}_{n\sigma}(t)$ and the advanced $G^{\text{A}}_{n\sigma}(t)$  which depend on the system correlation functions as
\begin{equation}
\begin{array}{lll}
G^{\text{R}}_{n\sigma}(t) &=& -i\Theta(t)
                            \{ C_{d_{n\sigma}d_{n\sigma}^{\dagger}}(t) + 
                                    C_{d_{n\sigma}^{\dagger}d_{n\sigma}}(-t) \}\\
G^{\text{A}}_{n\sigma}(t) &=&  i\Theta(-t)
                            \{ C_{d_{n\sigma}d_{n\sigma}^{\dagger}}(t) + 
                                    C_{d_{n\sigma}^{\dagger}d_{n\sigma}}(-t) \}.
                                    
\end{array}
\end{equation}
Here, the Heaviside function $\Theta(t)$ ensures that causality is properly accounted for. The correlation functions of the system are given by
\begin{equation}
\begin{aligned}
C_{d_{n\sigma}^{\nu}d_{n\sigma}^{\bar{\nu}}}(t)
=\text{Tr}\left\{d_{n\sigma}^{\nu}\mathcal{G}^{\text{o}}(t)\left\{d_{n\sigma}^{\bar{\nu}}\rho_{\text{T}}(\infty)\right\}\right\} , \label{sys_correlat_func}
\end{aligned}
\end{equation} 
where $\mathcal{G}^{\text{e}/\text{o}}(t)\{\hat{o}^{\text{e}/\text{o}}\}$ represents the propagator with arbitrary parity symmetry depending on whether it is applied on an even $\hat{o}^{\text{e}}$ or odd $\hat{o}^{\text{o}}$ parity operator. Here, the density operator $(\rho)$ referred to the total system (T) is considered to be in the long-time limit 
\begin{equation}
\begin{aligned}
\rho_{\text{T}}(\infty)=\lim_{\tau \to \infty}\mathcal{G}^{\text{e}}(\tau)\{\rho_{\text{T}}(0)\},
\end{aligned}
\end{equation} 
where the initial density operator is evolved into a correlated steady-state from an initially uncorrelated condition 
\begin{equation}
\begin{aligned}
\rho_{\text{T}}(0)=\rho_{\text{f}}^{\text{th}}\otimes\rho_{\text{b}}^{\text{th}}\otimes\rho_{\text{s}}(0), 
\end{aligned}
\end{equation} 
which depends on the thermal (th) equilibrium state of the fermionic (f) bath $(\rho_{\text{f}}^{\text{th}})$, the thermal equilibrium state of the bosonic (b) bath $(\rho_{\text{b}}^{\text{th}})$, and the initial state of the system (s) $\rho_{\text{s}}(0)$. The propagator $\mathcal{G}^{\text{e}/\text{o}}(t)\{\rho_{\text{T}}(0)\}$ describing the behavior of the quantum system can be obtained by solving the hierarchical equations of motion (HEOM) as follows.

Without loss of generality, we consider a $N$-level electronic system (ES) coupled to $n_\alpha$ metallic leads. In addition, the ES is coupled to a bosonic field, which, generically, can represent a cavity mode~\cite{Mauro2016,Stockklauser2017}, phonon mode~\cite{Gustafsson2014,Manenti2017,Iorsh2020} or surface plasmon~\cite{Benz2016,Po2020}.
We consider the Hamiltonian, 
\begin{equation}
\begin{aligned}
H_{\text{T}} = H_{\text{s}} + H_{\text{f}} + H_{\text{sf}} + H_{\text{sb}} + H_{\text{b}},
\label{H_total}
\end{aligned}
\end{equation}
where $H_{\text{s}}=H_{\text{e}}+H_{\text{c}}+H_{\text{ec}}$ is the Hamiltonian of the system (s) including the electronic (e) system ($H_{\text{e}}$), cavity (c) field ($H_{\text{c}}$), and their interaction $(H_{\text{ec}})$. Here, $H_{\text{b}}$ represents the Hamiltonian of the bosonic (b) bath. We further assume that the degrees of freedom of the system can also be coupled to their bosonic environment $H_{\text{sb}}$.
The Hamiltonian of the $N$-level ES can be written as 
\begin{equation}
\begin{aligned}
H_{\text{e}} = \sum_{n}^{N}\sum_{\sigma=\uparrow,\downarrow}
        \epsilon_{n}d_{n\sigma}^{\dagger}d_{n\sigma}
      + \sum_{n}^{N}U_{n}
         d_{n\uparrow}^{\dagger}d_{n\uparrow}d_{n\downarrow}^{\dagger}d_{n\downarrow},\label{H_e}
\end{aligned}
\end{equation}
in terms of the bare energies $\epsilon_n$ and Coulomb interactions $U_n$.
The interaction Hamiltonian between the ES and leads can therefore be written as
\begin{equation}
\begin{aligned}
H_{\text{sf}} = \sum_{\alpha,k}\sum_{n=1}^{N}\sum_{\sigma=\uparrow,\downarrow}
         \Gamma_{\alpha,k}(c_{\alpha,k}^{\dagger}d_{n\sigma}+d_{n\sigma}^{\dagger}c_{\alpha,k}).\label{H_ef}
\end{aligned}
\end{equation}
Additionally, the Hamiltonian describing the cavity (c) fields $a$ with energy $\omega_{\text{c}}$ is denoted as 
\begin{equation}
\begin{aligned}
H_{\text{c}} = \omega_{\text{c}}a^{\dagger}a\label{H_c}.
\end{aligned}
\end{equation}
the interaction between the ES and the cavity field with coupling strength $g_{\text{c}}$ can be characterized by 
\begin{equation}
\begin{aligned}
H_{\text{ec}} = 
        \sum_{\sigma=\uparrow,\downarrow}
         g_{\text{c}}Q_{\sigma}(a^{\dagger}+a),\label{H_ec}
\end{aligned}
\end{equation} 
where $Q_{\sigma}$ represents the fermionic interaction operator of the ES. The form of $Q_{\sigma}$ depends on the specific coupling type of the cavity field. For example, one may have longitudinal or transverse coupling \cite{Beaudoin2016,Neill2018}. The specific form of the interaction operator $Q_{\sigma}$ we consider here is a transverse (t) coupling $g_{\text{ct}}$. 
Moreover, the couplings of the bosonic (b) environment to the interior degrees of freedom of the system (s) with coupling strength $g_{\text{sb},j}$ ($j$ labels the mode number) can be modeled by 
\begin{equation}
\begin{aligned}
H_{\text{sb}} = \sum_{j} g_{\text{sb},j}V_{\text{s}}(a_{j}^{\dagger}+a_{j}),\label{H_cb}
\end{aligned}
\end{equation}
 where $b_{j}^{\dagger}$ $(b_{j})$ represents the creation (annihilation) operator of the bosonic environment,
\begin{equation}
\begin{aligned}
H_{\text{b}}=\sum_{j}\omega_{j}b_{j}^{\dagger}b_{j}.\label{H_b}
\end{aligned}
\end{equation}
 Note that $V_{\text{s}}$ refers to the Hermitian ES-interaction operators acting on fermionic degrees of freedom. In the fermionic case, $V_{\text{s}}$ must have even parity to be compatible with charge conservation. For the system-bath interactions, we can then write down the interaction Hamiltonian in the interaction picture as
\begin{equation}
\begin{aligned}
H_{\text{s},\text{env}}(t)&=H_{\text{sf}}(t)+H_{\text{sb}}(t)
\\&
=\sum_{\alpha}\sum_{n,\sigma}
\Big[\Gamma_{\alpha}\sum_{k}c_{\alpha k}^{\dagger}e^{i\epsilon_{\alpha k}t} U_{\text{s}}^{\dagger}(t)d_{n\sigma}U_{\text{s}}(t)
-\Gamma_{\alpha}\sum_{k}c_{\alpha k}e^{-i\epsilon_{\alpha k}t} U_{\text{s}}^{\dagger}(t)d_{n\sigma}^{\dagger}U_{\text{s}}(t)\Big]\\&
+\sum_{j}g_{\text{sb},j}\Big(b_{j}^{\dagger}e^{i\omega_{j}t}+b_{j}e^{-i\omega_{j}t}\Big)
U_{\text{s}}^{\dagger}(t)V_{\text{s}}U_{\text{s}}(t)
\\&
=\sum_{\alpha}\sum_{n,\sigma}
\Big[c_{\alpha}^{\dagger}(t)d_{n\sigma}(t)
-c_{\alpha}(t)d_{n\sigma}^{\dagger}(t)\Big]
+b(t)V_{\text{s}}(t).\label{H_int}
\end{aligned}
\end{equation}%
In this frame, the system (s) density operator ($\rho$) rotates (denoted by the tilde) as $\tilde{\rho}_{\text{s}}=U_{\text{s}}^{\dagger}(t)\rho_{\text{s}}(t)U_{\text{s}}(t)$, where $U_{\text{s}}(t)=e^{iH_{\text{s}}t}$. In order to derive the HEOM of the system, we begin with using the Liouville-von Neumann equation in the interaction frame
\begin{equation}
\frac{\partial\tilde{\rho}_{\text{s}}(t)}{\partial t} = -i[H_{\text{s},\text{env}}(t),\tilde{\rho}_{\text{s}}(t)],
\end{equation}
which can be integrated to obtain the formal solution 
\begin{equation}
\begin{aligned}
\tilde{\rho}_{\text{s}}(t) = \tilde{\rho}_{\text{s}}(0)
-i\int_{0}^{t}[H_{\text{s},\text{env}}(t_{1}),\tilde{\rho}_{\text{s}}(t_{1})]dt_{1}
.\label{integration_Neum_equ}
\end{aligned}
\end{equation}
By iteratively replacing $\tilde{\rho}_{\text{s}}(t_{1})$ with $\tilde{\rho}_{\text{s}}(t)$ in Eq.~(\ref{integration_Neum_equ}), one obtains the Dyson series of the von Neumann equation in terms of the time ordering superoperator $\hhat{T}$
\begin{equation}
\begin{aligned}
\tilde{\rho}_{\text{s}}(t)=
\sum_{n_{D}=0}^{\infty}\frac{(-i)^{n_{D}}}{n_{D}!}\hhat{T}
\int_{0}^{t}\left[\prod_{i=1}^{n_{D}}d t_{i}\hhat{H}^\times_{\text{s},\text{env}}(t_i)\right]\rho_{\text{s}}(0)
,\label{Dyson_series}
\end{aligned}
\end{equation}
where $\hhat{H}^{\times}_{\text{s},\text{env}}(t)=[H_{\text{s},\text{env}}(t),\cdot]_{-}$ and $[\cdot,\cdot]_{-}$ denotes the communtator. Here, the two hats ($\hhat{\cdot}$) refers to superoperator. For the reduced Dyson series to be solvable, we apply the canonical approach in Ref.~\cite{Mauro2022} which does not require any path integral. The formal expression of Eq.~(\ref{Dyson_series}) can be written as
\begin{equation}
\begin{aligned}
\tilde{\rho}_{\text{s}}(t)=\hhat{\mathcal{G}}(t)\left[\rho_{\text{s}}^{p}(0)\right]
,\label{G_series}
\end{aligned}
\end{equation}
where $\hhat{\mathcal{G}}(t)[\cdot]$ is a superoperator which propagates the even- ($p=+$) or odd-parity ($p=-$) density operator and can be used to calculate the DOS. Its explicit form~\cite{Mauro2022} is given by 
\begin{equation}
\begin{aligned}
\hhat{\mathcal{G}}(t)[\cdot]
=&\hhat{T}_{\text{s}}\exp\Big\{-\int_{0}^{t}d t_{1}\int_{0}^{t_{1}}d t_{2}
\Big[\hhat{W}_{\text{f}}(t_1,t_2)[\cdot]
+\hhat{W}_{\text{b}}(t_1,t_2)[\cdot]\Big]\Big\}
,\label{G_op}
\end{aligned}
\end{equation}
in terms of the following fermionic superoperator
\begin{equation}
\begin{aligned}
\hhat{W}_{\text{f}}(t_1,t_2)[\cdot]
=&\sum_{p=\pm}\sum_{\alpha nn'\sigma}\sum_{\nu=\pm 1}
\Big\{
C^{\nu}_{\alpha}(t_1,t_2)
\Big[d_{n'\sigma}^{\bar{\nu}}(t_2),d_{n\sigma}^{\nu}(t_1)\cdot\Big]_{-p}+C^{\nu}_{\alpha}(t_2,t_1)
\Big[\cdot d_{n'\sigma}^{\bar{\nu}}(t_2),d_{n\sigma}^{\nu}(t_1)\Big]_{-p}
\Big\}
,\label{W_f}
\end{aligned}
\end{equation}
and the following bosonic superoperator~\cite{Lambert2019,lambert2020bofinheom}
\begin{equation}
\begin{aligned}
\hhat{W}_{\text{b}}(t_1,t_2)[\cdot]
=\Big[V_{\text{s}}(t_1),\cdot\Big]_{-}\times
\Big\{
C^{\mathbb{R}}_{\text{b}}(t_1,t_2)\Big[a(t_2),\cdot\Big]_{-}
+i C^{\mathbb{I}}_{\text{b}}(t_1,t_2)\Big[a(t_2),\cdot\Big]_{+}
\Big\}
,\label{W_b}
\end{aligned}
\end{equation}
where $p=\mp$ represents the projection on the even or odd sector. Here, the $[\cdot,\cdot]_{-}$ and $[\cdot,\cdot]_{+}$ denote the commutator and anticommutator, respectively. For simplicity, we define $\nu$ to denote the presence ($\nu = 1$) or absence ($\nu = -1$) of a Hermitian conjugation. Moreover, we define $\bar{\nu}=-\nu$. 
As it can be seen from the previous expressions, the effects of the fermionic and bosonic environments on the system are completely encoded in the correlation functions which, in the fermionic case, depend on the spectral density $J_{\text{f}_{\alpha}}(\omega)=\pi\sum_{k} \Gamma_{\alpha,k}^{2}\delta(\omega-\omega_{k})$ and the Fermi–Dirac distribution $n_{\text{f}_{\alpha}}^{\text{eq}}(\omega)=\{\exp[(\omega-\mu_{\alpha})/k_{\text{B}}T_{\text{f}_{\alpha}}]+1\}^{-1}$ as
\begin{equation}
\begin{aligned}
C^{\nu}_{\alpha}(t_{1},t_{2})
&=\text{Tr}_{\text{f}}\left[\sum_{k}
\Gamma_{\alpha,k}^{2}c_{\alpha,k}^{\nu}c_{\alpha,k}^{\bar{\nu}}\rho_{\text{f}}(0)
\right]e^{\nu i\omega (t_{1}-t_{2})}\\
&=\frac{1}{2\pi}\int_{-\infty}^{\infty} d\omega 
J_{\text{f}_{\alpha}}(\omega)\Big[\frac{1-\nu}{2}+\nu n_{\text{f}}^{\text{eq}}(\omega)
\Big]e^{\nu i\omega (t_{1}-t_{2})}.
\end{aligned}
\end{equation}
Analogously, in the bosonic case, they depend on the spectral density $J_{\text{b}}(\omega)=2\pi\sum_{j} g_{\text{sb},j}^{2}\delta(\omega-\omega_{j})$ and the Bose–Einstein distribution $n_{\text{b}}^{\text{eq}}(\omega)=\{\exp[\omega/k_{\text{B}}T_{\text{b}}]-1\}^{-1}$ as
\begin{equation}
\begin{aligned}
C_{\text{b}}(t_{1},t_{2})
&=\text{Tr}_{\text{b}}\Big[\sum_{j}
g_{\text{cb},j}^{2}\Big(b_{j}^{\dagger}b_{j}e^{i\omega (t_{1}-t_{2})}
+b_{j}b_{j}^{\dagger}e^{-i\omega (t_{1}-t_{2})}\Big)\rho_{\text{b}}(0)
\Big]\\
&=\frac{1}{2\pi}\int_{-\infty}^{\infty} d\omega 
J_{\text{b}}(\omega)\Big[n_{\text{b}}^{\text{eq}}(\omega)e^{i\omega (t_{1}-t_{2})}
+(n_{\text{b}}^{\text{eq}}(\omega)+1)e^{-i\omega (t_{1}-t_{2})}
\Big],\label{Cb}
\end{aligned}
\end{equation}
where $k_{\text{B}}$ is the Boltzmann constant and $T_{\text{f}_{\alpha}}$ $(T_{\text{b}})$ represents the absolute temperature of the $\alpha$-fermionic (bosonic) bath. Non-zero chemical potential ($\mu_{\alpha}\neq 0$) in the $\alpha$-fermionic bath can account for non-equilibrium physics.

 To proceed in the derivation of the HEOM, the bath correlation functions will be expressed as a sum of exponential terms, which allows to define an iterative procedure. Specifically, based on some spectral decomposition schemes, such as the Matsubara spectral decomposition~\cite{Shi2009} or the  Pad\'{e} spectral decomposition~\cite{Jie2011}, the correlation functions of both fermionic and bosonic environments can be written as a sum of exponentials as
\begin{equation}
\begin{aligned}
C^{\nu}_{\alpha}(\tau)=\sum_{l=0}^{l_{\text{max}}}\eta_{l}^{\nu}\exp(-\gamma_{\alpha,\nu,l}\tau), 
\label{C_f_pade}
\end{aligned}
\end{equation}%
and
\begin{equation}
\begin{aligned}
C_{\text{b}}(\tau)=\sum_{m=0}^{m_{\text{max}}}\varepsilon_{m}\exp(-\chi_{m}\tau), 
\label{C_b_pade}
\end{aligned}
\end{equation}%
where $\tau=t_{1}-t_{2}$. However, in order  to obtain a closed form for the  HEOM, the bosonic correlation function has to be further decomposed into its real 
\begin{equation}
\begin{aligned}
C^{\mathbb{R}}_{\text{b}}(\tau)=\sum_{m=0}^{N_{\mathbb{R}}}\varepsilon_{m}^{\mathbb{R}}\exp(-\chi_{m}^{\mathbb{R}}\tau), 
\label{C_b_R}
\end{aligned}
\end{equation}%
and imaginary part 
\begin{equation}
\begin{aligned}
C^{\mathbb{I}}_{\text{b}}(\tau)=\sum_{m=0}^{N_{\mathbb{I}}}\varepsilon_{m}^{\mathbb{I}}\exp(-\chi_{m}^{\mathbb{I}}\tau), 
\label{C_b_I}
\end{aligned}
\end{equation}%
unless $\chi_{m}=\chi_{m}^{\ast}$. Here, $N_{\mathbb{R}}$ refers to the number of exponentials used to obtain $C^{\mathbb{R}}_{\text{b}}(\tau)$. Similarily, for $C^{\mathbb{I}}_{\text{b}}(\tau)$. By plugging the fermionic, [Eq.~(\ref{C_f_pade})], and bosonic correlation functions [Eq.~(\ref{C_b_R}) and Eq.~(\ref{C_b_I})] in Eq.~(\ref{G_op}) and taking the time derivative, one can obtain the explicit form of the superoperators $\hhat{\mathcal{A}}^{\nu}_{n\sigma}(t)$, $\hhat{\mathcal{B}}_{\alpha l n\sigma}^{\nu}(t)$, $\hhat{\mathcal{P}}_{m}(t)$ and $\hhat{K}(t)$ which leads to
\begin{equation}
\begin{aligned}
\partial_{t}\hhat{\mathcal{G}}(t)[\cdot]
&=-i\Big\{
\sum_{\alpha l n\sigma}\sum_{\nu=\pm 1}
\Big(d^{\bar{\nu}}_{n\sigma}(t)[\cdot]-\hhat{P}_{\text{s}}[[\cdot]d^{\bar{\nu}}_{n\sigma}(t)]\Big)
\times
(-i)\int_{0}^{t}d t_1 e^{-\gamma_{\alpha\nu l}(t-t_1)}
\Big(\eta^{\nu}_{l}d^{\nu}_{n\sigma}(t)[\cdot]
+\eta^{\ast\bar{\nu}}_{l}\hhat{P}_{\text{s}}[[\cdot]d^{\nu}_{n\sigma}(t)]\Big)
\\&
+\Big[V_{\text{s}}(t),\cdot\Big]_{-}\times
\int_{0}^{t}d t_1\Big(
-i\sum_{m=0}^{N_{\mathbb{R}}}\varepsilon_{m}^{\mathbb{R}}e^{-\chi_{m}^{\mathbb{R}}(t-t_1)}\Big[V_{\text{s}}(t_1),\cdot\Big]_{-}
+\sum_{m=0}^{N_{\mathbb{I}}}\varepsilon_{m}^{\mathbb{I}}e^{-\chi_{m}^{\mathbb{I}}(t-t_1)}\Big[V_{\text{s}}(t_1),\cdot\Big]_{+}
\Big)
\Big\}\hhat{\mathcal{G}}(t)[\cdot]
\\&
=-i\Big\{
\sum_{\alpha l n\sigma}\sum_{\nu=\pm 1}
\hhat{\mathcal{A}}^{\bar{\nu}}_{n\sigma}(t)\hhat{\mathcal{B}}_{\alpha l n\sigma}^{\nu}(t)
+\hhat{K}(t)\sum_{u=R,I}\sum_{m=0}^{N_{u}}\hhat{\mathcal{P}}^{u}_{m}(t)
\Big\}\hhat{\mathcal{G}}(t)[\cdot]
\\&
=-i\Big\{
\sum_{j}\hhat{\mathcal{A}}_{j}(t)\hhat{\mathcal{B}}_{j}(t)
+\hhat{K}(t)\sum_{q}\hhat{\mathcal{P}}_{q}(t)
\Big\}\hhat{\mathcal{G}}(t)[\cdot]
\\&
=-i\sum_{j}\hhat{\mathcal{A}}_{\bar{j}}(t)\hhat{\mathcal{G}}^{(1,0)}_{j\vert}(t)[\cdot]
-i\hhat{K}(t)\sum_{q}\hhat{\mathcal{G}}^{(0,1)}_{\vert q}(t)[\cdot], 
\label{dtg}
\end{aligned}
\end{equation}%
where $\hhat{P}_{\text{s}}[\cdot]=P_{\text{s}}[\cdot]P_{\text{s}}$ represents the parity superoperator of the system with 
\begin{equation}
\begin{aligned}
P_{\text{s}}=\prod_{n\sigma}\exp[i\pi d^{\dagger}_{n\sigma}d_{n\sigma}]. 
\end{aligned}
\end{equation}%
Here the symbols $j$ and $q$ involve the multi-index $\{\alpha, l, n, \sigma, \nu \}$ and $\{u, m\}$, respectively. Therefore, we can sequentially define the fermionic and bosonic part of the first-tier auxiliary superoperator propagator as 
\begin{equation}
\begin{aligned}
\hhat{\mathcal{G}}^{(1,0)}_{j\vert}(t)[\cdot] = \hhat{\mathcal{B}}_{j}(t)\hhat{\mathcal{G}}(t)[\cdot] 
\end{aligned}
\end{equation}%
and 
\begin{equation}
\begin{aligned}
\hhat{\mathcal{G}}^{(0,1)}_{\vert q}(t)[\cdot] = \hhat{\mathcal{P}}_{q}(t)\hhat{\mathcal{G}}(t)[\cdot]. 
\end{aligned}
\end{equation}%

To obtain the higher-tier auxiliary superoperator propagators, we must repeatedly take the time derivative of the different-tier auxiliary superoperator propagators.
First, the derivative of the first-fermionic-tier and first-bosonic-tier auxiliary superoperator propagators can be expressed, respectively, as  
\begin{equation}
\begin{aligned}
\partial_{t}\hhat{\mathcal{G}}^{(1,0)}_{j\vert}(t)[\cdot]
&=\partial_{t}\Big(\hhat{\mathcal{B}}_{j}(t)\hhat{\mathcal{G}}(t)[\cdot]\Big)
\\&
=\Big[-i\Big(
\eta^{\nu}_{l}d^{\nu}_{n\sigma}(t)[\cdot]
+\eta^{\ast\bar{\nu}}_{l}\hhat{P}_{\text{s}}[[\cdot]d^{\nu}_{n\sigma}(t)]\Big)\\&
-\gamma_{\alpha\nu l}\hhat{\mathcal{B}}_{j}(t)
-i\hhat{\mathcal{B}}_{j}(t)
\Big[
\sum_{j'}\hhat{\mathcal{A}}_{\bar{j'}}(t)\hhat{\mathcal{B}}_{j'}(t)
+\hhat{K}(t)\sum_{q}\hhat{\mathcal{P}}_{q}(t)\hhat{\mathcal{B}}_{j}(t)
\Big]\hhat{\mathcal{G}}(t)[\cdot]
\\&
=\Big[
-i\hhat{\mathcal{C}}^{\nu}_{l n\sigma}(t)-\gamma_{\alpha\nu l}\hhat{\mathcal{B}}_{j}(t)
\Big]\hhat{\mathcal{G}}(t)[\cdot]
-i\Big[
\sum_{j'}\hhat{\mathcal{A}}_{\bar{j'}}(t)(-1)\hhat{\mathcal{B}}_{j}(t)\hhat{\mathcal{B}}_{j'}(t)
+\hhat{K}(t)\sum_{q}\hhat{\mathcal{P}}_{q}(t)\hhat{\mathcal{B}}_{j}(t)
\Big]\hhat{\mathcal{G}}(t)[\cdot]
\\&
=-i\hhat{\mathcal{C}}_{j}\hhat{\mathcal{G}}(t)[\cdot]
-\gamma_{j}\hhat{\mathcal{G}}^{(1,0)}_{j\vert}(t)[\cdot]
-i\sum_{j'}(-1)\hhat{\mathcal{A}}_{\bar{j'}}(t)\hhat{\mathcal{G}}^{(2,0)}_{j'j\vert}(t)[\cdot]
-i\hhat{K}(t)\sum_{q}\hhat{\mathcal{G}}^{(1,1)}_{j\vert q}(t)[\cdot]
\label{dtg(1,0)}
\end{aligned}
\end{equation}%
and 
\begin{equation}
\begin{aligned}
\partial_{t}\hhat{\mathcal{G}}^{(0,1)}_{\vert q}(t)[\cdot]
&=\partial_{t}\Big(\hhat{\mathcal{P}}_{q}(t)\hhat{\mathcal{G}}(t)[\cdot]\Big)
\\&
=\Big[
-i\Big(
\delta_{u\mathbb{R}}c^\mathbb{R}_{m}\Big[V_{\text{s}}(t),\cdot\Big]_{-}
+i\delta_{u\mathbb{I}}c^\mathbb{I}_{m}\Big[V_{\text{s}}(t),\cdot\Big]_{+}
\Big)
-\chi^{u}_{m}\hhat{\mathcal{P}}_{q}(t)\\&
-i\hhat{\mathcal{P}}_{q}(t)
\Big[
\sum_{j'}\hhat{\mathcal{A}}_{\bar{j'}}(t)\hhat{\mathcal{B}}_{j'}(t)
+\hhat{K}(t)\sum_{q'}\hhat{\mathcal{P}}_{q'}(t)
\Big]\hhat{\mathcal{G}}(t)[\cdot]
\\&
=\Big(
-i\hhat{\mathcal{M}}^{u}_{m}-\chi^{u}_{m}\hhat{\mathcal{P}}_{q}(t)
\Big)\hhat{\mathcal{G}}(t)[\cdot]
-i\Big[
\sum_{j'}\hhat{\mathcal{A}}_{\bar{j'}}(t)\hhat{\mathcal{P}}_{q}(t)\hhat{\mathcal{B}}_{j'}(t)
+\hhat{K}(t)\sum_{q'}\hhat{\mathcal{P}}_{q}(t)\hhat{\mathcal{P}}_{q'}(t)
\Big]\hhat{\mathcal{G}}(t)[\cdot]
\\&
=-i\hhat{\mathcal{M}}_{q}\hhat{\mathcal{G}}(t)[\cdot]
-\chi_{q}\hhat{\mathcal{G}}^{(0,1)}_{\vert q}(t)[\cdot]
-i\sum_{j'}\hhat{\mathcal{A}}_{\bar{j'}}(t)\hhat{\mathcal{G}}^{(1,1)}_{j'\vert q}(t)[\cdot]
-i\hhat{K}(t)\sum_{q'}\hhat{\mathcal{G}}^{(0,2)}_{\vert q'q}(t)[\cdot].
\label{dtg(0,1r)}
\end{aligned}
\end{equation}%
The superoperator propagator can also be transformed back to the Schr\"{o}dinger (S) picture by using the transformation $\hhat{\mathcal{G}}^{\text{S}}(t)[\cdot]=U_{\text{s}}(t)\hhat{\mathcal{G}}(t)[\cdot]U_{\text{s}}^{\dagger}(t)$. By recursively taking the derivative of the $N$-tier auxiliary superoperator propagator and defining the cut-off parameters as $N=N_{\text{f}}+N_{\text{b}}$ and $N_{\text{b}}=N_{\mathbb{R}}+N_{\mathbb{I}}$, one can finally obtain the following HEOM in the Schr\"{o}dinger picture
\begin{equation}
\begin{aligned}
&\partial_{t}\hhat{\mathcal{G}}^{\text{S}(N_{\text{f}},N_{\text{b}})}_{j_{1}...\vert q_{1}...}(t)[\cdot]
=\partial_{t}\Big(
\hhat{\mathcal{B}}_{j_{n_{\text{f}}}}...\hhat{\mathcal{B}}_{j_{1}}
\hhat{\mathcal{P}}_{q_{N_{\text{b}}}}...\hhat{\mathcal{P}}_{q_{1}}\hhat{\mathcal{G}}(t)[\cdot]\Big)
\\&
= -\Big(
i\mathcal{L}_{\text{s}}+\sum_{r=1}^{N_{\text{f}}}\gamma_{j_{r}}+\sum_{\text{w}=1}^{N_{\text{b}}}\chi_{q_{\text{w}}}
\Big)\hhat{\mathcal{G}}^{\text{S}(N_{\text{f}},N_{\text{b}})}_{j_{1}...\vert q_{1}...}(t)[\cdot]
-i\sum_{r=1}^{N_{\text{f}}}(-1)^{N_{\text{f}}-r+1}\hhat{\mathcal{C}}_{j_{r}}
\hhat{\mathcal{G}}^{\text{S}(N_{\text{f}}-1,N_{\text{b}})}_{...j_{r-1}j_{r+1}...\vert q_{1}...}(t)[\cdot]\\&
-i\sum_{\text{w}=1}^{N_{\text{b}}}\hhat{M}_{q_{\text{w}}}
\hhat{\mathcal{G}}^{\text{S}(N_{\text{f}},N_{\text{b}}-1)}_{j_{1}...\vert ...q_{w-1}q_{w+1}...}(t)[\cdot]
-i\sum_{j'}(-1)^{N_{\text{f}}}\hhat{\mathcal{A}}_{\bar{j'}}
\hhat{\mathcal{G}}^{\text{S}(N_{\text{f}}+1,N_{\text{b}})}_{j'j_{1}...\vert q_{1}...}(t)[\cdot]
-i\hhat{K}\sum_{q'}
\hhat{\mathcal{G}}^{\text{S}(N_{\text{f}},N_{\text{b}}+1)}_{j_{1}...\vert q'q_{1}...}(t)[\cdot],
\label{HEOM}
\end{aligned}
\end{equation}%
where $j_{r}$ and $q_{\text{w}}$ represent the $r^{\text{th}}$ and $\text{w}^{\text{th}}$ term of multi-index ensembles in terms of $\{\alpha, l, n, \sigma, \nu \}$ and $\{u, m\}$, respectively. Here, we further define the superoperator $\hhat{\mathcal{L}}_{\text{s}}[\cdot]=\left[H_{\text{s}},\cdot\right]$. As we mentioned in the main text, we suppose the cavity field to be in a high-quality cavity so that we can neglect the interaction to the external electromagnetic fields. In this case, Eq.~(\ref{HEOM}) further simplifies to 
\begin{equation}
\begin{aligned}
\partial_{t}\Big[\hhat{\mathcal{G}}^{\text{S}(N)}_{j_{1}...\vert q_{1}...}(t)[\hat{o}^{\text{e}/\text{o}}]\Big]
&= -\Big(
i\hhat{\mathcal{L}}_{\text{s}}+\sum_{r=1}^{N_{\text{f}}}\gamma_{j_{r}}
\Big)\hhat{\mathcal{G}}^{\text{S}(N)}_{j_{1}...\vert q_{1}...}(t)[\hat{o}^{\text{e}/\text{o}}]
\\&
-i\sum_{r=1}^{N_{\text{f}}}(-1)^{N_{\text{f}}-r+1}\hhat{\mathcal{C}}_{j_{r}}
\hhat{\mathcal{G}}^{\text{S}(N-1)}_{...j_{r-1}j_{r+1}...\vert q_{1}...}(t)[\hat{o}^{\text{e}/\text{o}}]
\\&
-i\sum_{j'}(-1)^{N_{\text{f}}}\hhat{\mathcal{A}}_{\bar{j'}}
\hhat{\mathcal{G}}^{\text{S}(N+1)}_{j'j_{1}...\vert q_{1}...}(t)[\hat{o}^{\text{e}/\text{o}}].\label{HEOM_ODD}
\end{aligned}
\end{equation}%

\begin{figure}[]
\includegraphics[width = \columnwidth]{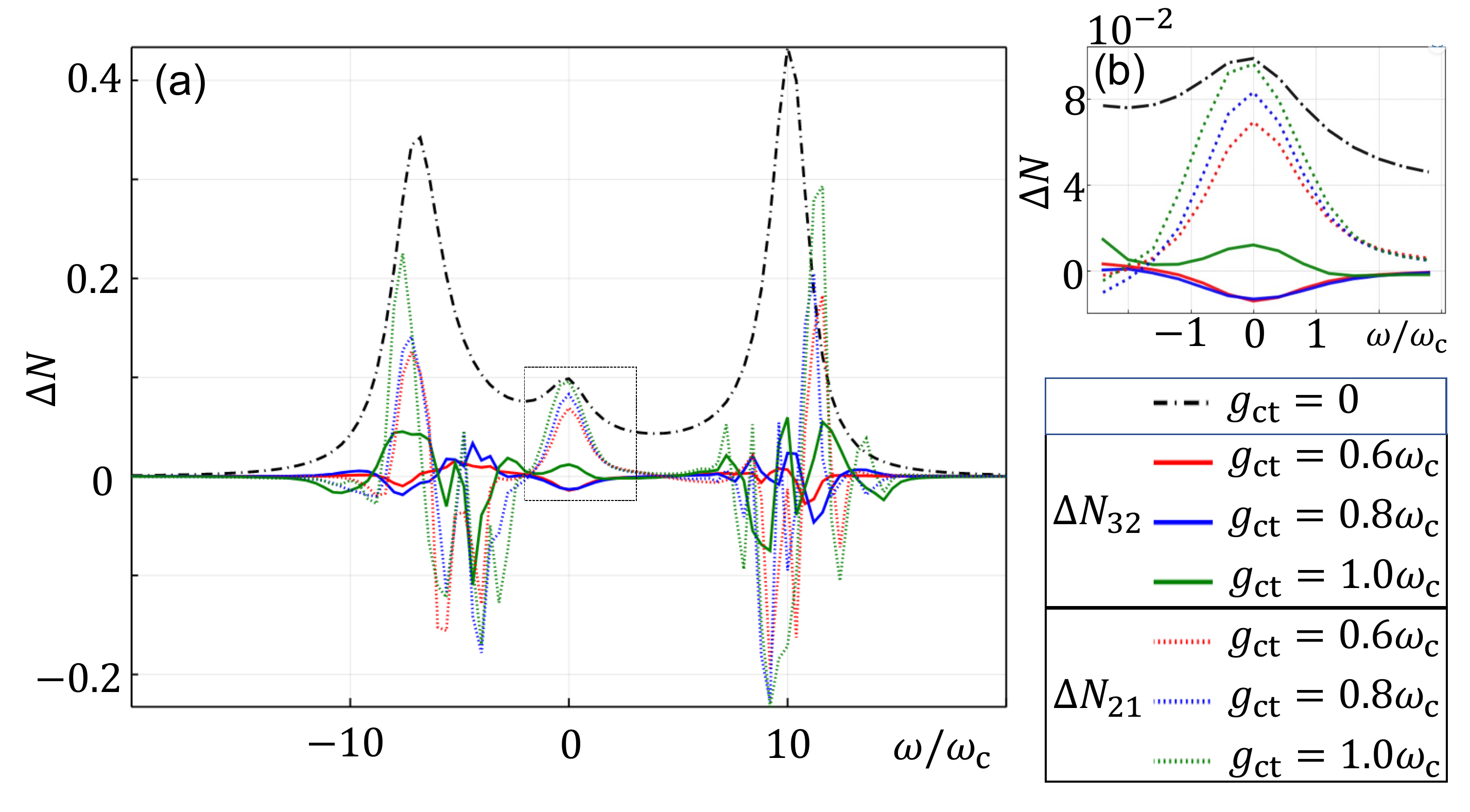}
\centering
\caption{(a) The deviations $\Delta N$ of the DOSs computed in $N=1$ from $N=2$ ($\Delta N_{21}$: dotted curve) and $N=2$ from $N=3$ ($\Delta N_{32}$ showed in solid curve) when $g_{\text{ct}}=0.6\omega_{\text{c}}$ (red), $g_{\text{ct}}=0.8\omega_{\text{c}}$ (blue), and $g_{\text{ct}}=1.0\omega_{\text{c}}$ (green). The black dash-dotted curve represents the case with $g_{\text{ct}}=0$. (b) Magnified plot in the vicinity of zero-frequency DOS.}\label{TLC_delta_N_convergence}
\end{figure}

\begin{figure}[]
\includegraphics[width = \columnwidth]{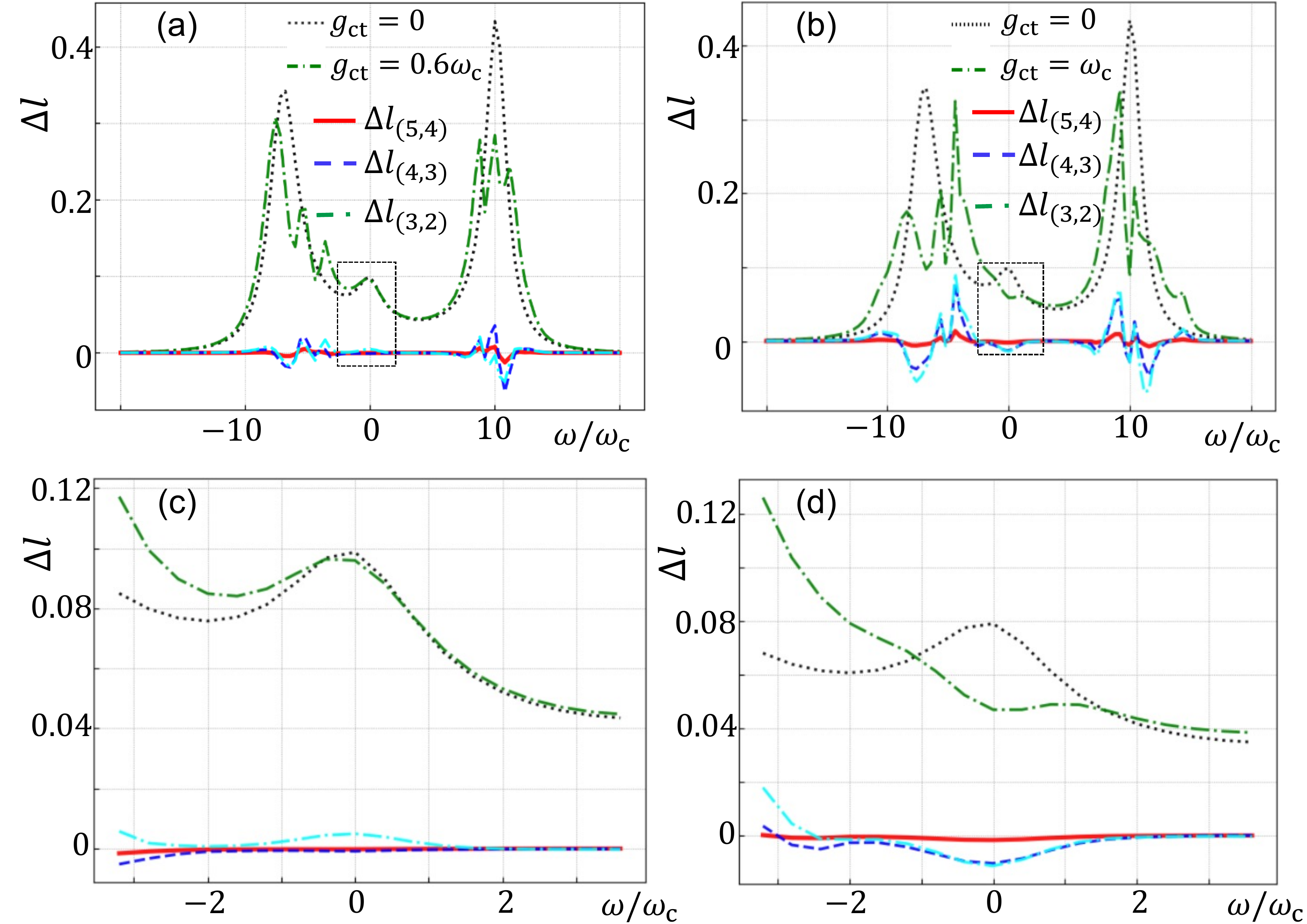}
\centering
\caption{The deviations $\Delta l$ of DOSs computed in $l_{\text{max}}=2$ from $l_{\text{max}}=3$, $l_{\text{max}}=3$ from $l_{\text{max}}=4$ and $l_{\text{max}}=4$ from $l_{\text{max}}=5$. The difference between five terms and four terms $\Delta l_{(5,4)}$, four terms and three terms $\Delta l_{(4,3)}$, three terms and two terms $\Delta l_{(3,2)}$ of the Pad\'{e} expansion for the DOS results are presented in red solid, blue dashed, and cyan dotdashed curves, respectively. It also shows that the results are convergent with $l_{\text{max}}=5$. (a)(c) and (b)(d) correspond to $g_{\text{ct}}=0.6\omega_{\text{c}}$ and $\omega_{\text{c}}$, respectively. Meanwhile, (c) and (d) display magnified plots near the zero-frequency DOS.}\label{TLC_delta_N_f_convergence}
\end{figure}
\begin{figure}[]
\includegraphics[width = \columnwidth]{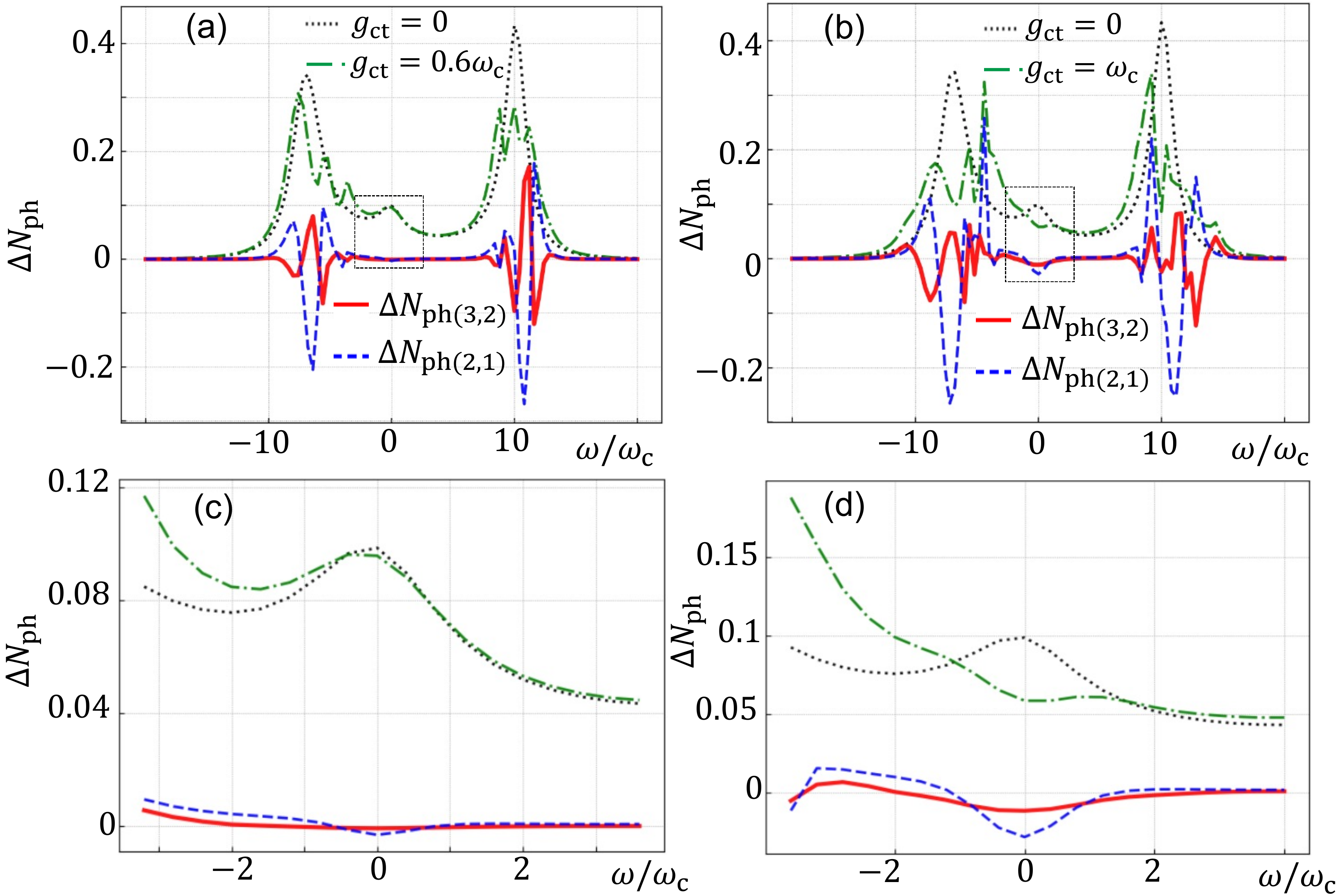}
\centering
\caption{The differences $\Delta N_{\text{ph}}$ between $N_{\text{ph}}=3$ and $N_{\text{ph}}=2$ ($\Delta N_{\text{ph}(3,2)}$) as well as $N_{\text{ph}}=2$ and $N_{\text{ph}}=1$ ($\Delta N_{\text{ph}(2,1)}$) for the DOSs are presented in red solid, blue dashed curves, respectively. It shows that the result is convergent around $\omega/\omega_{\text{c}} = 0$ with $N_{\text{ph}}=3$. (a)(c) and (b)(d) correspond to $g_{\text{ct}}=0.6\omega_{\text{c}}$ and $g_{\text{ct}}=\omega_{\text{c}}$, respectively. Furthermore, (c) and (d) showcase the magnified plots in the vicinity of the zero-frequency DOS.}\label{TLC_delta_N_ph_convergence}
\end{figure}

\begin{figure}[]
\includegraphics[width = \columnwidth]{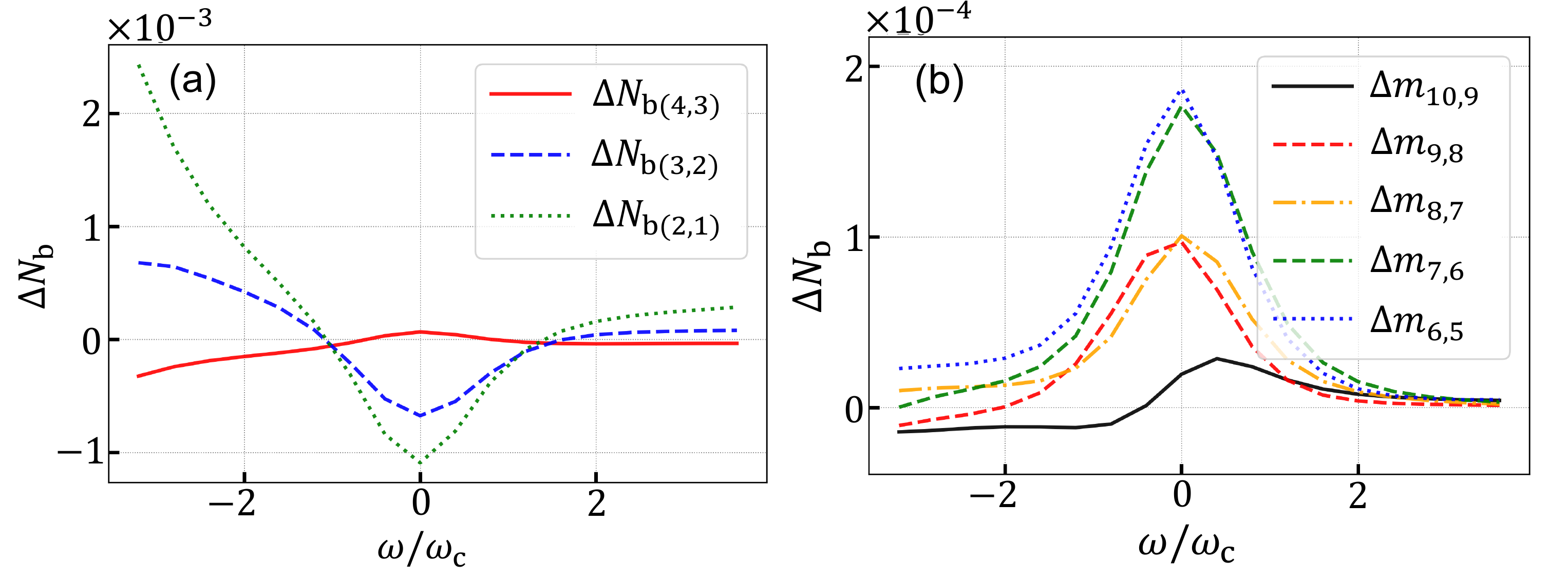}
\centering
\caption{(a) The deviations $\Delta N_{\text{b}}$ of DOSs computed at $N_{\text{b}}=4$ from $N_{\text{b}}=3$ ($\Delta N_{\text{b}(4,3)}$), $N_{\text{b}}=3$ from $N_{\text{b}}=2$ ($\Delta N_{\text{b}(3,2)}$), and $N_{\text{b}}=2$ from $N_{\text{b}}=1$ ($\Delta N_{\text{b}(2,1)}$) are presented in red solid, blue dashed, and green dotted curves, respectively. This demonstrates that the result converges around $\omega/\omega_{\text{c}} = 0$ with $N_{\text{b}}=4$. (b) The differences of the DOSs, from between $m_{\text{max}}=10$ and $N_{\text{ph}}=9$ ($\Delta m_{10,9}$) to between $m_{\text{max}}=6$ and $m_{\text{max}}=5$ ($\Delta m_{6,5}$), are shown for the analysis of possible convergence.}\label{TLC_delta_N_b_convergence}
\end{figure}

Note that the index $j_{r}\equiv\{\alpha,\nu,l,n,\sigma\}$ represents an electron on the $n$ level of the ES having spin $\sigma$ coming from $(\nu=+)$ or entering into  $(\nu=-)$ in the $\alpha$ noninteracting fermionic bath. Its Pad\'{e} bath correlation function is $C^{\nu}_{\alpha}(\tau)=\sum_{l=0}^{l_{\text{max}}}\eta_{\alpha,l}^{\nu}\exp(-\gamma_{\alpha,\nu,l}\tau)$. Here, the total fermionic cut-off $N=N_{\text{f}}$ is chosen to ensure the convergence of the HEOM. In addition, Eq.~(\ref{HEOM_ODD}) also encodes information about the system-bath interactions via the superoperators $\hhat{\mathcal{C}}_{j_{r}}$ and $\hhat{\mathcal{A}}_{\bar{j}}$ which have arbitrary parity symmetry. By solving these coupled differential equations involving the $(N+2)$ independent variables, $ \hhat{\mathcal{G}}^{\text{S}(N+1)}_{j'j_{1}...}\left[\hat{o}^{\text{e}/\text{o}}\right]$, ... , $\hhat{\mathcal{G}}^{\text{S}(1)}_{j_{1}}\left[\hat{o}^{\text{e}/\text{o}}\right]$, and $\hhat{\mathcal{G}}^{\text{S}(0)}\left[\hat{o}^{\text{e}/\text{o}}\right]$, one can obtain the required propagator $\hhat{\mathcal{G}}(t)\left[\hat{o}^{\text{e}/\text{o}}\right]=\hhat{\mathcal{G}}^{\text{S}(0)}\left[\hat{o}^{\text{e}/\text{o}}\right]$ to compute the system correlation functions in Eq.~(\ref{sys_correlat_func}), which can then be plugged in Eq~(\ref{sys_DOS}) to retrieve the DOS of the system.

\section{The convergence of the density of states}\label{DOS_conv}

To examine the degree of convergence, we compare the differences of DOSs between a certain tier and its neighboring tier of the HEOM. We take $g_{\text{ct}}=0.6\omega_{\text{c}}$ for example, as depicted in Fig.~\ref{TLC_delta_N_convergence}. The difference of DOSs between the second-tier hierarchy and first-tier hierarchy $(\Delta N_{2,1})$ is shown by the solid blue curve. The solid red curve represents the difference of the DOSs between the third-tier hierarchy and second-tier hierarchy $(\Delta N_{3,2})$. The smaller $\Delta N_{3,2}$ shows that the results are more convergent by applying the third-tier hierarchy of the HEOM. 

In addition to the truncation tier, the convergence of Pad$\acute{e}$ approximants to HEOM is shown in Fig.~\ref{TLC_delta_N_f_convergence}. By taking two Pad$\acute{e}$ terms $(l_{\text{max}}=2)$ and three Pad$\acute{e}$ terms $(l_{\text{max}}=3)$ into account for DOSs calculations, the light blue curve, labeled as $\Delta l_{3,2}$, stands for the difference between the results of $(l_{\text{max}}=2)$ and $(l_{\text{max}}=3)$. And the $\Delta l_{4,3}$ and $\Delta l_{5,4}$ are represented by blue solid and red solid curves, respectively. The relatively small deviation of $\Delta l_{5,4}$ indicates that it converges well when $l_{\text{max}}=5$. 

Furthermore, a sufficient cavity photon number should be taken into account when the electron-photon interaction reaches the USC limit. This is crucial especially for the cavity power spectrum. However, the DOS is not so sensitive to the cavity photon numbers. We thus examine how the cavity photon number affects the fermionic DOS as shown in Fig.~\ref{TLC_delta_N_ph_convergence}. Recall that the subindex ph refers to photons. It shows that a higher cavity photon number is required for achieving accurate peak height in splitting the Hubbard bands. 

The influence of different cavity photon numbers on the Kondo effect is subtle. In our simulations we found that a truncation in the Fock space at  $N_{\text{ph}}=3$ photons achieves a good convergence accuracy at the price of a reasonable computational cost.


In the bosonic continuum scenario, we also evaluate the convergence of the DOS using the same approach. We examine the difference $\Delta N_{\text{b}(i,i+1)}$ between the DOSs of the $i$th tier and its neighboring tier $(i+1)$th in the bosonic HEOM, and the difference $\Delta m_{j,j+1}$ between the $j$th and its neighboring $(j+1)$th Padé approximants of the bosonic continuum correlation function in the bosonic HEOM. The extremely small deviation values of $\Delta N_{\text{b}(4,3)}$ and $\Delta m_{(10,9)}$ indicate a high level of convergence when $N_{\text{b}}=4$ and $m_{\text{max}}=10$ as shown in Fig.~\ref{TLC_delta_N_b_convergence}.


\section{Modified master equations for degenerate systems}\label{BMME_dr}
To model a degenerate system in the USC regime, we here use the fermionic influence superoperator to derive a modified master equation, which takes the degenerate energy levels into account in its form for the Lindblad dissipators. We first decompose all operators into the eigenbasis $\ket{\varphi_{k}}$ and $\ket{\varphi_{l}}$ of the system Hilbert space in the interaction picture and define the operator $A^\nu_{\sigma,kl}(\omega)$ as
\begin{equation}
\begin{aligned}
&\sum_{\epsilon_{k}-\epsilon_{l}=\omega}\sum_{k,l}A^\nu_{\sigma,kl}(\omega)
e^{i\omega t}
=\sum_{\omega}\sum_{k,l}\sum_{n}
\bra{\varphi_{k}}d_{n\sigma}^{\nu}\ket{\varphi_{l}}\ketbra{\varphi_{k}}{\varphi_{l}},
\end{aligned}
\end{equation}%
where $\epsilon_{k}$ and $\epsilon_{l}$ are the eigenenergies corresponding to the states $\ket{\varphi_k}$ and $\ket{\varphi_l}$, respectively. With the definition of $\tau=t_{1}-t_{2}$, the expression in Eq.~(\ref{G_series}) for the related density operator $\tilde{\rho}_{\text{s}}(t)$ can be rewritten as 
\begin{equation}
\begin{aligned}
\tilde{\rho}_{\text{s}}(t)=
\hhat{T}_{\text{s}}\exp\Big\{
-\int_{0}^{t}d t_{1}\int_{0}^{\infty}d \tau
\sum_{p=\pm}\hhat{W}_{\text{f}_{p}}(t_1,\tau)[\cdot]
\Big\}\tilde{\rho}_{\text{s}}(t),\label{modified_ME}
\end{aligned}
\end{equation}%
where the fermionic influence superoperator is given by
\begin{equation}
\begin{aligned}
\hhat{W}_{\text{f}_{\pm}}(t_1,\tau)[\cdot]=
\sum_{\alpha\sigma}\sum_{\nu\omega\bar{\omega}}\sum_{klk'l'}&
\Big\{
e^{\bar{\nu}i\Delta\omega t_{1}}
\Big[
C^{\nu}_{\alpha}(\tau)e^{\nu i\omega\tau}
A^{\bar{\nu}}_{\sigma,k'l'}(\bar{\omega})
A^{\nu}_{\sigma,kl}(\omega)[\cdot]
+
C^{\nu}_{\alpha}(-\tau)e^{\bar{\nu} i\bar{\omega}\tau}
[\cdot]
A^{\bar{\nu}}_{\sigma,k'l'}(\bar{\omega})
A^{\nu}_{\sigma,kl}(\omega)\Big]\\& 
\mp e^{\nu i\Delta\omega t_{1}}\Big[
C^{\nu}_{\alpha}(\tau)e^{\nu i\bar{\omega}\tau}
A^{\nu}_{\sigma,k'l'}(\bar{\omega})
[\cdot]
A^{\bar{\nu}}_{\sigma,kl}(\omega)
+
C^{\nu}_{\alpha}(-\tau)e^{\bar{\nu} i\omega\tau}
A^{\nu}_{\sigma,k'l'}(\bar{\omega})
[\cdot]
A^{\bar{\nu}}_{\sigma,kl}(\omega)\Big]
\Big\}.
\end{aligned}
\end{equation}%
We now define $\Delta\omega=\omega-\bar{\omega}$ as the difference between the eigenenergies $\omega$ and $\bar{\omega}$. The nonsecular terms in terms of $\exp(\pm i\Delta\omega t_{1})$ can be neglected due to their fast oscillations when $\omega\neq\bar{\omega}$, Hence, by applying this secular approximation on the influence superoperator, one obtains
\begin{equation}
\begin{aligned}
\hhat{W}_{\text{f}_{\pm}}(\tau)[\cdot]=
\sum_{\alpha\sigma\nu}\sum_{\omega}\sum_{klk'l'}&
\Big\{
e^{\nu i\omega\tau}C^{\nu}_{\alpha}(\tau)
\Big[
A^{\bar{\nu}}_{\sigma,k'l'}(\omega)
A^{\nu}_{\sigma,kl}(\omega)[\cdot]
\mp
A^{\nu}_{\sigma,k'l'}(\omega)
[\cdot]
A^{\bar{\nu}}_{\sigma,kl}(\omega)
\Big]
\\&
+e^{\bar{\nu} i\omega\tau}C^{\nu}_{\alpha}(-\tau)
\Big[
[\cdot]
A^{\bar{\nu}}_{\sigma,k'l'}(\omega)
A^{\nu}_{\sigma,kl}(\omega)
\mp
A^{\nu}_{\sigma,k'l'}(\omega)
[\cdot]
A^{\bar{\nu}}_{\sigma,kl}(\omega)
\Big]
\Big\}.
\end{aligned}
\end{equation}%
Using the expression above, we write the time derivative of Eq.~(\ref{modified_ME}) in the Schr\"{o}dinger frame as
\begin{equation}
\begin{aligned}
\partial_t\rho_{\text{s}}(t)&=-i[H_{\text{s}},\rho_{\text{s}}(t)]
-\sum_{p=\pm}\int_{0}^{\infty}d\tau\hhat{W}_{\text{f}_{p}}(\tau)\rho^{p}_{\text{s}}(t)\\&
=-i[H_{\text{s}},\rho_{\text{s}}(t)]
+\sum_{\alpha\sigma\nu}\sum_{\omega,p=\pm}
\Big\{
\Gamma^\nu_{\alpha}(\omega)
[-A^{\bar{\nu}}_{\sigma}({\omega}),A^{\nu}_{\sigma}({\omega})\rho_{\text{s}}^{p}]_{-p}
+\bar{\Gamma}^{\nu}_{\sigma}(\omega)
[-\rho_{\text{s}}^{p}A^{\bar{\nu}}_{\sigma}({\omega}), A^{\nu}_{\sigma}({\omega})]_{-p}
\Big\},\label{context_ME}
\end{aligned}
\end{equation}%
where
\begin{equation}
\begin{aligned}
\Gamma^\nu_{\alpha}(\omega)&=&\int_0^\infty d\tau
~C^\nu_{\alpha}(\tau)\exp(\nu i\omega \tau),
\end{aligned}
\end{equation}
and 
\begin{equation}
\begin{aligned}
\bar{\Gamma}^{\nu}_{\alpha}(\omega)&=&\int_0^\infty d\tau
~C^\nu_{\alpha}(-\tau)\exp(-\nu i\omega \tau).
\end{aligned}
\end{equation}
Here, the information about degenerate transitions is encoded in the decomposed operator, $A^{\nu}_{\sigma}({\omega})=\sum_{kl}A^{\nu}_{\sigma,kl}({\omega})$. By neglecting the Lamb shift due to its small value in Eq.~(\ref{context_ME}), one can obtain the modified master equation in Lindblad form as
\begin{equation}
\begin{aligned}
&\partial_t\rho_{\text{s}}(t)=-i[H_{\text{s}},\rho_{\text{s}}(t)]
+\sum_{\alpha\sigma\nu}\sum_{\omega,p=\pm}
\Big\{
\gamma^\nu_{\alpha}(\omega)
\Big[
\pm A^{\nu}_{\sigma}({\omega})\rho_{\text{s}}^{p}A^{\bar{\nu}}_{\sigma}({\omega})
-\frac{1}{2}[A^{\bar{\nu}}_{\sigma}({\omega})A^{\nu}_{\sigma}({\omega}),\rho_{\text{s}}^{p}]_{+}
\Big]
\Big\},
\end{aligned}
\end{equation}%
where 
\begin{equation}
\begin{aligned}
\gamma_{\alpha}^{\nu}=\Gamma^\nu_{\alpha}(\omega)+\bar{\Gamma}^{\nu}_{\alpha}(\omega)
=2\pi J_{\text{f}}(-\omega)n^{\text{eq}}_\text{f}(-\nu\omega).
\end{aligned}
\end{equation}%
The superindex eq refers to equilibrium. In this section we have derived the modified master equations for degenerate system via the canonical approach~\cite{Mauro2022}.

\begin{table}
  \caption{The composition of the dressed states when $g_{\text{ct}}=0.6\omega_{\text{c}}$.}
  \label{component_gc06}
  \includegraphics[width=\linewidth]{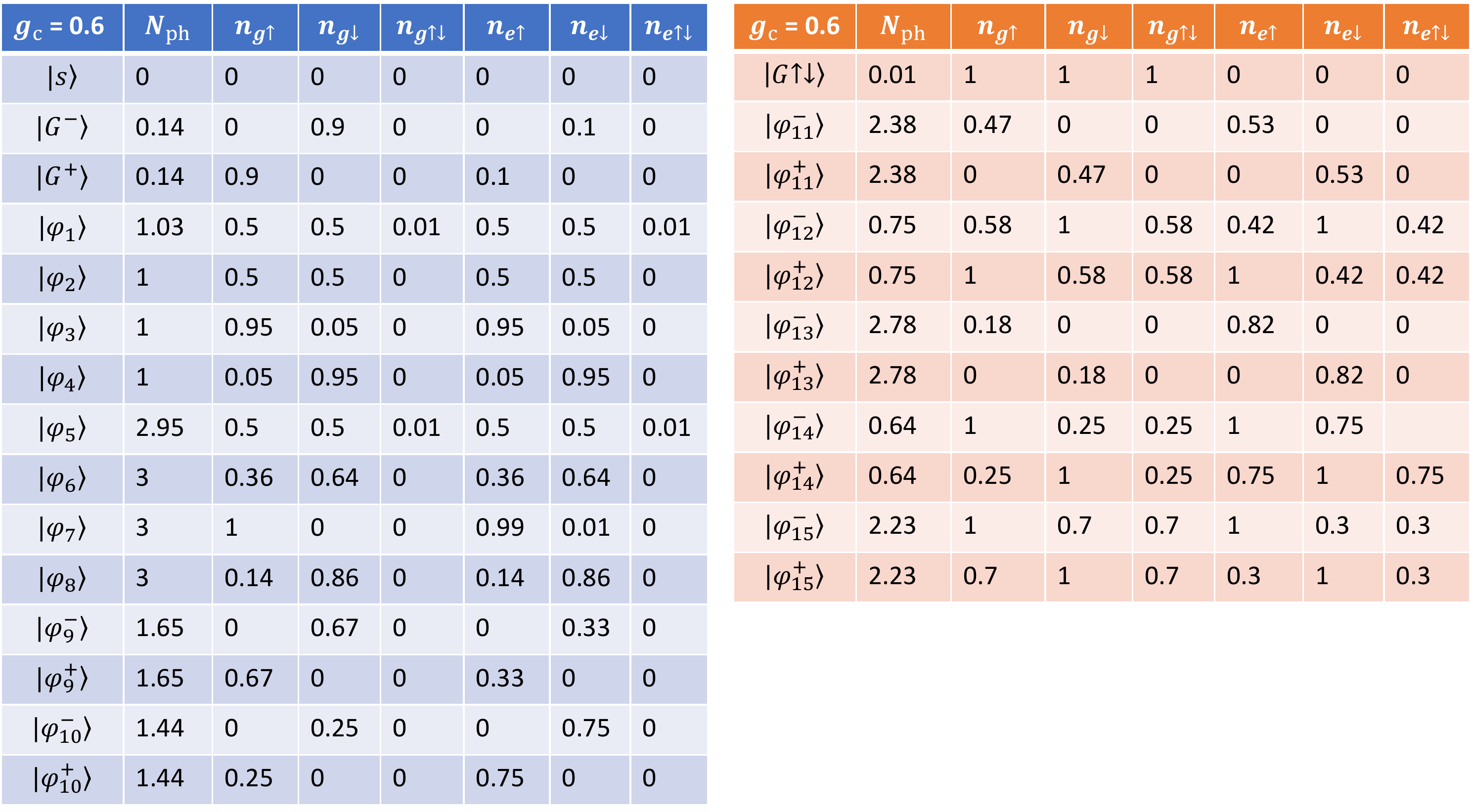}
\end{table}

\begin{table}
  \caption{The composition of the dressed states when $g_{\text{ct}}=\omega_{\text{c}}$}
  \label{component_gc1}
  \includegraphics[width=\linewidth]{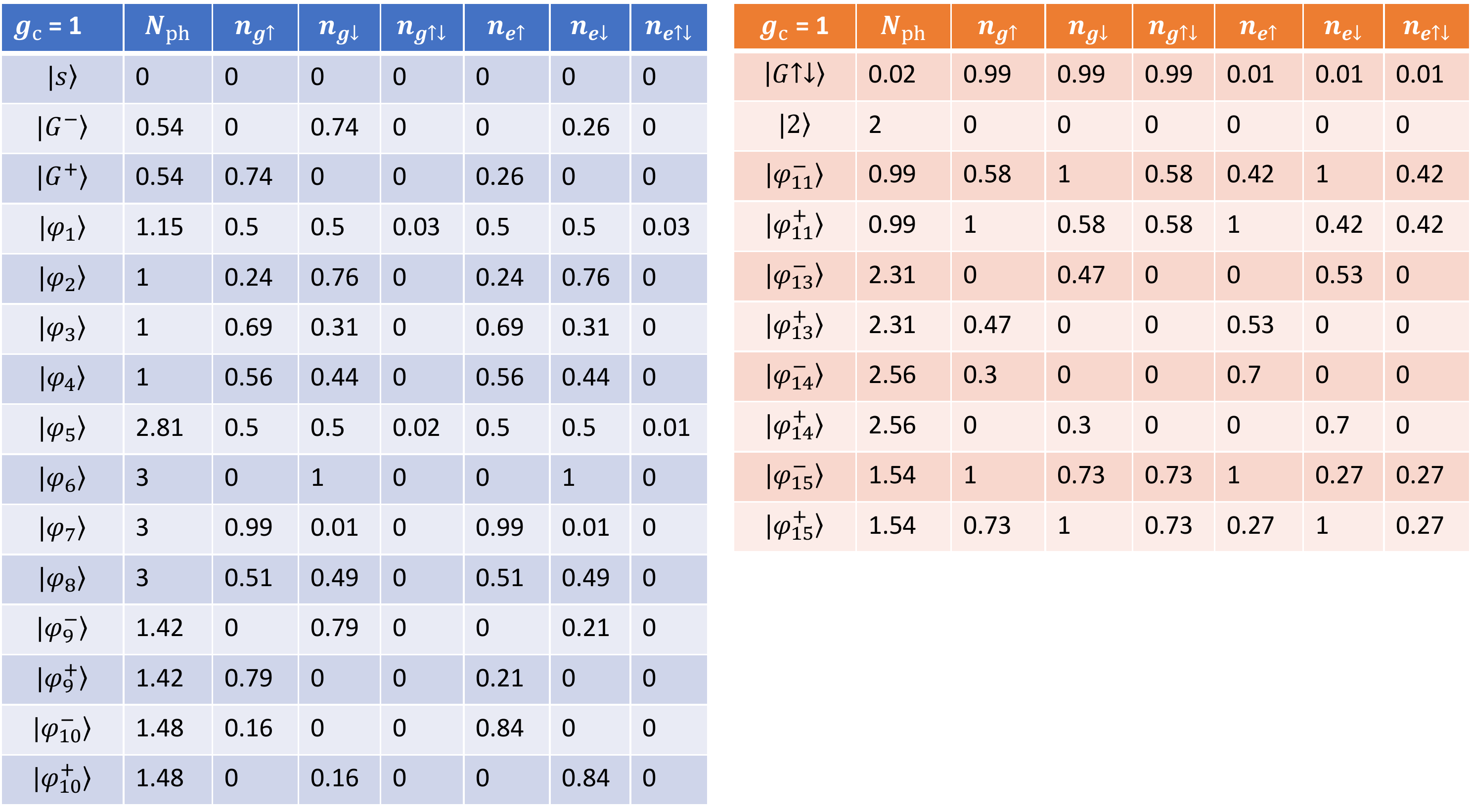}
\end{table}

\section{The composition of the dressed states}\label{composition_dr}
To evaluate how the quantum system evolves, it is useful to investigate the evolution of its eigenstates (with a coupled cavity in dressed states). For the case analyzed in the main text, there are 64 eigenstates in total since we truncate the cavity Fock space to three states. However, here we only show the eigenstates which contribute more to the quantum behavior. Each contains different bare components as presented by the different expectation values shown in Table.~\ref{component_gc06} ($g_{\text{ct}}=0.6\omega_{\text{c}}$) and Table.~\ref{component_gc1} ($g_{\text{ct}}=\omega_{\text{c}}$). Here, the number of occupied electrons on states $\ket{i}$ ($i=e,g$), with spin configuration $\sigma=\uparrow,\downarrow$ or double occupation $\sigma=\uparrow\downarrow$, can be determined by $n_{i\sigma}=\text{Tr}(\hat{n}_{i\sigma}\rho_{\text{s}})$. The average photon number of each eigenstate is given by $N_{\text{ph}}=\text{Tr}(a^{\dagger}a\rho_{\text{s}})$.

\section{Reabsorption of the Diamagnetic term}\label{a_diamagnetic}
In this section, we show that the addition of a diamagnetic term to the Hamiltonian describing the light-matter interaction  can be effectively modeled as a renormalization of the light frequency, the light-matter interaction, and the zero-point energy \cite{Anton2019}.  Note that this term commonly appears in the standard formulation of the light-matter Hamiltonian in the Coulomb Gauge, but recent works have shown that a better approximation to the full untruncated light-matter interaction can be found using the dipole gauge ~\cite{Bernardis2018,DiStefano2019,Garziano2020,Yoshihara2022}. Following~\cite{Bernardis2018}, the transformation to the dipole Gauge leads to a different renormalization of the parameters than the ones given below, and alters the system-cavity interaction to be proportional to a different cavity quadrature $P=i(a^\dagger-a)$. However, this does not change the physics of our model. One might consider the influence of the transformation on the system-lead coupling, but arguably this should only induce a local renormalization of the matter potential where it is interacting with the cavity, and not cause photonic-dressing of the coupling.
To deal with the diamagnetic term in the Coulomb gauge, we focus on the bosonic sector of an abstract Hamiltonian describing the interaction between matter and a bosonic mode $a$ with the additional presence of a $(a+a^\dagger)^2$ diamagnetic (D) energy, i.e., 
\begin{equation}
    H_{\text{D}} =\omega a^\dagger a+D(a+a^\dagger)^2+g(a+a^\dagger)\hat{s}\;.
\end{equation}
Here, $\hat{s}$ is the matter coupling-operator, while $\omega$, $D$, and $g$ represent, respectively, the bosonic frequency, the strength of the diamagnetic potential, and the  strength of the light-matter interaction. It is useful to start by rewriting this Hamiltonian as
\begin{equation}
\label{eq:D}
    H_{\text{D}} =(\omega+2D) a^\dagger a+D[a^2+(a^\dagger)^2]+g(a+a^\dagger)\hat{s}+D\;.
\end{equation}
To make progress, we consider a change of variables by defining the mode $\tilde{a}$ through the following Bogoliubov trasformation
\begin{equation}
\begin{array}{lll}
\tilde{a}&=&\cosh(\lambda)a+\sinh(\lambda)a^\dagger\\
\tilde{a}^\dagger&=&\cosh(\lambda)a^\dagger+\sinh(\lambda)a\;,
\end{array}
\end{equation}
designed to satisfy the constraint  $[\tilde{a},\tilde{a}^\dagger]=\cosh^2(\lambda)-\sinh^2(\lambda)=1$. This transformation allows to write the Hamiltonian $H_\text{D}$  as
\begin{equation}
\label{eq:bog}
    H_{\text{D}} =\tilde{\omega} \tilde{a}^\dagger \tilde{a}+\tilde{g}(\tilde{a}+\tilde{a}^\dagger)\hat{s}+\tilde{E}_0\;,
\end{equation}
in terms of the renormalized frequency $\tilde{\omega}$, renormalized coupling $\tilde{g}$, and renormalized zero-point energy $\tilde{E}_0$. The equivalence in Eq.~(\ref{eq:bog}) can be shown explicitly by rewriting it as
\begin{equation}
\begin{aligned}
\label{eq:toCompare}
    H_{\text{D}} &=\tilde{\omega}[\cosh^2(\lambda)+\sinh^2(\lambda)] \tilde{a}^\dagger \tilde{a}
    +\tilde{\omega}[\cosh(\lambda)\sinh(\lambda)][\tilde{a}^2+(\tilde{a}^\dagger)^2]\hat{s}\\&
    +\tilde{g}[\cosh(\lambda)+\sinh(\lambda)](\tilde{a}+\tilde{a}^\dagger)
    +\sinh^2(\lambda)\tilde{\omega}+\tilde{E}_0,
\end{aligned}
\end{equation}
and then comparing it to Eq.~(\ref{eq:D}) term by term. This produces the following constraints
\begin{equation}
\label{eq:toCompare2}
\begin{array}{lll}
\tilde{\omega}\cosh(2\lambda)&=&\omega+2D\\
    \tilde{\omega}\sinh(2\lambda)&=&2D\\
    
    \tilde{g}[\cosh(\lambda)+\sinh(\lambda)]&=&g\\
    \tilde{E}_0+\sinh^2(\lambda)\tilde{\omega}&=&D\;,
    \end{array}
\end{equation}
where we used the identities $[\cosh^2(\lambda)+\sinh^2(\lambda)]=\cosh(2\lambda)$ and $2\cosh(\lambda)\sinh(\lambda)=\sinh(2\lambda)$. The ratio between the first two lines of Eq.~(\ref{eq:toCompare2}) implies that $2\lambda=\text{arctanh}[2D/(\omega+2D)]$, which, inserted in the second row of Eq.~(\ref{eq:toCompare2}) gives 
\begin{equation}
\tilde{\omega}\sinh\left\{\text{arctanh}\left[\frac{2D}{\omega+2D}\right]\right\}=2D\;.
\end{equation}
Using the identity $\sinh\left[\text{arctanh}(x)\right]=x/\sqrt{1-x^2}$, we finally obtain the expression for the renormalized frequency as
\begin{equation}
\label{eq:renFreq}
    \tilde{\omega}=\sqrt{\omega^2+4\omega D}\;.
\end{equation}
We can now take the first and second lines in Eq.~(\ref{eq:toCompare2}), and write them as
\begin{equation}
\begin{array}{lll}
\tilde{\omega}[\cosh^2(\lambda)+\sinh^2(\lambda)]&=&\omega+2D\\
    2\tilde{\omega}\cosh(\lambda)\sinh(\lambda)&=&2D\;,
    \end{array}
\end{equation}
and then sum them to find the relation
\begin{equation}
\label{eq:toCompare3}
\begin{array}{lll}
\displaystyle\frac{1}{[\cosh(\lambda)+\sinh(\lambda)]^2}&=&\displaystyle\frac{\tilde{\omega}}{\omega+4D}\;.
    \end{array}
\end{equation}
This expression can be used in the third row of Eq.~(\ref{eq:toCompare2}) which, divided by $\tilde{\omega}$, results in
\begin{equation}
\begin{array}{lll}
\displaystyle\frac{\tilde{g}}{\tilde{\omega}}&=&\displaystyle\frac{g}{\tilde{\omega}}\frac{1}{\cosh(\lambda)+\sinh(\lambda)}=\displaystyle\frac{g}{\sqrt{\tilde{\omega}}\sqrt{\omega+4D}}\;.
\end{array}
\end{equation}
Using Eq.~(\ref{eq:renFreq}) for the renormalized frequency, we finally obtain the expression for the normalized light-matter coupling
\begin{equation}
\begin{array}{lll}
\displaystyle\frac{\tilde{g}}{\tilde{\omega}}&=&\displaystyle\frac{g}{\omega}\left(1+\frac{4D}{\omega}\right)^{-3/4}\;.
\end{array}
\end{equation}
We can finish by deriving the expression for the renormalized zero-point energy. This can be done by considering the identities $\sinh(\lambda)=[\cosh(2\lambda)-1]/2$ and $\cosh[\text{arctanh(x)}]=1/\sqrt{1-x^2}$ which, inserted in Eq.~(\ref{eq:toCompare2}), lead to
\begin{equation}
\begin{array}{lll}
    \tilde{E}_0&=&\displaystyle D-\frac{\tilde{\omega}}{2}\left(\frac{\omega+2D}{\sqrt{\omega^2+4\omega D}}-1\right)=\displaystyle\frac{\tilde{\omega}-\omega}{2}\;,
    \end{array}
\end{equation}
where we used the expression for the renormalized frequency in Eq.~(\ref{eq:renFreq}).\\

In summary, the renormalized parameters needed to reabsorb the diamagnetic potential are given by
\begin{equation}
\begin{aligned}
    \tilde{\omega}=\sqrt{\omega^2+4\omega D}\;,~~~~\frac{\tilde{g}}{\tilde{\omega}}=\frac{g}{\omega}\left(1+\frac{4D}{\omega}\right)^{-3/4}\;,~~~~
        \tilde{E}_0= (\tilde{\omega}-\omega)/2\;.
\end{aligned}
\end{equation}
It has been shown that the diamagnetic term can cause a shift in the energy levels of the system, as well as changes in the coupling strengths between the light and the matter.
\end{widetext}
\nocite{apsrev41Control}
\bibliographystyle{apsrev4-2}
%


\begin{thebibliography}{94}%
\makeatletter
\providecommand \@ifxundefined [1]{%
 \@ifx{#1\undefined}
}%
\providecommand \@ifnum [1]{%
 \ifnum #1\expandafter \@firstoftwo
 \else \expandafter \@secondoftwo
 \fi
}%
\providecommand \@ifx [1]{%
 \ifx #1\expandafter \@firstoftwo
 \else \expandafter \@secondoftwo
 \fi
}%
\providecommand \natexlab [1]{#1}%
\providecommand \enquote  [1]{``#1''}%
\providecommand \bibnamefont  [1]{#1}%
\providecommand \bibfnamefont [1]{#1}%
\providecommand \citenamefont [1]{#1}%
\providecommand \href@noop [0]{\@secondoftwo}%
\providecommand \href [0]{\begingroup \@sanitize@url \@href}%
\providecommand \@href[1]{\@@startlink{#1}\@@href}%
\providecommand \@@href[1]{\endgroup#1\@@endlink}%
\providecommand \@sanitize@url [0]{\catcode `\\12\catcode `\$12\catcode
  `\&12\catcode `\#12\catcode `\^12\catcode `\_12\catcode `\%12\relax}%
\providecommand \@@startlink[1]{}%
\providecommand \@@endlink[0]{}%
\providecommand \url  [0]{\begingroup\@sanitize@url \@url }%
\providecommand \@url [1]{\endgroup\@href {#1}{\urlprefix }}%
\providecommand \urlprefix  [0]{URL }%
\providecommand \Eprint [0]{\href }%
\providecommand \doibase [0]{https://doi.org/}%
\providecommand \selectlanguage [0]{\@gobble}%
\providecommand \bibinfo  [0]{\@secondoftwo}%
\providecommand \bibfield  [0]{\@secondoftwo}%
\providecommand \translation [1]{[#1]}%
\providecommand \BibitemOpen [0]{}%
\providecommand \bibitemStop [0]{}%
\providecommand \bibitemNoStop [0]{.\EOS\space}%
\providecommand \EOS [0]{\spacefactor3000\relax}%
\providecommand \BibitemShut  [1]{\csname bibitem#1\endcsname}%
\let\auto@bib@innerbib\@empty
\bibitem [{\citenamefont {Smith}\ \emph {et~al.}(2019)\citenamefont {Smith},
  \citenamefont {Kim}, \citenamefont {Pollmann},\ and\ \citenamefont
  {Knolle}}]{Smith2019}%
  \BibitemOpen
  \bibfield  {author} {\bibinfo {author} {\bibfnamefont {A.}~\bibnamefont
  {Smith}}, \bibinfo {author} {\bibfnamefont {M.~S.}\ \bibnamefont {Kim}},
  \bibinfo {author} {\bibfnamefont {F.}~\bibnamefont {Pollmann}},\ and\
  \bibinfo {author} {\bibfnamefont {J.}~\bibnamefont {Knolle}},\ }\bibfield
  {title} {\bibinfo {title} {Simulating quantum many-body dynamics on a current
  digital quantum computer},\ }\href
  {https://doi.org/10.1038/s41534-019-0217-0} {\bibfield  {journal} {\bibinfo
  {journal} {npj Quantum Inf.}\ }\textbf {\bibinfo {volume} {5}},\ \bibinfo
  {pages} {106} (\bibinfo {year} {2019})}\BibitemShut {NoStop}%
\bibitem [{\citenamefont {Carleo}\ and\ \citenamefont
  {Troyer}(2017)}]{Carleo2017}%
  \BibitemOpen
  \bibfield  {author} {\bibinfo {author} {\bibfnamefont {G.}~\bibnamefont
  {Carleo}}\ and\ \bibinfo {author} {\bibfnamefont {M.}~\bibnamefont
  {Troyer}},\ }\bibfield  {title} {\bibinfo {title} {Solving the quantum
  many-body problem with artificial neural networks},\ }\href
  {https://doi.org/10.1126/science.aag2302} {\bibfield  {journal} {\bibinfo
  {journal} {Science}\ }\textbf {\bibinfo {volume} {355}},\ \bibinfo {pages}
  {602} (\bibinfo {year} {2017})}\BibitemShut {NoStop}%
\bibitem [{\citenamefont {Gorshkov}\ \emph {et~al.}(2013)\citenamefont
  {Gorshkov}, \citenamefont {Nath},\ and\ \citenamefont {Pohl}}]{Thomas2013}%
  \BibitemOpen
  \bibfield  {author} {\bibinfo {author} {\bibfnamefont {A.~V.}\ \bibnamefont
  {Gorshkov}}, \bibinfo {author} {\bibfnamefont {R.}~\bibnamefont {Nath}},\
  and\ \bibinfo {author} {\bibfnamefont {T.}~\bibnamefont {Pohl}},\ }\bibfield
  {title} {\bibinfo {title} {Dissipative many-body quantum optics in {R}ydberg
  media},\ }\href {https://doi.org/10.1103/PhysRevLett.110.153601} {\bibfield
  {journal} {\bibinfo  {journal} {Phys. Rev. Lett.}\ }\textbf {\bibinfo
  {volume} {110}},\ \bibinfo {pages} {153601} (\bibinfo {year}
  {2013})}\BibitemShut {NoStop}%
\bibitem [{\citenamefont {Mart{\'{\i}}nez}\ \emph {et~al.}(2019)\citenamefont
  {Mart{\'{\i}}nez}, \citenamefont {L{\'{e}}ger}, \citenamefont {Gheeraert},
  \citenamefont {Dassonneville}, \citenamefont {Planat}, \citenamefont
  {Foroughi}, \citenamefont {Krupko}, \citenamefont {Buisson}, \citenamefont
  {Naud}, \citenamefont {Hasch-Guichard}, \citenamefont {Florens},
  \citenamefont {Snyman},\ and\ \citenamefont {Roch}}]{PuertasMartnez2019}%
  \BibitemOpen
  \bibfield  {author} {\bibinfo {author} {\bibfnamefont {J.~P.}\ \bibnamefont
  {Mart{\'{\i}}nez}}, \bibinfo {author} {\bibfnamefont {S.}~\bibnamefont
  {L{\'{e}}ger}}, \bibinfo {author} {\bibfnamefont {N.}~\bibnamefont
  {Gheeraert}}, \bibinfo {author} {\bibfnamefont {R.}~\bibnamefont
  {Dassonneville}}, \bibinfo {author} {\bibfnamefont {L.}~\bibnamefont
  {Planat}}, \bibinfo {author} {\bibfnamefont {F.}~\bibnamefont {Foroughi}},
  \bibinfo {author} {\bibfnamefont {Y.}~\bibnamefont {Krupko}}, \bibinfo
  {author} {\bibfnamefont {O.}~\bibnamefont {Buisson}}, \bibinfo {author}
  {\bibfnamefont {C.}~\bibnamefont {Naud}}, \bibinfo {author} {\bibfnamefont
  {W.}~\bibnamefont {Hasch-Guichard}}, \bibinfo {author} {\bibfnamefont
  {S.}~\bibnamefont {Florens}}, \bibinfo {author} {\bibfnamefont
  {I.}~\bibnamefont {Snyman}},\ and\ \bibinfo {author} {\bibfnamefont
  {N.}~\bibnamefont {Roch}},\ }\bibfield  {title} {\bibinfo {title} {A tunable
  {J}osephson platform to explore many-body quantum optics in circuit-{QED}},\
  }\href {https://doi.org/10.1038/s41534-018-0104-0} {\bibfield  {journal}
  {\bibinfo  {journal} {npj Quantum Inf.}\ }\textbf {\bibinfo {volume} {5}},\
  \bibinfo {pages} {19} (\bibinfo {year} {2019})}\BibitemShut {NoStop}%
\bibitem [{\citenamefont {Bruus}\ and\ \citenamefont
  {Flensberg}(2004)}]{bruus2004}%
  \BibitemOpen
  \bibfield  {author} {\bibinfo {author} {\bibfnamefont {H.}~\bibnamefont
  {Bruus}}\ and\ \bibinfo {author} {\bibfnamefont {K.}~\bibnamefont
  {Flensberg}},\ }\href@noop {} {\emph {\bibinfo {title} {Many-body quantum
  theory in condensed matter physics: an introduction}}}\ (\bibinfo
  {publisher} {Oxford University Press},\ \bibinfo {year} {2004})\BibitemShut
  {NoStop}%
\bibitem [{\citenamefont {Kouwenhoven}\ and\ \citenamefont
  {Glazman}(2001)}]{Kouwenhoven2001}%
  \BibitemOpen
  \bibfield  {author} {\bibinfo {author} {\bibfnamefont {L.}~\bibnamefont
  {Kouwenhoven}}\ and\ \bibinfo {author} {\bibfnamefont {L.}~\bibnamefont
  {Glazman}},\ }\bibfield  {title} {\bibinfo {title} {Revival of the {K}ondo
  effect},\ }\href {https://doi.org/10.1088/2058-7058/14/1/28} {\bibfield
  {journal} {\bibinfo  {journal} {Physics World}\ }\textbf {\bibinfo {volume}
  {14}},\ \bibinfo {pages} {33} (\bibinfo {year} {2001})}\BibitemShut {NoStop}%
\bibitem [{\citenamefont {Scott}\ and\ \citenamefont
  {Natelson}(2010)}]{Scott2010}%
  \BibitemOpen
  \bibfield  {author} {\bibinfo {author} {\bibfnamefont {G.~D.}\ \bibnamefont
  {Scott}}\ and\ \bibinfo {author} {\bibfnamefont {D.}~\bibnamefont
  {Natelson}},\ }\bibfield  {title} {\bibinfo {title} {Kondo resonances in
  molecular devices},\ }\href {https://doi.org/10.1021/nn100793s} {\bibfield
  {journal} {\bibinfo  {journal} {{ACS} Nano}\ }\textbf {\bibinfo {volume}
  {4}},\ \bibinfo {pages} {3560} (\bibinfo {year} {2010})}\BibitemShut
  {NoStop}%
\bibitem [{\citenamefont {Jeong}\ \emph {et~al.}(2001)\citenamefont {Jeong},
  \citenamefont {Chang},\ and\ \citenamefont {Melloch}}]{Jeong2001}%
  \BibitemOpen
  \bibfield  {author} {\bibinfo {author} {\bibfnamefont {H.}~\bibnamefont
  {Jeong}}, \bibinfo {author} {\bibfnamefont {A.~M.}\ \bibnamefont {Chang}},\
  and\ \bibinfo {author} {\bibfnamefont {M.~R.}\ \bibnamefont {Melloch}},\
  }\bibfield  {title} {\bibinfo {title} {The {K}ondo effect in an artificial
  quantum dot molecule},\ }\href {https://doi.org/10.1126/science.1063182}
  {\bibfield  {journal} {\bibinfo  {journal} {Science}\ }\textbf {\bibinfo
  {volume} {293}},\ \bibinfo {pages} {2221} (\bibinfo {year}
  {2001})}\BibitemShut {NoStop}%
\bibitem [{\citenamefont {Le~Hur}(2015)}]{Hur2015}%
  \BibitemOpen
  \bibfield  {author} {\bibinfo {author} {\bibfnamefont {K.}~\bibnamefont
  {Le~Hur}},\ }\bibfield  {title} {\bibinfo {title} {Quantum dots and the
  {K}ondo effect},\ }\href {https://doi.org/10.1038/526203a} {\bibfield
  {journal} {\bibinfo  {journal} {Nature}\ }\textbf {\bibinfo {volume} {526}},\
  \bibinfo {pages} {203} (\bibinfo {year} {2015})}\BibitemShut {NoStop}%
\bibitem [{\citenamefont {Avinun-Kalish}\ \emph {et~al.}(2004)\citenamefont
  {Avinun-Kalish}, \citenamefont {Heiblum}, \citenamefont {Silva},
  \citenamefont {Mahalu},\ and\ \citenamefont {Umansky}}]{Kalish2004}%
  \BibitemOpen
  \bibfield  {author} {\bibinfo {author} {\bibfnamefont {M.}~\bibnamefont
  {Avinun-Kalish}}, \bibinfo {author} {\bibfnamefont {M.}~\bibnamefont
  {Heiblum}}, \bibinfo {author} {\bibfnamefont {A.}~\bibnamefont {Silva}},
  \bibinfo {author} {\bibfnamefont {D.}~\bibnamefont {Mahalu}},\ and\ \bibinfo
  {author} {\bibfnamefont {V.}~\bibnamefont {Umansky}},\ }\bibfield  {title}
  {\bibinfo {title} {Controlled dephasing of a quantum dot in the {K}ondo
  regime},\ }\href {https://doi.org/10.1103/PhysRevLett.92.156801} {\bibfield
  {journal} {\bibinfo  {journal} {Phys. Rev. Lett.}\ }\textbf {\bibinfo
  {volume} {92}},\ \bibinfo {pages} {156801} (\bibinfo {year}
  {2004})}\BibitemShut {NoStop}%
\bibitem [{\citenamefont {Sprinzak}\ \emph {et~al.}(2002)\citenamefont
  {Sprinzak}, \citenamefont {Ji}, \citenamefont {Heiblum}, \citenamefont
  {Mahalu},\ and\ \citenamefont {Shtrikman}}]{Sprinzak2002}%
  \BibitemOpen
  \bibfield  {author} {\bibinfo {author} {\bibfnamefont {D.}~\bibnamefont
  {Sprinzak}}, \bibinfo {author} {\bibfnamefont {Y.}~\bibnamefont {Ji}},
  \bibinfo {author} {\bibfnamefont {M.}~\bibnamefont {Heiblum}}, \bibinfo
  {author} {\bibfnamefont {D.}~\bibnamefont {Mahalu}},\ and\ \bibinfo {author}
  {\bibfnamefont {H.}~\bibnamefont {Shtrikman}},\ }\bibfield  {title} {\bibinfo
  {title} {Charge distribution in a {K}ondo-correlated quantum dot},\ }\href
  {https://doi.org/10.1103/PhysRevLett.88.176805} {\bibfield  {journal}
  {\bibinfo  {journal} {Phys. Rev. Lett.}\ }\textbf {\bibinfo {volume} {88}},\
  \bibinfo {pages} {176805} (\bibinfo {year} {2002})}\BibitemShut {NoStop}%
\bibitem [{\citenamefont {Keller}\ \emph {et~al.}(2013)\citenamefont {Keller},
  \citenamefont {Amasha}, \citenamefont {Weymann}, \citenamefont {Moca},
  \citenamefont {Rau}, \citenamefont {Katine}, \citenamefont {Shtrikman},
  \citenamefont {Zar{\'{a}}nd},\ and\ \citenamefont
  {Goldhaber-Gordon}}]{Keller2013}%
  \BibitemOpen
  \bibfield  {author} {\bibinfo {author} {\bibfnamefont {A.~J.}\ \bibnamefont
  {Keller}}, \bibinfo {author} {\bibfnamefont {S.}~\bibnamefont {Amasha}},
  \bibinfo {author} {\bibfnamefont {I.}~\bibnamefont {Weymann}}, \bibinfo
  {author} {\bibfnamefont {C.~P.}\ \bibnamefont {Moca}}, \bibinfo {author}
  {\bibfnamefont {I.~G.}\ \bibnamefont {Rau}}, \bibinfo {author} {\bibfnamefont
  {J.~A.}\ \bibnamefont {Katine}}, \bibinfo {author} {\bibfnamefont
  {H.}~\bibnamefont {Shtrikman}}, \bibinfo {author} {\bibfnamefont
  {G.}~\bibnamefont {Zar{\'{a}}nd}},\ and\ \bibinfo {author} {\bibfnamefont
  {D.}~\bibnamefont {Goldhaber-Gordon}},\ }\bibfield  {title} {\bibinfo {title}
  {Emergent {SU}(4) {K}ondo physics in a spin{\textendash}charge-entangled
  double quantum dot},\ }\href {https://doi.org/10.1038/nphys2844} {\bibfield
  {journal} {\bibinfo  {journal} {Nat. Phys.}\ }\textbf {\bibinfo {volume}
  {10}},\ \bibinfo {pages} {145} (\bibinfo {year} {2013})}\BibitemShut
  {NoStop}%
\bibitem [{\citenamefont {Shang}\ \emph {et~al.}(2018)\citenamefont {Shang},
  \citenamefont {Zhang}, \citenamefont {Cao}, \citenamefont {Li}, \citenamefont
  {Xiao}, \citenamefont {Guo},\ and\ \citenamefont {Guo}}]{Shang2018}%
  \BibitemOpen
  \bibfield  {author} {\bibinfo {author} {\bibfnamefont {R.-N.}\ \bibnamefont
  {Shang}}, \bibinfo {author} {\bibfnamefont {T.}~\bibnamefont {Zhang}},
  \bibinfo {author} {\bibfnamefont {G.}~\bibnamefont {Cao}}, \bibinfo {author}
  {\bibfnamefont {H.-O.}\ \bibnamefont {Li}}, \bibinfo {author} {\bibfnamefont
  {M.}~\bibnamefont {Xiao}}, \bibinfo {author} {\bibfnamefont {G.-C.}\
  \bibnamefont {Guo}},\ and\ \bibinfo {author} {\bibfnamefont {G.-P.}\
  \bibnamefont {Guo}},\ }\bibfield  {title} {\bibinfo {title} {Direct
  observation of the orbital spin {K}ondo effect in gallium arsenide quantum
  dots},\ }\href {https://doi.org/10.1103/PhysRevB.97.085307} {\bibfield
  {journal} {\bibinfo  {journal} {Phys. Rev. B}\ }\textbf {\bibinfo {volume}
  {97}},\ \bibinfo {pages} {085307} (\bibinfo {year} {2018})}\BibitemShut
  {NoStop}%
\bibitem [{\citenamefont {Pustilnik}\ and\ \citenamefont
  {Glazman}(2004)}]{Pustilnik2004}%
  \BibitemOpen
  \bibfield  {author} {\bibinfo {author} {\bibfnamefont {M.}~\bibnamefont
  {Pustilnik}}\ and\ \bibinfo {author} {\bibfnamefont {L.}~\bibnamefont
  {Glazman}},\ }\bibfield  {title} {\bibinfo {title} {Kondo effect in quantum
  dots},\ }\href {https://doi.org/10.1088/0953-8984/16/16/R01} {\bibfield
  {journal} {\bibinfo  {journal} {J. Condens. Matter Phys.}\ }\textbf {\bibinfo
  {volume} {16}},\ \bibinfo {pages} {R513} (\bibinfo {year}
  {2004})}\BibitemShut {NoStop}%
\bibitem [{\citenamefont {Wingreen}(2004)}]{Wingreen2004}%
  \BibitemOpen
  \bibfield  {author} {\bibinfo {author} {\bibfnamefont {N.~S.}\ \bibnamefont
  {Wingreen}},\ }\bibfield  {title} {\bibinfo {title} {Quantum many-body
  effects in a single-electron transistor},\ }\href
  {https://doi.org/10.1126/science.1098302} {\bibfield  {journal} {\bibinfo
  {journal} {Science}\ }\textbf {\bibinfo {volume} {304}},\ \bibinfo {pages}
  {1258} (\bibinfo {year} {2004})}\BibitemShut {NoStop}%
\bibitem [{\citenamefont {Park}\ \emph {et~al.}(2002)\citenamefont {Park},
  \citenamefont {Pasupathy}, \citenamefont {Goldsmith}, \citenamefont {Chang},
  \citenamefont {Yaish}, \citenamefont {Petta}, \citenamefont {Rinkoski},
  \citenamefont {Sethna}, \citenamefont {Abru{\~{n}}a}, \citenamefont
  {McEuen},\ and\ \citenamefont {Ralph}}]{Park2002}%
  \BibitemOpen
  \bibfield  {author} {\bibinfo {author} {\bibfnamefont {J.}~\bibnamefont
  {Park}}, \bibinfo {author} {\bibfnamefont {A.~N.}\ \bibnamefont {Pasupathy}},
  \bibinfo {author} {\bibfnamefont {J.~I.}\ \bibnamefont {Goldsmith}}, \bibinfo
  {author} {\bibfnamefont {C.}~\bibnamefont {Chang}}, \bibinfo {author}
  {\bibfnamefont {Y.}~\bibnamefont {Yaish}}, \bibinfo {author} {\bibfnamefont
  {J.~R.}\ \bibnamefont {Petta}}, \bibinfo {author} {\bibfnamefont
  {M.}~\bibnamefont {Rinkoski}}, \bibinfo {author} {\bibfnamefont {J.~P.}\
  \bibnamefont {Sethna}}, \bibinfo {author} {\bibfnamefont {H.~D.}\
  \bibnamefont {Abru{\~{n}}a}}, \bibinfo {author} {\bibfnamefont {P.~L.}\
  \bibnamefont {McEuen}},\ and\ \bibinfo {author} {\bibfnamefont {D.~C.}\
  \bibnamefont {Ralph}},\ }\bibfield  {title} {\bibinfo {title} {Coulomb
  blockade and the {K}ondo effect in single-atom transistors},\ }\href
  {https://doi.org/10.1038/nature00791} {\bibfield  {journal} {\bibinfo
  {journal} {Nature}\ }\textbf {\bibinfo {volume} {417}},\ \bibinfo {pages}
  {722} (\bibinfo {year} {2002})}\BibitemShut {NoStop}%
\bibitem [{\citenamefont {Yu}\ \emph {et~al.}(2004)\citenamefont {Yu},
  \citenamefont {Keane}, \citenamefont {Ciszek}, \citenamefont {Cheng},
  \citenamefont {Stewart}, \citenamefont {Tour},\ and\ \citenamefont
  {Natelson}}]{Natelson2004}%
  \BibitemOpen
  \bibfield  {author} {\bibinfo {author} {\bibfnamefont {L.~H.}\ \bibnamefont
  {Yu}}, \bibinfo {author} {\bibfnamefont {Z.~K.}\ \bibnamefont {Keane}},
  \bibinfo {author} {\bibfnamefont {J.~W.}\ \bibnamefont {Ciszek}}, \bibinfo
  {author} {\bibfnamefont {L.}~\bibnamefont {Cheng}}, \bibinfo {author}
  {\bibfnamefont {M.~P.}\ \bibnamefont {Stewart}}, \bibinfo {author}
  {\bibfnamefont {J.~M.}\ \bibnamefont {Tour}},\ and\ \bibinfo {author}
  {\bibfnamefont {D.}~\bibnamefont {Natelson}},\ }\bibfield  {title} {\bibinfo
  {title} {Inelastic electron tunneling via molecular vibrations in
  single-molecule transistors},\ }\href
  {https://doi.org/10.1103/PhysRevLett.93.266802} {\bibfield  {journal}
  {\bibinfo  {journal} {Phys. Rev. Lett.}\ }\textbf {\bibinfo {volume} {93}},\
  \bibinfo {pages} {266802} (\bibinfo {year} {2004})}\BibitemShut {NoStop}%
\bibitem [{\citenamefont {Frey}\ \emph {et~al.}(2012)\citenamefont {Frey},
  \citenamefont {Leek}, \citenamefont {Beck}, \citenamefont {Blais},
  \citenamefont {Ihn}, \citenamefont {Ensslin},\ and\ \citenamefont
  {Wallraff}}]{Frey2012}%
  \BibitemOpen
  \bibfield  {author} {\bibinfo {author} {\bibfnamefont {T.}~\bibnamefont
  {Frey}}, \bibinfo {author} {\bibfnamefont {P.~J.}\ \bibnamefont {Leek}},
  \bibinfo {author} {\bibfnamefont {M.}~\bibnamefont {Beck}}, \bibinfo {author}
  {\bibfnamefont {A.}~\bibnamefont {Blais}}, \bibinfo {author} {\bibfnamefont
  {T.}~\bibnamefont {Ihn}}, \bibinfo {author} {\bibfnamefont {K.}~\bibnamefont
  {Ensslin}},\ and\ \bibinfo {author} {\bibfnamefont {A.}~\bibnamefont
  {Wallraff}},\ }\bibfield  {title} {\bibinfo {title} {Dipole coupling of a
  double quantum dot to a microwave resonator},\ }\href
  {https://doi.org/10.1103/PhysRevLett.108.046807} {\bibfield  {journal}
  {\bibinfo  {journal} {Phys. Rev. Lett.}\ }\textbf {\bibinfo {volume} {108}},\
  \bibinfo {pages} {046807} (\bibinfo {year} {2012})}\BibitemShut {NoStop}%
\bibitem [{\citenamefont {Bruhat}\ \emph {et~al.}(2016)\citenamefont {Bruhat},
  \citenamefont {Viennot}, \citenamefont {Dartiailh}, \citenamefont
  {Desjardins}, \citenamefont {Kontos},\ and\ \citenamefont
  {Cottet}}]{Bruhat2016}%
  \BibitemOpen
  \bibfield  {author} {\bibinfo {author} {\bibfnamefont {L.~E.}\ \bibnamefont
  {Bruhat}}, \bibinfo {author} {\bibfnamefont {J.~J.}\ \bibnamefont {Viennot}},
  \bibinfo {author} {\bibfnamefont {M.~C.}\ \bibnamefont {Dartiailh}}, \bibinfo
  {author} {\bibfnamefont {M.~M.}\ \bibnamefont {Desjardins}}, \bibinfo
  {author} {\bibfnamefont {T.}~\bibnamefont {Kontos}},\ and\ \bibinfo {author}
  {\bibfnamefont {A.}~\bibnamefont {Cottet}},\ }\bibfield  {title} {\bibinfo
  {title} {Cavity photons as a probe for charge relaxation resistance and
  photon emission in a quantum dot coupled to normal and superconducting
  continua},\ }\href {https://doi.org/10.1103/PhysRevX.6.021014} {\bibfield
  {journal} {\bibinfo  {journal} {Phys. Rev. X}\ }\textbf {\bibinfo {volume}
  {6}},\ \bibinfo {pages} {021014} (\bibinfo {year} {2016})}\BibitemShut
  {NoStop}%
\bibitem [{\citenamefont {Halbhuber}\ \emph {et~al.}(2020)\citenamefont
  {Halbhuber}, \citenamefont {Mornhinweg}, \citenamefont {Zeller},
  \citenamefont {Ciuti}, \citenamefont {Bougeard}, \citenamefont {Huber},\ and\
  \citenamefont {Lange}}]{Halbhuber2020}%
  \BibitemOpen
  \bibfield  {author} {\bibinfo {author} {\bibfnamefont {M.}~\bibnamefont
  {Halbhuber}}, \bibinfo {author} {\bibfnamefont {J.}~\bibnamefont
  {Mornhinweg}}, \bibinfo {author} {\bibfnamefont {V.}~\bibnamefont {Zeller}},
  \bibinfo {author} {\bibfnamefont {C.}~\bibnamefont {Ciuti}}, \bibinfo
  {author} {\bibfnamefont {D.}~\bibnamefont {Bougeard}}, \bibinfo {author}
  {\bibfnamefont {R.}~\bibnamefont {Huber}},\ and\ \bibinfo {author}
  {\bibfnamefont {C.}~\bibnamefont {Lange}},\ }\bibfield  {title} {\bibinfo
  {title} {Non-adiabatic stripping of a cavity field from electrons in the
  deep-strong coupling regime},\ }\href
  {https://doi.org/10.1038/s41566-020-0673-2} {\bibfield  {journal} {\bibinfo
  {journal} {Nat. Photonics}\ }\textbf {\bibinfo {volume} {14}},\ \bibinfo
  {pages} {675} (\bibinfo {year} {2020})}\BibitemShut {NoStop}%
\bibitem [{\citenamefont {van Woerkom}\ \emph {et~al.}(2018)\citenamefont {van
  Woerkom}, \citenamefont {Scarlino}, \citenamefont {Ungerer}, \citenamefont
  {M\"uller}, \citenamefont {Koski}, \citenamefont {Landig}, \citenamefont
  {Reichl}, \citenamefont {Wegscheider}, \citenamefont {Ihn}, \citenamefont
  {Ensslin},\ and\ \citenamefont {Wallraff}}]{Woerkom2018}%
  \BibitemOpen
  \bibfield  {author} {\bibinfo {author} {\bibfnamefont {D.~J.}\ \bibnamefont
  {van Woerkom}}, \bibinfo {author} {\bibfnamefont {P.}~\bibnamefont
  {Scarlino}}, \bibinfo {author} {\bibfnamefont {J.~H.}\ \bibnamefont
  {Ungerer}}, \bibinfo {author} {\bibfnamefont {C.}~\bibnamefont {M\"uller}},
  \bibinfo {author} {\bibfnamefont {J.~V.}\ \bibnamefont {Koski}}, \bibinfo
  {author} {\bibfnamefont {A.~J.}\ \bibnamefont {Landig}}, \bibinfo {author}
  {\bibfnamefont {C.}~\bibnamefont {Reichl}}, \bibinfo {author} {\bibfnamefont
  {W.}~\bibnamefont {Wegscheider}}, \bibinfo {author} {\bibfnamefont
  {T.}~\bibnamefont {Ihn}}, \bibinfo {author} {\bibfnamefont {K.}~\bibnamefont
  {Ensslin}},\ and\ \bibinfo {author} {\bibfnamefont {A.}~\bibnamefont
  {Wallraff}},\ }\bibfield  {title} {\bibinfo {title} {Microwave
  photon-mediated interactions between semiconductor qubits},\ }\href
  {https://doi.org/10.1103/PhysRevX.8.041018} {\bibfield  {journal} {\bibinfo
  {journal} {Phys. Rev. X}\ }\textbf {\bibinfo {volume} {8}},\ \bibinfo {pages}
  {041018} (\bibinfo {year} {2018})}\BibitemShut {NoStop}%
\bibitem [{\citenamefont {Cottet}\ \emph {et~al.}(2015)\citenamefont {Cottet},
  \citenamefont {Kontos},\ and\ \citenamefont {Dou\ifmmode~\mbox{\c{c}}\else
  \c{c}\fi{}ot}}]{Cottet2015}%
  \BibitemOpen
  \bibfield  {author} {\bibinfo {author} {\bibfnamefont {A.}~\bibnamefont
  {Cottet}}, \bibinfo {author} {\bibfnamefont {T.}~\bibnamefont {Kontos}},\
  and\ \bibinfo {author} {\bibfnamefont {B.}~\bibnamefont
  {Dou\ifmmode~\mbox{\c{c}}\else \c{c}\fi{}ot}},\ }\bibfield  {title} {\bibinfo
  {title} {Electron-photon coupling in mesoscopic quantum electrodynamics},\
  }\href {https://doi.org/10.1103/PhysRevB.91.205417} {\bibfield  {journal}
  {\bibinfo  {journal} {Phys. Rev. B}\ }\textbf {\bibinfo {volume} {91}},\
  \bibinfo {pages} {205417} (\bibinfo {year} {2015})}\BibitemShut {NoStop}%
\bibitem [{\citenamefont {Rokaj}\ \emph {et~al.}(2022)\citenamefont {Rokaj},
  \citenamefont {Ruggenthaler}, \citenamefont {Eich},\ and\ \citenamefont
  {Rubio}}]{Rubio2022}%
  \BibitemOpen
  \bibfield  {author} {\bibinfo {author} {\bibfnamefont {V.}~\bibnamefont
  {Rokaj}}, \bibinfo {author} {\bibfnamefont {M.}~\bibnamefont {Ruggenthaler}},
  \bibinfo {author} {\bibfnamefont {F.~G.}\ \bibnamefont {Eich}},\ and\
  \bibinfo {author} {\bibfnamefont {A.}~\bibnamefont {Rubio}},\ }\bibfield
  {title} {\bibinfo {title} {Free electron gas in cavity quantum
  electrodynamics},\ }\href {https://doi.org/10.1103/PhysRevResearch.4.013012}
  {\bibfield  {journal} {\bibinfo  {journal} {Phys. Rev. Res.}\ }\textbf
  {\bibinfo {volume} {4}},\ \bibinfo {pages} {013012} (\bibinfo {year}
  {2022})}\BibitemShut {NoStop}%
\bibitem [{\citenamefont {Desjardins}\ \emph {et~al.}(2017)\citenamefont
  {Desjardins}, \citenamefont {Viennot}, \citenamefont {Dartiailh},
  \citenamefont {Bruhat}, \citenamefont {Delbecq}, \citenamefont {Lee},
  \citenamefont {Choi}, \citenamefont {Cottet},\ and\ \citenamefont
  {Kontos}}]{Desjardins2017}%
  \BibitemOpen
  \bibfield  {author} {\bibinfo {author} {\bibfnamefont {M.~M.}\ \bibnamefont
  {Desjardins}}, \bibinfo {author} {\bibfnamefont {J.~J.}\ \bibnamefont
  {Viennot}}, \bibinfo {author} {\bibfnamefont {M.~C.}\ \bibnamefont
  {Dartiailh}}, \bibinfo {author} {\bibfnamefont {L.~E.}\ \bibnamefont
  {Bruhat}}, \bibinfo {author} {\bibfnamefont {M.~R.}\ \bibnamefont {Delbecq}},
  \bibinfo {author} {\bibfnamefont {M.}~\bibnamefont {Lee}}, \bibinfo {author}
  {\bibfnamefont {M.-S.}\ \bibnamefont {Choi}}, \bibinfo {author}
  {\bibfnamefont {A.}~\bibnamefont {Cottet}},\ and\ \bibinfo {author}
  {\bibfnamefont {T.}~\bibnamefont {Kontos}},\ }\bibfield  {title} {\bibinfo
  {title} {Observation of the frozen charge of a {K}ondo resonance},\ }\href
  {https://doi.org/10.1038/nature21704} {\bibfield  {journal} {\bibinfo
  {journal} {Nature}\ }\textbf {\bibinfo {volume} {545}},\ \bibinfo {pages}
  {71} (\bibinfo {year} {2017})}\BibitemShut {NoStop}%
\bibitem [{\citenamefont {Deng}\ \emph {et~al.}(2021)\citenamefont {Deng},
  \citenamefont {Henriet}, \citenamefont {Wei}, \citenamefont {Li},
  \citenamefont {Li}, \citenamefont {Cao}, \citenamefont {Xiao}, \citenamefont
  {Guo}, \citenamefont {Schir\'o}, \citenamefont {Le~Hur},\ and\ \citenamefont
  {Guo}}]{GuoPing2021}%
  \BibitemOpen
  \bibfield  {author} {\bibinfo {author} {\bibfnamefont {G.-W.}\ \bibnamefont
  {Deng}}, \bibinfo {author} {\bibfnamefont {L.}~\bibnamefont {Henriet}},
  \bibinfo {author} {\bibfnamefont {D.}~\bibnamefont {Wei}}, \bibinfo {author}
  {\bibfnamefont {S.-X.}\ \bibnamefont {Li}}, \bibinfo {author} {\bibfnamefont
  {H.-O.}\ \bibnamefont {Li}}, \bibinfo {author} {\bibfnamefont
  {G.}~\bibnamefont {Cao}}, \bibinfo {author} {\bibfnamefont {M.}~\bibnamefont
  {Xiao}}, \bibinfo {author} {\bibfnamefont {G.-C.}\ \bibnamefont {Guo}},
  \bibinfo {author} {\bibfnamefont {M.}~\bibnamefont {Schir\'o}}, \bibinfo
  {author} {\bibfnamefont {K.}~\bibnamefont {Le~Hur}},\ and\ \bibinfo {author}
  {\bibfnamefont {G.-P.}\ \bibnamefont {Guo}},\ }\bibfield  {title} {\bibinfo
  {title} {Kondo induced $\ensuremath{\pi}$-phase shift of microwave photons in
  a circuit quantum electrodynamics architecture},\ }\href
  {https://doi.org/10.1103/PhysRevB.104.125407} {\bibfield  {journal} {\bibinfo
   {journal} {Phys. Rev. B}\ }\textbf {\bibinfo {volume} {104}},\ \bibinfo
  {pages} {125407} (\bibinfo {year} {2021})}\BibitemShut {NoStop}%
\bibitem [{\citenamefont {Kockum}\ \emph {et~al.}(2019)\citenamefont {Kockum},
  \citenamefont {Miranowicz}, \citenamefont {De~Liberato}, \citenamefont
  {Savasta},\ and\ \citenamefont {Nori}}]{Anton2019}%
  \BibitemOpen
  \bibfield  {author} {\bibinfo {author} {\bibfnamefont {A.~F.}\ \bibnamefont
  {Kockum}}, \bibinfo {author} {\bibfnamefont {A.}~\bibnamefont {Miranowicz}},
  \bibinfo {author} {\bibfnamefont {S.}~\bibnamefont {De~Liberato}}, \bibinfo
  {author} {\bibfnamefont {S.}~\bibnamefont {Savasta}},\ and\ \bibinfo {author}
  {\bibfnamefont {F.}~\bibnamefont {Nori}},\ }\bibfield  {title} {\bibinfo
  {title} {Ultrastrong coupling between light and matter},\ }\href
  {https://doi.org/10.1038/s42254-018-0006-2} {\bibfield  {journal} {\bibinfo
  {journal} {Nat. Rev. Phys.}\ }\textbf {\bibinfo {volume} {1}},\ \bibinfo
  {pages} {19} (\bibinfo {year} {2019})}\BibitemShut {NoStop}%
\bibitem [{\citenamefont {Forn-D\'{i}az}\ \emph {et~al.}(2019)\citenamefont
  {Forn-D\'{i}az}, \citenamefont {Lamata}, \citenamefont {Rico}, \citenamefont
  {Kono},\ and\ \citenamefont {Solano}}]{Solano2019}%
  \BibitemOpen
  \bibfield  {author} {\bibinfo {author} {\bibfnamefont {P.}~\bibnamefont
  {Forn-D\'{i}az}}, \bibinfo {author} {\bibfnamefont {L.}~\bibnamefont
  {Lamata}}, \bibinfo {author} {\bibfnamefont {E.}~\bibnamefont {Rico}},
  \bibinfo {author} {\bibfnamefont {J.}~\bibnamefont {Kono}},\ and\ \bibinfo
  {author} {\bibfnamefont {E.}~\bibnamefont {Solano}},\ }\bibfield  {title}
  {\bibinfo {title} {Ultrastrong coupling regimes of light-matter
  interaction},\ }\href {https://doi.org/10.1103/RevModPhys.91.025005}
  {\bibfield  {journal} {\bibinfo  {journal} {Rev. Mod. Phys.}\ }\textbf
  {\bibinfo {volume} {91}},\ \bibinfo {pages} {025005} (\bibinfo {year}
  {2019})}\BibitemShut {NoStop}%
\bibitem [{\citenamefont {Stockklauser}\ \emph {et~al.}(2017)\citenamefont
  {Stockklauser}, \citenamefont {Scarlino}, \citenamefont {Koski},
  \citenamefont {Gasparinetti}, \citenamefont {Andersen}, \citenamefont
  {Reichl}, \citenamefont {Wegscheider}, \citenamefont {Ihn}, \citenamefont
  {Ensslin},\ and\ \citenamefont {Wallraff}}]{Stockklauser2017}%
  \BibitemOpen
  \bibfield  {author} {\bibinfo {author} {\bibfnamefont {A.}~\bibnamefont
  {Stockklauser}}, \bibinfo {author} {\bibfnamefont {P.}~\bibnamefont
  {Scarlino}}, \bibinfo {author} {\bibfnamefont {J.~V.}\ \bibnamefont {Koski}},
  \bibinfo {author} {\bibfnamefont {S.}~\bibnamefont {Gasparinetti}}, \bibinfo
  {author} {\bibfnamefont {C.~K.}\ \bibnamefont {Andersen}}, \bibinfo {author}
  {\bibfnamefont {C.}~\bibnamefont {Reichl}}, \bibinfo {author} {\bibfnamefont
  {W.}~\bibnamefont {Wegscheider}}, \bibinfo {author} {\bibfnamefont
  {T.}~\bibnamefont {Ihn}}, \bibinfo {author} {\bibfnamefont {K.}~\bibnamefont
  {Ensslin}},\ and\ \bibinfo {author} {\bibfnamefont {A.}~\bibnamefont
  {Wallraff}},\ }\bibfield  {title} {\bibinfo {title} {Strong coupling cavity
  {QED} with gate-defined double quantum dots enabled by a high impedance
  resonator},\ }\href {https://doi.org/10.1103/PhysRevX.7.011030} {\bibfield
  {journal} {\bibinfo  {journal} {Phys. Rev. X}\ }\textbf {\bibinfo {volume}
  {7}},\ \bibinfo {pages} {011030} (\bibinfo {year} {2017})}\BibitemShut
  {NoStop}%
\bibitem [{\citenamefont {Stassi}\ \emph {et~al.}(2013)\citenamefont {Stassi},
  \citenamefont {Ridolfo}, \citenamefont {Di~Stefano}, \citenamefont
  {Hartmann},\ and\ \citenamefont {Savasta}}]{Stassi2013}%
  \BibitemOpen
  \bibfield  {author} {\bibinfo {author} {\bibfnamefont {R.}~\bibnamefont
  {Stassi}}, \bibinfo {author} {\bibfnamefont {A.}~\bibnamefont {Ridolfo}},
  \bibinfo {author} {\bibfnamefont {O.}~\bibnamefont {Di~Stefano}}, \bibinfo
  {author} {\bibfnamefont {M.~J.}\ \bibnamefont {Hartmann}},\ and\ \bibinfo
  {author} {\bibfnamefont {S.}~\bibnamefont {Savasta}},\ }\bibfield  {title}
  {\bibinfo {title} {Spontaneous conversion from virtual to real photons in the
  ultrastrong-coupling regime},\ }\href
  {https://doi.org/10.1103/PhysRevLett.110.243601} {\bibfield  {journal}
  {\bibinfo  {journal} {Phys. Rev. Lett.}\ }\textbf {\bibinfo {volume} {110}},\
  \bibinfo {pages} {243601} (\bibinfo {year} {2013})}\BibitemShut {NoStop}%
\bibitem [{\citenamefont {Kockum}\ \emph {et~al.}(2017)\citenamefont {Kockum},
  \citenamefont {Miranowicz}, \citenamefont {Macr\`{\i}}, \citenamefont
  {Savasta},\ and\ \citenamefont {Nori}}]{Anton2017}%
  \BibitemOpen
  \bibfield  {author} {\bibinfo {author} {\bibfnamefont {A.~F.}\ \bibnamefont
  {Kockum}}, \bibinfo {author} {\bibfnamefont {A.}~\bibnamefont {Miranowicz}},
  \bibinfo {author} {\bibfnamefont {V.}~\bibnamefont {Macr\`{\i}}}, \bibinfo
  {author} {\bibfnamefont {S.}~\bibnamefont {Savasta}},\ and\ \bibinfo {author}
  {\bibfnamefont {F.}~\bibnamefont {Nori}},\ }\bibfield  {title} {\bibinfo
  {title} {Deterministic quantum nonlinear optics with single atoms and virtual
  photons},\ }\href {https://doi.org/10.1103/PhysRevA.95.063849} {\bibfield
  {journal} {\bibinfo  {journal} {Phys. Rev. A}\ }\textbf {\bibinfo {volume}
  {95}},\ \bibinfo {pages} {063849} (\bibinfo {year} {2017})}\BibitemShut
  {NoStop}%
\bibitem [{\citenamefont {Garziano}\ \emph {et~al.}(2016)\citenamefont
  {Garziano}, \citenamefont {Macr\`{\i}}, \citenamefont {Stassi}, \citenamefont
  {Di~Stefano}, \citenamefont {Nori},\ and\ \citenamefont
  {Savasta}}]{Garziano2016}%
  \BibitemOpen
  \bibfield  {author} {\bibinfo {author} {\bibfnamefont {L.}~\bibnamefont
  {Garziano}}, \bibinfo {author} {\bibfnamefont {V.}~\bibnamefont
  {Macr\`{\i}}}, \bibinfo {author} {\bibfnamefont {R.}~\bibnamefont {Stassi}},
  \bibinfo {author} {\bibfnamefont {O.}~\bibnamefont {Di~Stefano}}, \bibinfo
  {author} {\bibfnamefont {F.}~\bibnamefont {Nori}},\ and\ \bibinfo {author}
  {\bibfnamefont {S.}~\bibnamefont {Savasta}},\ }\bibfield  {title} {\bibinfo
  {title} {One photon can simultaneously excite two or more atoms},\ }\href
  {https://doi.org/10.1103/PhysRevLett.117.043601} {\bibfield  {journal}
  {\bibinfo  {journal} {Phys. Rev. Lett.}\ }\textbf {\bibinfo {volume} {117}},\
  \bibinfo {pages} {043601} (\bibinfo {year} {2016})}\BibitemShut {NoStop}%
\bibitem [{\citenamefont {Cirio}\ \emph {et~al.}(2016)\citenamefont {Cirio},
  \citenamefont {De~Liberato}, \citenamefont {Lambert},\ and\ \citenamefont
  {Nori}}]{Mauro2016}%
  \BibitemOpen
  \bibfield  {author} {\bibinfo {author} {\bibfnamefont {M.}~\bibnamefont
  {Cirio}}, \bibinfo {author} {\bibfnamefont {S.}~\bibnamefont {De~Liberato}},
  \bibinfo {author} {\bibfnamefont {N.}~\bibnamefont {Lambert}},\ and\ \bibinfo
  {author} {\bibfnamefont {F.}~\bibnamefont {Nori}},\ }\bibfield  {title}
  {\bibinfo {title} {Ground state electroluminescence},\ }\href
  {https://doi.org/10.1103/PhysRevLett.116.113601} {\bibfield  {journal}
  {\bibinfo  {journal} {Phys. Rev. Lett.}\ }\textbf {\bibinfo {volume} {116}},\
  \bibinfo {pages} {113601} (\bibinfo {year} {2016})}\BibitemShut {NoStop}%
\bibitem [{\citenamefont {Cirio}\ \emph {et~al.}(2019)\citenamefont {Cirio},
  \citenamefont {Shammah}, \citenamefont {Lambert}, \citenamefont
  {De~Liberato},\ and\ \citenamefont {Nori}}]{Cirio2019}%
  \BibitemOpen
  \bibfield  {author} {\bibinfo {author} {\bibfnamefont {M.}~\bibnamefont
  {Cirio}}, \bibinfo {author} {\bibfnamefont {N.}~\bibnamefont {Shammah}},
  \bibinfo {author} {\bibfnamefont {N.}~\bibnamefont {Lambert}}, \bibinfo
  {author} {\bibfnamefont {S.}~\bibnamefont {De~Liberato}},\ and\ \bibinfo
  {author} {\bibfnamefont {F.}~\bibnamefont {Nori}},\ }\bibfield  {title}
  {\bibinfo {title} {Multielectron ground state electroluminescence},\ }\href
  {https://doi.org/10.1103/PhysRevLett.122.190403} {\bibfield  {journal}
  {\bibinfo  {journal} {Phys. Rev. Lett.}\ }\textbf {\bibinfo {volume} {122}},\
  \bibinfo {pages} {190403} (\bibinfo {year} {2019})}\BibitemShut {NoStop}%
\bibitem [{\citenamefont {Herrera}\ and\ \citenamefont
  {Spano}(2016)}]{Felipe2016}%
  \BibitemOpen
  \bibfield  {author} {\bibinfo {author} {\bibfnamefont {F.}~\bibnamefont
  {Herrera}}\ and\ \bibinfo {author} {\bibfnamefont {F.~C.}\ \bibnamefont
  {Spano}},\ }\bibfield  {title} {\bibinfo {title} {Cavity-controlled chemistry
  in molecular ensembles},\ }\href
  {https://doi.org/10.1103/PhysRevLett.116.238301} {\bibfield  {journal}
  {\bibinfo  {journal} {Phys. Rev. Lett.}\ }\textbf {\bibinfo {volume} {116}},\
  \bibinfo {pages} {238301} (\bibinfo {year} {2016})}\BibitemShut {NoStop}%
\bibitem [{\citenamefont {Mart{\'{\i}}nez-Mart{\'{\i}}nez}\ \emph
  {et~al.}(2018)\citenamefont {Mart{\'{\i}}nez-Mart{\'{\i}}nez}, \citenamefont
  {Ribeiro}, \citenamefont {Campos-Gonz{\'{a}}lez-Angulo},\ and\ \citenamefont
  {Yuen-Zhou}}]{MartnezMartnez2018}%
  \BibitemOpen
  \bibfield  {author} {\bibinfo {author} {\bibfnamefont {L.~A.}\ \bibnamefont
  {Mart{\'{\i}}nez-Mart{\'{\i}}nez}}, \bibinfo {author} {\bibfnamefont {R.~F.}\
  \bibnamefont {Ribeiro}}, \bibinfo {author} {\bibfnamefont {J.}~\bibnamefont
  {Campos-Gonz{\'{a}}lez-Angulo}},\ and\ \bibinfo {author} {\bibfnamefont
  {J.}~\bibnamefont {Yuen-Zhou}},\ }\bibfield  {title} {\bibinfo {title} {Can
  ultrastrong coupling change ground-state chemical reactions?},\ }\href
  {https://doi.org/10.1021/acsphotonics.7b00610} {\bibfield  {journal}
  {\bibinfo  {journal} {{ACS} Photonics}\ }\textbf {\bibinfo {volume} {5}},\
  \bibinfo {pages} {167} (\bibinfo {year} {2018})}\BibitemShut {NoStop}%
\bibitem [{\citenamefont {Sch\"{a}fer}\ \emph {et~al.}(2022)\citenamefont
  {Sch\"{a}fer}, \citenamefont {Flick}, \citenamefont {Ronca}, \citenamefont
  {Narang},\ and\ \citenamefont {Rubio}}]{Schfer2022}%
  \BibitemOpen
  \bibfield  {author} {\bibinfo {author} {\bibfnamefont {C.}~\bibnamefont
  {Sch\"{a}fer}}, \bibinfo {author} {\bibfnamefont {J.}~\bibnamefont {Flick}},
  \bibinfo {author} {\bibfnamefont {E.}~\bibnamefont {Ronca}}, \bibinfo
  {author} {\bibfnamefont {P.}~\bibnamefont {Narang}},\ and\ \bibinfo {author}
  {\bibfnamefont {A.}~\bibnamefont {Rubio}},\ }\bibfield  {title} {\bibinfo
  {title} {Shining light on the microscopic resonant mechanism responsible for
  cavity-mediated chemical reactivity},\ }\href
  {https://doi.org/10.1038/s41467-022-35363-6} {\bibfield  {journal} {\bibinfo
  {journal} {Nat. Commun.}\ }\textbf {\bibinfo {volume} {13}},\ \bibinfo
  {pages} {7817} (\bibinfo {year} {2022})}\BibitemShut {NoStop}%
\bibitem [{\citenamefont {Garcia-Vidal}\ \emph {et~al.}(2021)\citenamefont
  {Garcia-Vidal}, \citenamefont {Ciuti},\ and\ \citenamefont
  {Ebbesen}}]{GarciaVidal2021}%
  \BibitemOpen
  \bibfield  {author} {\bibinfo {author} {\bibfnamefont {F.~J.}\ \bibnamefont
  {Garcia-Vidal}}, \bibinfo {author} {\bibfnamefont {C.}~\bibnamefont
  {Ciuti}},\ and\ \bibinfo {author} {\bibfnamefont {T.~W.}\ \bibnamefont
  {Ebbesen}},\ }\bibfield  {title} {\bibinfo {title} {Manipulating matter by
  strong coupling to vacuum fields},\ }\href
  {https://doi.org/10.1126/science.abd0336} {\bibfield  {journal} {\bibinfo
  {journal} {Science}\ }\textbf {\bibinfo {volume} {373}},\ \bibinfo {pages}
  {eabd0336} (\bibinfo {year} {2021})}\BibitemShut {NoStop}%
\bibitem [{\citenamefont {Bloch}\ \emph {et~al.}(2022)\citenamefont {Bloch},
  \citenamefont {Cavalleri}, \citenamefont {Galitski}, \citenamefont {Hafezi},\
  and\ \citenamefont {Rubio}}]{Bloch2022}%
  \BibitemOpen
  \bibfield  {author} {\bibinfo {author} {\bibfnamefont {J.}~\bibnamefont
  {Bloch}}, \bibinfo {author} {\bibfnamefont {A.}~\bibnamefont {Cavalleri}},
  \bibinfo {author} {\bibfnamefont {V.}~\bibnamefont {Galitski}}, \bibinfo
  {author} {\bibfnamefont {M.}~\bibnamefont {Hafezi}},\ and\ \bibinfo {author}
  {\bibfnamefont {A.}~\bibnamefont {Rubio}},\ }\bibfield  {title} {\bibinfo
  {title} {Strongly correlated electron{\textendash}photon systems},\ }\href
  {https://doi.org/10.1038/s41586-022-04726-w} {\bibfield  {journal} {\bibinfo
  {journal} {Nature}\ }\textbf {\bibinfo {volume} {606}},\ \bibinfo {pages}
  {41} (\bibinfo {year} {2022})}\BibitemShut {NoStop}%
\bibitem [{\citenamefont {Lambert}\ \emph {et~al.}(2019)\citenamefont
  {Lambert}, \citenamefont {Ahmed}, \citenamefont {Cirio},\ and\ \citenamefont
  {Nori}}]{Lambert2019}%
  \BibitemOpen
  \bibfield  {author} {\bibinfo {author} {\bibfnamefont {N.}~\bibnamefont
  {Lambert}}, \bibinfo {author} {\bibfnamefont {S.}~\bibnamefont {Ahmed}},
  \bibinfo {author} {\bibfnamefont {M.}~\bibnamefont {Cirio}},\ and\ \bibinfo
  {author} {\bibfnamefont {F.}~\bibnamefont {Nori}},\ }\bibfield  {title}
  {\bibinfo {title} {Modelling the ultra-strongly coupled spin-boson model with
  unphysical modes},\ }\href {https://doi.org/10.1038/s41467-019-11656-1}
  {\bibfield  {journal} {\bibinfo  {journal} {Nat. Commun.}\ }\textbf {\bibinfo
  {volume} {10}},\ \bibinfo {pages} {3721} (\bibinfo {year}
  {2019})}\BibitemShut {NoStop}%
\bibitem [{\citenamefont {Nataf}\ and\ \citenamefont
  {Ciuti}(2011)}]{Pierre2011}%
  \BibitemOpen
  \bibfield  {author} {\bibinfo {author} {\bibfnamefont {P.}~\bibnamefont
  {Nataf}}\ and\ \bibinfo {author} {\bibfnamefont {C.}~\bibnamefont {Ciuti}},\
  }\bibfield  {title} {\bibinfo {title} {Protected quantum computation with
  multiple resonators in ultrastrong coupling circuit qed},\ }\href
  {https://doi.org/10.1103/PhysRevLett.107.190402} {\bibfield  {journal}
  {\bibinfo  {journal} {Phys. Rev. Lett.}\ }\textbf {\bibinfo {volume} {107}},\
  \bibinfo {pages} {190402} (\bibinfo {year} {2011})}\BibitemShut {NoStop}%
\bibitem [{\citenamefont {Wendin}(2017)}]{Wendin2017}%
  \BibitemOpen
  \bibfield  {author} {\bibinfo {author} {\bibfnamefont {G.}~\bibnamefont
  {Wendin}},\ }\bibfield  {title} {\bibinfo {title} {Quantum information
  processing with superconducting circuits: a review},\ }\href
  {https://doi.org/10.1088/1361-6633/aa7e1a} {\bibfield  {journal} {\bibinfo
  {journal} {Reports on Progress in Physics}\ }\textbf {\bibinfo {volume}
  {80}},\ \bibinfo {pages} {106001} (\bibinfo {year} {2017})}\BibitemShut
  {NoStop}%
\bibitem [{\citenamefont {Stassi}\ and\ \citenamefont
  {Nori}(2018)}]{PhysRevA.97.033823}%
  \BibitemOpen
  \bibfield  {author} {\bibinfo {author} {\bibfnamefont {R.}~\bibnamefont
  {Stassi}}\ and\ \bibinfo {author} {\bibfnamefont {F.}~\bibnamefont {Nori}},\
  }\bibfield  {title} {\bibinfo {title} {Long-lasting quantum memories:
  Extending the coherence time of superconducting artificial atoms in the
  ultrastrong-coupling regime},\ }\href
  {https://doi.org/10.1103/PhysRevA.97.033823} {\bibfield  {journal} {\bibinfo
  {journal} {Phys. Rev. A}\ }\textbf {\bibinfo {volume} {97}},\ \bibinfo
  {pages} {033823} (\bibinfo {year} {2018})}\BibitemShut {NoStop}%
\bibitem [{\citenamefont {Tame}\ \emph {et~al.}(2013)\citenamefont {Tame},
  \citenamefont {McEnery}, \citenamefont {\"{O}zdemir}, \citenamefont {Lee},
  \citenamefont {Maier},\ and\ \citenamefont {Kim}}]{Tame2013}%
  \BibitemOpen
  \bibfield  {author} {\bibinfo {author} {\bibfnamefont {M.~S.}\ \bibnamefont
  {Tame}}, \bibinfo {author} {\bibfnamefont {K.~R.}\ \bibnamefont {McEnery}},
  \bibinfo {author} {\bibfnamefont {{\c{S}}.~K.}\ \bibnamefont {\"{O}zdemir}},
  \bibinfo {author} {\bibfnamefont {J.}~\bibnamefont {Lee}}, \bibinfo {author}
  {\bibfnamefont {S.~A.}\ \bibnamefont {Maier}},\ and\ \bibinfo {author}
  {\bibfnamefont {M.~S.}\ \bibnamefont {Kim}},\ }\bibfield  {title} {\bibinfo
  {title} {Quantum plasmonics},\ }\href {https://doi.org/10.1038/nphys2615}
  {\bibfield  {journal} {\bibinfo  {journal} {Nat. Phys.}\ }\textbf {\bibinfo
  {volume} {9}},\ \bibinfo {pages} {329} (\bibinfo {year} {2013})}\BibitemShut
  {NoStop}%
\bibitem [{\citenamefont {Seah}\ \emph {et~al.}(2018)\citenamefont {Seah},
  \citenamefont {Nimmrichter},\ and\ \citenamefont {Scarani}}]{Stella2018}%
  \BibitemOpen
  \bibfield  {author} {\bibinfo {author} {\bibfnamefont {S.}~\bibnamefont
  {Seah}}, \bibinfo {author} {\bibfnamefont {S.}~\bibnamefont {Nimmrichter}},\
  and\ \bibinfo {author} {\bibfnamefont {V.}~\bibnamefont {Scarani}},\
  }\bibfield  {title} {\bibinfo {title} {Refrigeration beyond weak internal
  coupling},\ }\href {https://doi.org/10.1103/PhysRevE.98.012131} {\bibfield
  {journal} {\bibinfo  {journal} {Phys. Rev. E}\ }\textbf {\bibinfo {volume}
  {98}},\ \bibinfo {pages} {012131} (\bibinfo {year} {2018})}\BibitemShut
  {NoStop}%
\bibitem [{\citenamefont {Ivander}\ \emph {et~al.}(2022)\citenamefont
  {Ivander}, \citenamefont {Anto-Sztrikacs},\ and\ \citenamefont
  {Segal}}]{Ivander2022}%
  \BibitemOpen
  \bibfield  {author} {\bibinfo {author} {\bibfnamefont {F.}~\bibnamefont
  {Ivander}}, \bibinfo {author} {\bibfnamefont {N.}~\bibnamefont
  {Anto-Sztrikacs}},\ and\ \bibinfo {author} {\bibfnamefont {D.}~\bibnamefont
  {Segal}},\ }\bibfield  {title} {\bibinfo {title} {Strong system-bath coupling
  effects in quantum absorption refrigerators},\ }\href
  {https://doi.org/10.1103/PhysRevE.105.034112} {\bibfield  {journal} {\bibinfo
   {journal} {Phys. Rev. E}\ }\textbf {\bibinfo {volume} {105}},\ \bibinfo
  {pages} {034112} (\bibinfo {year} {2022})}\BibitemShut {NoStop}%
\bibitem [{\citenamefont {Hewson}(1993)}]{Hewson1993}%
  \BibitemOpen
  \bibfield  {author} {\bibinfo {author} {\bibfnamefont {A.~C.}\ \bibnamefont
  {Hewson}},\ }\href {https://doi.org/10.1017/cbo9780511470752} {\emph
  {\bibinfo {title} {The {K}ondo Problem to Heavy Fermions}}}\ (\bibinfo
  {publisher} {Cambridge University Press},\ \bibinfo {year}
  {1993})\BibitemShut {NoStop}%
\bibitem [{\citenamefont {Bulla}\ \emph {et~al.}(2008)\citenamefont {Bulla},
  \citenamefont {Costi},\ and\ \citenamefont {Pruschke}}]{Ralf2008}%
  \BibitemOpen
  \bibfield  {author} {\bibinfo {author} {\bibfnamefont {R.}~\bibnamefont
  {Bulla}}, \bibinfo {author} {\bibfnamefont {T.~A.}\ \bibnamefont {Costi}},\
  and\ \bibinfo {author} {\bibfnamefont {T.}~\bibnamefont {Pruschke}},\
  }\bibfield  {title} {\bibinfo {title} {Numerical renormalization group method
  for quantum impurity systems},\ }\href
  {https://doi.org/10.1103/RevModPhys.80.395} {\bibfield  {journal} {\bibinfo
  {journal} {Rev. Mod. Phys.}\ }\textbf {\bibinfo {volume} {80}},\ \bibinfo
  {pages} {395} (\bibinfo {year} {2008})}\BibitemShut {NoStop}%
\bibitem [{\citenamefont {Roch}\ \emph {et~al.}(2009)\citenamefont {Roch},
  \citenamefont {Florens}, \citenamefont {Costi}, \citenamefont {Wernsdorfer},\
  and\ \citenamefont {Balestro}}]{Nicolas2009}%
  \BibitemOpen
  \bibfield  {author} {\bibinfo {author} {\bibfnamefont {N.}~\bibnamefont
  {Roch}}, \bibinfo {author} {\bibfnamefont {S.}~\bibnamefont {Florens}},
  \bibinfo {author} {\bibfnamefont {T.~A.}\ \bibnamefont {Costi}}, \bibinfo
  {author} {\bibfnamefont {W.}~\bibnamefont {Wernsdorfer}},\ and\ \bibinfo
  {author} {\bibfnamefont {F.}~\bibnamefont {Balestro}},\ }\bibfield  {title}
  {\bibinfo {title} {Observation of the underscreened {K}ondo effect in a
  molecular transistor},\ }\href
  {https://doi.org/10.1103/PhysRevLett.103.197202} {\bibfield  {journal}
  {\bibinfo  {journal} {Phys. Rev. Lett.}\ }\textbf {\bibinfo {volume} {103}},\
  \bibinfo {pages} {197202} (\bibinfo {year} {2009})}\BibitemShut {NoStop}%
\bibitem [{\citenamefont {Li}\ \emph {et~al.}(2012)\citenamefont {Li},
  \citenamefont {Tong}, \citenamefont {Zheng}, \citenamefont {Hou},
  \citenamefont {Wei}, \citenamefont {Hu},\ and\ \citenamefont
  {Yan}}]{Yan2012}%
  \BibitemOpen
  \bibfield  {author} {\bibinfo {author} {\bibfnamefont {Z.~H.}\ \bibnamefont
  {Li}}, \bibinfo {author} {\bibfnamefont {N.~H.}\ \bibnamefont {Tong}},
  \bibinfo {author} {\bibfnamefont {X.}~\bibnamefont {Zheng}}, \bibinfo
  {author} {\bibfnamefont {D.}~\bibnamefont {Hou}}, \bibinfo {author}
  {\bibfnamefont {J.~H.}\ \bibnamefont {Wei}}, \bibinfo {author} {\bibfnamefont
  {J.}~\bibnamefont {Hu}},\ and\ \bibinfo {author} {\bibfnamefont {Y.~J.}\
  \bibnamefont {Yan}},\ }\bibfield  {title} {\bibinfo {title} {Hierarchical
  {L}iouville-space approach for accurate and universal characterization of
  quantum impurity systems},\ }\href
  {https://doi.org/10.1103/PhysRevLett.109.266403} {\bibfield  {journal}
  {\bibinfo  {journal} {Phys. Rev. Lett.}\ }\textbf {\bibinfo {volume} {109}},\
  \bibinfo {pages} {266403} (\bibinfo {year} {2012})}\BibitemShut {NoStop}%
\bibitem [{\citenamefont {Kolesnychenko}\ \emph {et~al.}(2005)\citenamefont
  {Kolesnychenko}, \citenamefont {Heijnen}, \citenamefont {Zhuravlev},
  \citenamefont {de~Kort}, \citenamefont {Katsnelson}, \citenamefont
  {Lichtenstein},\ and\ \citenamefont {van Kempen}}]{STM}%
  \BibitemOpen
  \bibfield  {author} {\bibinfo {author} {\bibfnamefont {O.~Y.}\ \bibnamefont
  {Kolesnychenko}}, \bibinfo {author} {\bibfnamefont {G.~M.~M.}\ \bibnamefont
  {Heijnen}}, \bibinfo {author} {\bibfnamefont {A.~K.}\ \bibnamefont
  {Zhuravlev}}, \bibinfo {author} {\bibfnamefont {R.}~\bibnamefont {de~Kort}},
  \bibinfo {author} {\bibfnamefont {M.~I.}\ \bibnamefont {Katsnelson}},
  \bibinfo {author} {\bibfnamefont {A.~I.}\ \bibnamefont {Lichtenstein}},\ and\
  \bibinfo {author} {\bibfnamefont {H.}~\bibnamefont {van Kempen}},\ }\bibfield
   {title} {\bibinfo {title} {Surface electronic structure of {C}r(001):
  Experiment and theory},\ }\href {https://doi.org/10.1103/PhysRevB.72.085456}
  {\bibfield  {journal} {\bibinfo  {journal} {Phys. Rev. B}\ }\textbf {\bibinfo
  {volume} {72}},\ \bibinfo {pages} {085456} (\bibinfo {year}
  {2005})}\BibitemShut {NoStop}%
\bibitem [{\citenamefont {Jin}\ \emph {et~al.}(2008)\citenamefont {Jin},
  \citenamefont {Zheng},\ and\ \citenamefont {Yan}}]{Jin2008}%
  \BibitemOpen
  \bibfield  {author} {\bibinfo {author} {\bibfnamefont {J.}~\bibnamefont
  {Jin}}, \bibinfo {author} {\bibfnamefont {X.}~\bibnamefont {Zheng}},\ and\
  \bibinfo {author} {\bibfnamefont {Y.~J.}\ \bibnamefont {Yan}},\ }\bibfield
  {title} {\bibinfo {title} {Exact dynamics of dissipative electronic systems
  and quantum transport: {H}ierarchical equations of motion approach},\ }\href
  {https://doi.org/10.1063/1.2938087} {\bibfield  {journal} {\bibinfo
  {journal} {J. Chem. Phys}\ }\textbf {\bibinfo {volume} {128}},\ \bibinfo
  {pages} {234703} (\bibinfo {year} {2008})}\BibitemShut {NoStop}%
\bibitem [{\citenamefont {Kato}\ and\ \citenamefont
  {Tanimura}(2016)}]{Kato2016}%
  \BibitemOpen
  \bibfield  {author} {\bibinfo {author} {\bibfnamefont {A.}~\bibnamefont
  {Kato}}\ and\ \bibinfo {author} {\bibfnamefont {Y.}~\bibnamefont
  {Tanimura}},\ }\bibfield  {title} {\bibinfo {title} {Quantum heat current
  under non-perturbative and non-markovian conditions: Applications to heat
  machines},\ }\href {https://doi.org/10.1063/1.4971370} {\bibfield  {journal}
  {\bibinfo  {journal} {J. Chem. Phys}\ }\textbf {\bibinfo {volume} {145}},\
  \bibinfo {pages} {224105} (\bibinfo {year} {2016})}\BibitemShut {NoStop}%
\bibitem [{\citenamefont {Tanimura}(2020)}]{Numericallyexact2020}%
  \BibitemOpen
  \bibfield  {author} {\bibinfo {author} {\bibfnamefont {Y.}~\bibnamefont
  {Tanimura}},\ }\bibfield  {title} {\bibinfo {title} {Numerically
  {\textquotedblleft}exact{\textquotedblright} approach to open quantum
  dynamics: The hierarchical equations of motion ({HEOM})},\ }\href
  {https://doi.org/10.1063/5.0011599} {\bibfield  {journal} {\bibinfo
  {journal} {J. Chem. Phys}\ }\textbf {\bibinfo {volume} {153}},\ \bibinfo
  {pages} {020901} (\bibinfo {year} {2020})}\BibitemShut {NoStop}%
\bibitem [{\citenamefont {Lambert}\ \emph {et~al.}(2023)\citenamefont
  {Lambert}, \citenamefont {Raheja}, \citenamefont {Cross}, \citenamefont
  {Menczel}, \citenamefont {Ahmed}, \citenamefont {Pitchford}, \citenamefont
  {Burgarth},\ and\ \citenamefont {Nori}}]{lambert2020bofinheom}%
  \BibitemOpen
  \bibfield  {author} {\bibinfo {author} {\bibfnamefont {N.}~\bibnamefont
  {Lambert}}, \bibinfo {author} {\bibfnamefont {T.}~\bibnamefont {Raheja}},
  \bibinfo {author} {\bibfnamefont {S.}~\bibnamefont {Cross}}, \bibinfo
  {author} {\bibfnamefont {P.}~\bibnamefont {Menczel}}, \bibinfo {author}
  {\bibfnamefont {S.}~\bibnamefont {Ahmed}}, \bibinfo {author} {\bibfnamefont
  {A.}~\bibnamefont {Pitchford}}, \bibinfo {author} {\bibfnamefont
  {D.}~\bibnamefont {Burgarth}},\ and\ \bibinfo {author} {\bibfnamefont
  {F.}~\bibnamefont {Nori}},\ }\bibfield  {title} {\bibinfo {title}
  {Qutip-bofin: A bosonic and fermionic numerical
  hierarchical-equations-of-motion library with applications in
  light-harvesting, quantum control, and single-molecule electronics},\ }\href
  {https://doi.org/10.1103/PhysRevResearch.5.013181} {\bibfield  {journal}
  {\bibinfo  {journal} {Phys. Rev. Res.}\ }\textbf {\bibinfo {volume} {5}},\
  \bibinfo {pages} {013181} (\bibinfo {year} {2023})}\BibitemShut {NoStop}%
\bibitem [{\citenamefont {Koyanagi}\ and\ \citenamefont
  {Tanimura}(2022)}]{Koyanagi2022}%
  \BibitemOpen
  \bibfield  {author} {\bibinfo {author} {\bibfnamefont {S.}~\bibnamefont
  {Koyanagi}}\ and\ \bibinfo {author} {\bibfnamefont {Y.}~\bibnamefont
  {Tanimura}},\ }\bibfield  {title} {\bibinfo {title} {Numerically
  {\textquotedblleft}exact{\textquotedblright} simulations of a quantum carnot
  cycle: Analysis using thermodynamic work diagrams},\ }\href
  {https://doi.org/10.1063/5.0107305} {\bibfield  {journal} {\bibinfo
  {journal} {J. Chem. Phys}\ }\textbf {\bibinfo {volume} {157}},\ \bibinfo
  {pages} {084110} (\bibinfo {year} {2022})}\BibitemShut {NoStop}%
\bibitem [{\citenamefont {Childress}\ \emph {et~al.}(2004)\citenamefont
  {Childress}, \citenamefont {S\o{}rensen},\ and\ \citenamefont
  {Lukin}}]{Childress2004}%
  \BibitemOpen
  \bibfield  {author} {\bibinfo {author} {\bibfnamefont {L.}~\bibnamefont
  {Childress}}, \bibinfo {author} {\bibfnamefont {A.~S.}\ \bibnamefont
  {S\o{}rensen}},\ and\ \bibinfo {author} {\bibfnamefont {M.~D.}\ \bibnamefont
  {Lukin}},\ }\bibfield  {title} {\bibinfo {title} {Mesoscopic cavity quantum
  electrodynamics with quantum dots},\ }\href
  {https://doi.org/10.1103/PhysRevA.69.042302} {\bibfield  {journal} {\bibinfo
  {journal} {Phys. Rev. A}\ }\textbf {\bibinfo {volume} {69}},\ \bibinfo
  {pages} {042302} (\bibinfo {year} {2004})}\BibitemShut {NoStop}%
\bibitem [{\citenamefont {Hagenm\"uller}\ \emph {et~al.}(2017)\citenamefont
  {Hagenm\"uller}, \citenamefont {Schachenmayer}, \citenamefont {Sch\"utz},
  \citenamefont {Genes},\ and\ \citenamefont {Pupillo}}]{Hagenm2017}%
  \BibitemOpen
  \bibfield  {author} {\bibinfo {author} {\bibfnamefont {D.}~\bibnamefont
  {Hagenm\"uller}}, \bibinfo {author} {\bibfnamefont {J.}~\bibnamefont
  {Schachenmayer}}, \bibinfo {author} {\bibfnamefont {S.}~\bibnamefont
  {Sch\"utz}}, \bibinfo {author} {\bibfnamefont {C.}~\bibnamefont {Genes}},\
  and\ \bibinfo {author} {\bibfnamefont {G.}~\bibnamefont {Pupillo}},\
  }\bibfield  {title} {\bibinfo {title} {Cavity-enhanced transport of charge},\
  }\href {https://doi.org/10.1103/PhysRevLett.119.223601} {\bibfield  {journal}
  {\bibinfo  {journal} {Phys. Rev. Lett.}\ }\textbf {\bibinfo {volume} {119}},\
  \bibinfo {pages} {223601} (\bibinfo {year} {2017})}\BibitemShut {NoStop}%
\bibitem [{\citenamefont {Hagenm\"uller}\ \emph {et~al.}(2018)\citenamefont
  {Hagenm\"uller}, \citenamefont {Sch\"utz}, \citenamefont {Schachenmayer},
  \citenamefont {Genes},\ and\ \citenamefont {Pupillo}}]{Hagenm2018}%
  \BibitemOpen
  \bibfield  {author} {\bibinfo {author} {\bibfnamefont {D.}~\bibnamefont
  {Hagenm\"uller}}, \bibinfo {author} {\bibfnamefont {S.}~\bibnamefont
  {Sch\"utz}}, \bibinfo {author} {\bibfnamefont {J.}~\bibnamefont
  {Schachenmayer}}, \bibinfo {author} {\bibfnamefont {C.}~\bibnamefont
  {Genes}},\ and\ \bibinfo {author} {\bibfnamefont {G.}~\bibnamefont
  {Pupillo}},\ }\bibfield  {title} {\bibinfo {title} {Cavity-assisted
  mesoscopic transport of fermions: Coherent and dissipative dynamics},\ }\href
  {https://doi.org/10.1103/PhysRevB.97.205303} {\bibfield  {journal} {\bibinfo
  {journal} {Phys. Rev. B}\ }\textbf {\bibinfo {volume} {97}},\ \bibinfo
  {pages} {205303} (\bibinfo {year} {2018})}\BibitemShut {NoStop}%
\bibitem [{\citenamefont {Bartolo}\ and\ \citenamefont
  {Ciuti}(2018)}]{Bartolo2018}%
  \BibitemOpen
  \bibfield  {author} {\bibinfo {author} {\bibfnamefont {N.}~\bibnamefont
  {Bartolo}}\ and\ \bibinfo {author} {\bibfnamefont {C.}~\bibnamefont
  {Ciuti}},\ }\bibfield  {title} {\bibinfo {title} {Vacuum-dressed cavity
  magnetotransport of a two-dimensional electron gas},\ }\href
  {https://doi.org/10.1103/PhysRevB.98.205301} {\bibfield  {journal} {\bibinfo
  {journal} {Phys. Rev. B}\ }\textbf {\bibinfo {volume} {98}},\ \bibinfo
  {pages} {205301} (\bibinfo {year} {2018})}\BibitemShut {NoStop}%
\bibitem [{\citenamefont {Xiang}\ \emph {et~al.}(2013)\citenamefont {Xiang},
  \citenamefont {Ashhab}, \citenamefont {You},\ and\ \citenamefont
  {Nori}}]{Xiang2013}%
  \BibitemOpen
  \bibfield  {author} {\bibinfo {author} {\bibfnamefont {Z.-L.}\ \bibnamefont
  {Xiang}}, \bibinfo {author} {\bibfnamefont {S.}~\bibnamefont {Ashhab}},
  \bibinfo {author} {\bibfnamefont {J.~Q.}\ \bibnamefont {You}},\ and\ \bibinfo
  {author} {\bibfnamefont {F.}~\bibnamefont {Nori}},\ }\bibfield  {title}
  {\bibinfo {title} {Hybrid quantum circuits: Superconducting circuits
  interacting with other quantum systems},\ }\href
  {https://doi.org/10.1103/RevModPhys.85.623} {\bibfield  {journal} {\bibinfo
  {journal} {Rev. Mod. Phys.}\ }\textbf {\bibinfo {volume} {85}},\ \bibinfo
  {pages} {623} (\bibinfo {year} {2013})}\BibitemShut {NoStop}%
\bibitem [{\citenamefont {Geiser}\ \emph {et~al.}(2012)\citenamefont {Geiser},
  \citenamefont {Castellano}, \citenamefont {Scalari}, \citenamefont {Beck},
  \citenamefont {Nevou},\ and\ \citenamefont {Faist}}]{Geiser2012}%
  \BibitemOpen
  \bibfield  {author} {\bibinfo {author} {\bibfnamefont {M.}~\bibnamefont
  {Geiser}}, \bibinfo {author} {\bibfnamefont {F.}~\bibnamefont {Castellano}},
  \bibinfo {author} {\bibfnamefont {G.}~\bibnamefont {Scalari}}, \bibinfo
  {author} {\bibfnamefont {M.}~\bibnamefont {Beck}}, \bibinfo {author}
  {\bibfnamefont {L.}~\bibnamefont {Nevou}},\ and\ \bibinfo {author}
  {\bibfnamefont {J.}~\bibnamefont {Faist}},\ }\bibfield  {title} {\bibinfo
  {title} {Ultrastrong coupling regime and plasmon polaritons in parabolic
  semiconductor quantum wells},\ }\href
  {https://doi.org/10.1103/PhysRevLett.108.106402} {\bibfield  {journal}
  {\bibinfo  {journal} {Phys. Rev. Lett.}\ }\textbf {\bibinfo {volume} {108}},\
  \bibinfo {pages} {106402} (\bibinfo {year} {2012})}\BibitemShut {NoStop}%
\bibitem [{\citenamefont {Gubbin}\ \emph {et~al.}(2014)\citenamefont {Gubbin},
  \citenamefont {Maier},\ and\ \citenamefont {K{\'{e}}na-Cohen}}]{Gubbin2014}%
  \BibitemOpen
  \bibfield  {author} {\bibinfo {author} {\bibfnamefont {C.~R.}\ \bibnamefont
  {Gubbin}}, \bibinfo {author} {\bibfnamefont {S.~A.}\ \bibnamefont {Maier}},\
  and\ \bibinfo {author} {\bibfnamefont {S.}~\bibnamefont {K{\'{e}}na-Cohen}},\
  }\bibfield  {title} {\bibinfo {title} {Low-voltage polariton
  electroluminescence from an ultrastrongly coupled organic light-emitting
  diode},\ }\href {https://doi.org/10.1063/1.4871271} {\bibfield  {journal}
  {\bibinfo  {journal} {Applied Physics Letters}\ }\textbf {\bibinfo {volume}
  {104}},\ \bibinfo {pages} {233302} (\bibinfo {year} {2014})}\BibitemShut
  {NoStop}%
\bibitem [{\citenamefont {Delbecq}\ \emph {et~al.}(2011)\citenamefont
  {Delbecq}, \citenamefont {Schmitt}, \citenamefont {Parmentier}, \citenamefont
  {Roch}, \citenamefont {Viennot}, \citenamefont {F\`eve}, \citenamefont
  {Huard}, \citenamefont {Mora}, \citenamefont {Cottet},\ and\ \citenamefont
  {Kontos}}]{Delbecq2011}%
  \BibitemOpen
  \bibfield  {author} {\bibinfo {author} {\bibfnamefont {M.~R.}\ \bibnamefont
  {Delbecq}}, \bibinfo {author} {\bibfnamefont {V.}~\bibnamefont {Schmitt}},
  \bibinfo {author} {\bibfnamefont {F.~D.}\ \bibnamefont {Parmentier}},
  \bibinfo {author} {\bibfnamefont {N.}~\bibnamefont {Roch}}, \bibinfo {author}
  {\bibfnamefont {J.~J.}\ \bibnamefont {Viennot}}, \bibinfo {author}
  {\bibfnamefont {G.}~\bibnamefont {F\`eve}}, \bibinfo {author} {\bibfnamefont
  {B.}~\bibnamefont {Huard}}, \bibinfo {author} {\bibfnamefont
  {C.}~\bibnamefont {Mora}}, \bibinfo {author} {\bibfnamefont {A.}~\bibnamefont
  {Cottet}},\ and\ \bibinfo {author} {\bibfnamefont {T.}~\bibnamefont
  {Kontos}},\ }\bibfield  {title} {\bibinfo {title} {Coupling a quantum dot,
  fermionic leads, and a microwave cavity on a chip},\ }\href
  {https://doi.org/10.1103/PhysRevLett.107.256804} {\bibfield  {journal}
  {\bibinfo  {journal} {Phys. Rev. Lett.}\ }\textbf {\bibinfo {volume} {107}},\
  \bibinfo {pages} {256804} (\bibinfo {year} {2011})}\BibitemShut {NoStop}%
\bibitem [{\citenamefont {Toida}\ \emph {et~al.}(2013)\citenamefont {Toida},
  \citenamefont {Nakajima},\ and\ \citenamefont {Komiyama}}]{Toida2013}%
  \BibitemOpen
  \bibfield  {author} {\bibinfo {author} {\bibfnamefont {H.}~\bibnamefont
  {Toida}}, \bibinfo {author} {\bibfnamefont {T.}~\bibnamefont {Nakajima}},\
  and\ \bibinfo {author} {\bibfnamefont {S.}~\bibnamefont {Komiyama}},\
  }\bibfield  {title} {\bibinfo {title} {Vacuum {R}abi splitting in a
  semiconductor circuit {QED} system},\ }\href
  {https://doi.org/10.1103/PhysRevLett.110.066802} {\bibfield  {journal}
  {\bibinfo  {journal} {Phys. Rev. Lett.}\ }\textbf {\bibinfo {volume} {110}},\
  \bibinfo {pages} {066802} (\bibinfo {year} {2013})}\BibitemShut {NoStop}%
\bibitem [{\citenamefont {Mi}\ \emph {et~al.}(2017)\citenamefont {Mi},
  \citenamefont {Cady}, \citenamefont {Zajac}, \citenamefont {Deelman},\ and\
  \citenamefont {Petta}}]{Mi2017}%
  \BibitemOpen
  \bibfield  {author} {\bibinfo {author} {\bibfnamefont {X.}~\bibnamefont
  {Mi}}, \bibinfo {author} {\bibfnamefont {J.~V.}\ \bibnamefont {Cady}},
  \bibinfo {author} {\bibfnamefont {D.~M.}\ \bibnamefont {Zajac}}, \bibinfo
  {author} {\bibfnamefont {P.~W.}\ \bibnamefont {Deelman}},\ and\ \bibinfo
  {author} {\bibfnamefont {J.~R.}\ \bibnamefont {Petta}},\ }\bibfield  {title}
  {\bibinfo {title} {Strong coupling of a single electron in silicon to a
  microwave photon},\ }\href {https://doi.org/10.1126/science.aal2469}
  {\bibfield  {journal} {\bibinfo  {journal} {Science}\ }\textbf {\bibinfo
  {volume} {355}},\ \bibinfo {pages} {156--158} (\bibinfo {year}
  {2017})}\BibitemShut {NoStop}%
\bibitem [{\citenamefont {Delbecq}\ \emph {et~al.}(2013)\citenamefont
  {Delbecq}, \citenamefont {Bruhat}, \citenamefont {Viennot}, \citenamefont
  {Datta}, \citenamefont {Cottet},\ and\ \citenamefont {Kontos}}]{Delbecq2013}%
  \BibitemOpen
  \bibfield  {author} {\bibinfo {author} {\bibfnamefont {M.~R.}\ \bibnamefont
  {Delbecq}}, \bibinfo {author} {\bibfnamefont {L.~E.}\ \bibnamefont {Bruhat}},
  \bibinfo {author} {\bibfnamefont {J.~J.}\ \bibnamefont {Viennot}}, \bibinfo
  {author} {\bibfnamefont {S.}~\bibnamefont {Datta}}, \bibinfo {author}
  {\bibfnamefont {A.}~\bibnamefont {Cottet}},\ and\ \bibinfo {author}
  {\bibfnamefont {T.}~\bibnamefont {Kontos}},\ }\bibfield  {title} {\bibinfo
  {title} {Photon-mediated interaction between distant quantum dot circuits},\
  }\href {https://doi.org/10.1038/ncomms2407} {\bibfield  {journal} {\bibinfo
  {journal} {Nat. Commun.}\ }\textbf {\bibinfo {volume} {4}},\ \bibinfo {pages}
  {1400} (\bibinfo {year} {2013})}\BibitemShut {NoStop}%
\bibitem [{\citenamefont {Viennot}\ \emph {et~al.}(2014)\citenamefont
  {Viennot}, \citenamefont {Delbecq}, \citenamefont {Dartiailh}, \citenamefont
  {Cottet},\ and\ \citenamefont {Kontos}}]{Viennot2014}%
  \BibitemOpen
  \bibfield  {author} {\bibinfo {author} {\bibfnamefont {J.~J.}\ \bibnamefont
  {Viennot}}, \bibinfo {author} {\bibfnamefont {M.~R.}\ \bibnamefont
  {Delbecq}}, \bibinfo {author} {\bibfnamefont {M.~C.}\ \bibnamefont
  {Dartiailh}}, \bibinfo {author} {\bibfnamefont {A.}~\bibnamefont {Cottet}},\
  and\ \bibinfo {author} {\bibfnamefont {T.}~\bibnamefont {Kontos}},\
  }\bibfield  {title} {\bibinfo {title} {Out-of-equilibrium charge dynamics in
  a hybrid circuit quantum electrodynamics architecture},\ }\href
  {https://doi.org/10.1103/PhysRevB.89.165404} {\bibfield  {journal} {\bibinfo
  {journal} {Phys. Rev. B}\ }\textbf {\bibinfo {volume} {89}},\ \bibinfo
  {pages} {165404} (\bibinfo {year} {2014})}\BibitemShut {NoStop}%
\bibitem [{\citenamefont {Brandes}\ and\ \citenamefont
  {Lambert}(2003)}]{Neill2003}%
  \BibitemOpen
  \bibfield  {author} {\bibinfo {author} {\bibfnamefont {T.}~\bibnamefont
  {Brandes}}\ and\ \bibinfo {author} {\bibfnamefont {N.}~\bibnamefont
  {Lambert}},\ }\bibfield  {title} {\bibinfo {title} {Steering of a bosonic
  mode with a double quantum dot},\ }\href
  {https://doi.org/10.1103/PhysRevB.67.125323} {\bibfield  {journal} {\bibinfo
  {journal} {Phys. Rev. B}\ }\textbf {\bibinfo {volume} {67}},\ \bibinfo
  {pages} {125323} (\bibinfo {year} {2003})}\BibitemShut {NoStop}%
\bibitem [{\citenamefont {Beaudoin}\ \emph {et~al.}(2016)\citenamefont
  {Beaudoin}, \citenamefont {Quirion}, \citenamefont {Coish},\ and\
  \citenamefont {Ladri{\`{e}}re}}]{Beaudoin2016}%
  \BibitemOpen
  \bibfield  {author} {\bibinfo {author} {\bibfnamefont {F.}~\bibnamefont
  {Beaudoin}}, \bibinfo {author} {\bibfnamefont {D.~L.}\ \bibnamefont
  {Quirion}}, \bibinfo {author} {\bibfnamefont {W.~A.}\ \bibnamefont {Coish}},\
  and\ \bibinfo {author} {\bibfnamefont {M.~P.}\ \bibnamefont
  {Ladri{\`{e}}re}},\ }\bibfield  {title} {\bibinfo {title} {Coupling a single
  electron spin to a microwave resonator: controlling transverse and
  longitudinal couplings},\ }\href
  {https://doi.org/10.1088/0957-4484/27/46/464003} {\bibfield  {journal}
  {\bibinfo  {journal} {Nanotechnology}\ }\textbf {\bibinfo {volume} {27}},\
  \bibinfo {pages} {464003} (\bibinfo {year} {2016})}\BibitemShut {NoStop}%
\bibitem [{\citenamefont {Lambert}\ \emph {et~al.}(2018)\citenamefont
  {Lambert}, \citenamefont {Cirio}, \citenamefont {Delbecq}, \citenamefont
  {Allison}, \citenamefont {Marx}, \citenamefont {Tarucha},\ and\ \citenamefont
  {Nori}}]{Neill2018}%
  \BibitemOpen
  \bibfield  {author} {\bibinfo {author} {\bibfnamefont {N.}~\bibnamefont
  {Lambert}}, \bibinfo {author} {\bibfnamefont {M.}~\bibnamefont {Cirio}},
  \bibinfo {author} {\bibfnamefont {M.}~\bibnamefont {Delbecq}}, \bibinfo
  {author} {\bibfnamefont {G.}~\bibnamefont {Allison}}, \bibinfo {author}
  {\bibfnamefont {M.}~\bibnamefont {Marx}}, \bibinfo {author} {\bibfnamefont
  {S.}~\bibnamefont {Tarucha}},\ and\ \bibinfo {author} {\bibfnamefont
  {F.}~\bibnamefont {Nori}},\ }\bibfield  {title} {\bibinfo {title} {Amplified
  and tunable transverse and longitudinal spin-photon coupling in hybrid
  circuit-{QED}},\ }\href {https://doi.org/10.1103/PhysRevB.97.125429}
  {\bibfield  {journal} {\bibinfo  {journal} {Phys. Rev. B}\ }\textbf {\bibinfo
  {volume} {97}},\ \bibinfo {pages} {125429} (\bibinfo {year}
  {2018})}\BibitemShut {NoStop}%
\bibitem [{\citenamefont {Abadillo-Uriel}\ \emph {et~al.}(2022)\citenamefont
  {Abadillo-Uriel}, \citenamefont {Eriksson}, \citenamefont {Coppersmith},\
  and\ \citenamefont {Friesen}}]{STqubit2019}%
  \BibitemOpen
  \bibfield  {author} {\bibinfo {author} {\bibfnamefont {J.~C.}\ \bibnamefont
  {Abadillo-Uriel}}, \bibinfo {author} {\bibfnamefont {M.~A.}\ \bibnamefont
  {Eriksson}}, \bibinfo {author} {\bibfnamefont {S.~N.}\ \bibnamefont
  {Coppersmith}},\ and\ \bibinfo {author} {\bibfnamefont {M.}~\bibnamefont
  {Friesen}},\ }\bibfield  {title} {\bibinfo {title} {Enhancing the dipolar
  coupling of a {S}-{T}$_0$ qubit with a transverse sweet spot},\ }\href
  {https://doi.org/10.1038/s41467-019-13548-w} {\bibfield  {journal} {\bibinfo
  {journal} {Nat. Commun.}\ }\textbf {\bibinfo {volume} {10}},\ \bibinfo
  {pages} {5641} (\bibinfo {year} {2022})}\BibitemShut {NoStop}%
\bibitem [{\citenamefont {Cirio}\ \emph {et~al.}(2022)\citenamefont {Cirio},
  \citenamefont {Kuo}, \citenamefont {Chen}, \citenamefont {Nori},\ and\
  \citenamefont {Lambert}}]{Mauro2022}%
  \BibitemOpen
  \bibfield  {author} {\bibinfo {author} {\bibfnamefont {M.}~\bibnamefont
  {Cirio}}, \bibinfo {author} {\bibfnamefont {P.~C.}\ \bibnamefont {Kuo}},
  \bibinfo {author} {\bibfnamefont {Y.~N.}\ \bibnamefont {Chen}}, \bibinfo
  {author} {\bibfnamefont {F.}~\bibnamefont {Nori}},\ and\ \bibinfo {author}
  {\bibfnamefont {N.}~\bibnamefont {Lambert}},\ }\bibfield  {title} {\bibinfo
  {title} {Canonical derivation of the fermionic influence superoperator},\
  }\href {https://doi.org/10.1103/PhysRevB.105.035121} {\bibfield  {journal}
  {\bibinfo  {journal} {Phys. Rev. B}\ }\textbf {\bibinfo {volume} {105}},\
  \bibinfo {pages} {035121} (\bibinfo {year} {2022})}\BibitemShut {NoStop}%
\bibitem [{\citenamefont {Huang}\ \emph {et~al.}(2023)\citenamefont {Huang},
  \citenamefont {Kuo}, \citenamefont {Lambert}, \citenamefont {Cirio},
  \citenamefont {Cross}, \citenamefont {Yang}, \citenamefont {Nori},\ and\
  \citenamefont {Chen}}]{huang2023heomjl}%
  \BibitemOpen
  \bibfield  {author} {\bibinfo {author} {\bibfnamefont {Y.-T.}\ \bibnamefont
  {Huang}}, \bibinfo {author} {\bibfnamefont {P.-C.}\ \bibnamefont {Kuo}},
  \bibinfo {author} {\bibfnamefont {N.}~\bibnamefont {Lambert}}, \bibinfo
  {author} {\bibfnamefont {M.}~\bibnamefont {Cirio}}, \bibinfo {author}
  {\bibfnamefont {S.}~\bibnamefont {Cross}}, \bibinfo {author} {\bibfnamefont
  {S.-L.}\ \bibnamefont {Yang}}, \bibinfo {author} {\bibfnamefont
  {F.}~\bibnamefont {Nori}},\ and\ \bibinfo {author} {\bibfnamefont {Y.-N.}\
  \bibnamefont {Chen}},\ }\href@noop {} {\bibinfo {title} {Hierarchicaleom.jl:
  An efficient julia framework for hierarchical equations of motion in open
  quantum systems}} (\bibinfo {year} {2023}),\ \Eprint
  {https://arxiv.org/abs/2306.07522} {arXiv:2306.07522} \BibitemShut {NoStop}%
\bibitem [{\citenamefont {Spinelli}\ \emph {et~al.}(2015)\citenamefont
  {Spinelli}, \citenamefont {Gerrits}, \citenamefont {Toskovic}, \citenamefont
  {Bryant}, \citenamefont {Ternes},\ and\ \citenamefont {Otte}}]{Spinelli2015}%
  \BibitemOpen
  \bibfield  {author} {\bibinfo {author} {\bibfnamefont {A.}~\bibnamefont
  {Spinelli}}, \bibinfo {author} {\bibfnamefont {M.}~\bibnamefont {Gerrits}},
  \bibinfo {author} {\bibfnamefont {R.}~\bibnamefont {Toskovic}}, \bibinfo
  {author} {\bibfnamefont {B.}~\bibnamefont {Bryant}}, \bibinfo {author}
  {\bibfnamefont {M.}~\bibnamefont {Ternes}},\ and\ \bibinfo {author}
  {\bibfnamefont {A.~F.}\ \bibnamefont {Otte}},\ }\bibfield  {title} {\bibinfo
  {title} {Exploring the phase diagram of the two-impurity {K}ondo problem},\
  }\href {https://doi.org/10.1038/ncomms10046} {\bibfield  {journal} {\bibinfo
  {journal} {Nat. Commun.}\ }\textbf {\bibinfo {volume} {6}},\ \bibinfo {pages}
  {10046} (\bibinfo {year} {2015})}\BibitemShut {NoStop}%
\bibitem [{\citenamefont {Shi}\ \emph {et~al.}(2016)\citenamefont {Shi},
  \citenamefont {Wu}, \citenamefont {Gonz\'alez-Tudela},\ and\ \citenamefont
  {Cirac}}]{Cirac2016}%
  \BibitemOpen
  \bibfield  {author} {\bibinfo {author} {\bibfnamefont {T.}~\bibnamefont
  {Shi}}, \bibinfo {author} {\bibfnamefont {Y.-H.}\ \bibnamefont {Wu}},
  \bibinfo {author} {\bibfnamefont {A.}~\bibnamefont {Gonz\'alez-Tudela}},\
  and\ \bibinfo {author} {\bibfnamefont {J.~I.}\ \bibnamefont {Cirac}},\
  }\bibfield  {title} {\bibinfo {title} {Bound states in boson impurity
  models},\ }\href {https://doi.org/10.1103/PhysRevX.6.021027} {\bibfield
  {journal} {\bibinfo  {journal} {Phys. Rev. X}\ }\textbf {\bibinfo {volume}
  {6}},\ \bibinfo {pages} {021027} (\bibinfo {year} {2016})}\BibitemShut
  {NoStop}%
\bibitem [{\citenamefont {S\'anchez-Burillo}\ \emph {et~al.}(2019)\citenamefont
  {S\'anchez-Burillo}, \citenamefont {Mart\'{\i}n-Moreno}, \citenamefont
  {Garc\'{\i}a-Ripoll},\ and\ \citenamefont {Zueco}}]{Zueco2019}%
  \BibitemOpen
  \bibfield  {author} {\bibinfo {author} {\bibfnamefont {E.}~\bibnamefont
  {S\'anchez-Burillo}}, \bibinfo {author} {\bibfnamefont {L.}~\bibnamefont
  {Mart\'{\i}n-Moreno}}, \bibinfo {author} {\bibfnamefont {J.~J.}\ \bibnamefont
  {Garc\'{\i}a-Ripoll}},\ and\ \bibinfo {author} {\bibfnamefont
  {D.}~\bibnamefont {Zueco}},\ }\bibfield  {title} {\bibinfo {title} {Single
  photons by quenching the vacuum},\ }\href
  {https://doi.org/10.1103/PhysRevLett.123.013601} {\bibfield  {journal}
  {\bibinfo  {journal} {Phys. Rev. Lett.}\ }\textbf {\bibinfo {volume} {123}},\
  \bibinfo {pages} {013601} (\bibinfo {year} {2019})}\BibitemShut {NoStop}%
\bibitem [{\citenamefont {Garc\'{\i}a-Elcano}\ \emph
  {et~al.}(2020)\citenamefont {Garc\'{\i}a-Elcano}, \citenamefont
  {Gonz\'alez-Tudela},\ and\ \citenamefont {Bravo-Abad}}]{Jorge2020}%
  \BibitemOpen
  \bibfield  {author} {\bibinfo {author} {\bibfnamefont {I.}~\bibnamefont
  {Garc\'{\i}a-Elcano}}, \bibinfo {author} {\bibfnamefont {A.}~\bibnamefont
  {Gonz\'alez-Tudela}},\ and\ \bibinfo {author} {\bibfnamefont
  {J.}~\bibnamefont {Bravo-Abad}},\ }\bibfield  {title} {\bibinfo {title}
  {Tunable and robust long-range coherent interactions between quantum emitters
  mediated by weyl bound states},\ }\href
  {https://doi.org/10.1103/PhysRevLett.125.163602} {\bibfield  {journal}
  {\bibinfo  {journal} {Phys. Rev. Lett.}\ }\textbf {\bibinfo {volume} {125}},\
  \bibinfo {pages} {163602} (\bibinfo {year} {2020})}\BibitemShut {NoStop}%
\bibitem [{\citenamefont {Anto-Sztrikacs}\ \emph {et~al.}(2023)\citenamefont
  {Anto-Sztrikacs}, \citenamefont {Nazir},\ and\ \citenamefont
  {Segal}}]{Segal2023}%
  \BibitemOpen
  \bibfield  {author} {\bibinfo {author} {\bibfnamefont {N.}~\bibnamefont
  {Anto-Sztrikacs}}, \bibinfo {author} {\bibfnamefont {A.}~\bibnamefont
  {Nazir}},\ and\ \bibinfo {author} {\bibfnamefont {D.}~\bibnamefont {Segal}},\
  }\bibfield  {title} {\bibinfo {title} {Effective-hamiltonian theory of open
  quantum systems at strong coupling},\ }\href
  {https://doi.org/10.1103/PRXQuantum.4.020307} {\bibfield  {journal} {\bibinfo
   {journal} {PRX Quantum}\ }\textbf {\bibinfo {volume} {4}},\ \bibinfo {pages}
  {020307} (\bibinfo {year} {2023})}\BibitemShut {NoStop}%
\bibitem [{\citenamefont {Iles-Smith}\ \emph {et~al.}(2014)\citenamefont
  {Iles-Smith}, \citenamefont {Lambert},\ and\ \citenamefont
  {Nazir}}]{IlesSmith2014}%
  \BibitemOpen
  \bibfield  {author} {\bibinfo {author} {\bibfnamefont {J.}~\bibnamefont
  {Iles-Smith}}, \bibinfo {author} {\bibfnamefont {N.}~\bibnamefont
  {Lambert}},\ and\ \bibinfo {author} {\bibfnamefont {A.}~\bibnamefont
  {Nazir}},\ }\bibfield  {title} {\bibinfo {title} {Environmental dynamics,
  correlations, and the emergence of noncanonical equilibrium states in open
  quantum systems},\ }\href {https://doi.org/10.1103/PhysRevA.90.032114}
  {\bibfield  {journal} {\bibinfo  {journal} {Phys. Rev. A}\ }\textbf {\bibinfo
  {volume} {90}},\ \bibinfo {pages} {032114} (\bibinfo {year}
  {2014})}\BibitemShut {NoStop}%
\bibitem [{\citenamefont {Iles-Smith}\ \emph {et~al.}(2016)\citenamefont
  {Iles-Smith}, \citenamefont {Dijkstra}, \citenamefont {Lambert},\ and\
  \citenamefont {Nazir}}]{IlesSmith2016}%
  \BibitemOpen
  \bibfield  {author} {\bibinfo {author} {\bibfnamefont {J.}~\bibnamefont
  {Iles-Smith}}, \bibinfo {author} {\bibfnamefont {A.~G.}\ \bibnamefont
  {Dijkstra}}, \bibinfo {author} {\bibfnamefont {N.}~\bibnamefont {Lambert}},\
  and\ \bibinfo {author} {\bibfnamefont {A.}~\bibnamefont {Nazir}},\ }\bibfield
   {title} {\bibinfo {title} {Energy transfer in structured and unstructured
  environments: Master equations beyond the born-markov approximations},\
  }\href {https://doi.org/10.1063/1.4940218} {\bibfield  {journal} {\bibinfo
  {journal} {J. Chem. Phys.}\ }\textbf {\bibinfo {volume} {144}},\ \bibinfo
  {pages} {044110} (\bibinfo {year} {2016})}\BibitemShut {NoStop}%
\bibitem [{\citenamefont {S.~Luo}(2023)}]{LuoSi}%
  \BibitemOpen
  \bibfield  {author} {\bibinfo {author} {\bibfnamefont {M.~C.}\ \bibnamefont
  {S.~Luo}, \bibfnamefont {N.~Lambert}},\ }\bibfield  {title} {\bibinfo {title}
  {A quantum-classical decomposition of gaussian quantum environments: a
  stochastic pseudomode model},\ }\href {https://arxiv.org/abs/2301.07554}
  {\bibfield  {journal} {\bibinfo  {journal} {arXiv:2301.07554v1}\ } (\bibinfo
  {year} {2023})}\BibitemShut {NoStop}%
\bibitem [{\citenamefont {De~Liberato}(2017)}]{DeLiberato2017}%
  \BibitemOpen
  \bibfield  {author} {\bibinfo {author} {\bibfnamefont {S.}~\bibnamefont
  {De~Liberato}},\ }\bibfield  {title} {\bibinfo {title} {Virtual photons in
  the ground state of a dissipative system},\ }\href
  {https://doi.org/10.1038/s41467-017-01504-5} {\bibfield  {journal} {\bibinfo
  {journal} {Nature Commun.}\ }\textbf {\bibinfo {volume} {8}},\ \bibinfo
  {pages} {1465} (\bibinfo {year} {2017})}\BibitemShut {NoStop}%
\bibitem [{\citenamefont {De~Bernardis}\ \emph {et~al.}(2018)\citenamefont
  {De~Bernardis}, \citenamefont {Pilar}, \citenamefont {Jaako}, \citenamefont
  {De~Liberato},\ and\ \citenamefont {Rabl}}]{Bernardis2018}%
  \BibitemOpen
  \bibfield  {author} {\bibinfo {author} {\bibfnamefont {D.}~\bibnamefont
  {De~Bernardis}}, \bibinfo {author} {\bibfnamefont {P.}~\bibnamefont {Pilar}},
  \bibinfo {author} {\bibfnamefont {T.}~\bibnamefont {Jaako}}, \bibinfo
  {author} {\bibfnamefont {S.}~\bibnamefont {De~Liberato}},\ and\ \bibinfo
  {author} {\bibfnamefont {P.}~\bibnamefont {Rabl}},\ }\bibfield  {title}
  {\bibinfo {title} {Breakdown of gauge invariance in ultrastrong-coupling
  cavity {QED}},\ }\href {https://doi.org/10.1103/PhysRevA.98.053819}
  {\bibfield  {journal} {\bibinfo  {journal} {Phys. Rev. A}\ }\textbf {\bibinfo
  {volume} {98}},\ \bibinfo {pages} {053819} (\bibinfo {year}
  {2018})}\BibitemShut {NoStop}%
\bibitem [{\citenamefont {Stefano}\ \emph {et~al.}(2019)\citenamefont
  {Stefano}, \citenamefont {Settineri}, \citenamefont {Macr{\`{\i}}},
  \citenamefont {Garziano}, \citenamefont {Stassi}, \citenamefont {Savasta},\
  and\ \citenamefont {Nori}}]{DiStefano2019}%
  \BibitemOpen
  \bibfield  {author} {\bibinfo {author} {\bibfnamefont {O.~D.}\ \bibnamefont
  {Stefano}}, \bibinfo {author} {\bibfnamefont {A.}~\bibnamefont {Settineri}},
  \bibinfo {author} {\bibfnamefont {V.}~\bibnamefont {Macr{\`{\i}}}}, \bibinfo
  {author} {\bibfnamefont {L.}~\bibnamefont {Garziano}}, \bibinfo {author}
  {\bibfnamefont {R.}~\bibnamefont {Stassi}}, \bibinfo {author} {\bibfnamefont
  {S.}~\bibnamefont {Savasta}},\ and\ \bibinfo {author} {\bibfnamefont
  {F.}~\bibnamefont {Nori}},\ }\bibfield  {title} {\bibinfo {title} {Resolution
  of gauge ambiguities in ultrastrong-coupling cavity quantum
  electrodynamics},\ }\href {https://doi.org/10.1038/s41567-019-0534-4}
  {\bibfield  {journal} {\bibinfo  {journal} {Nat. Phys.}\ }\textbf {\bibinfo
  {volume} {15}},\ \bibinfo {pages} {803--808} (\bibinfo {year}
  {2019})}\BibitemShut {NoStop}%
\bibitem [{\citenamefont {Smith}\ \emph {et~al.}(2022)\citenamefont {Smith},
  \citenamefont {Chen}, \citenamefont {Chang}, \citenamefont {Wu},
  \citenamefont {Lo}, \citenamefont {Chao}, \citenamefont {Farrer},
  \citenamefont {Beere}, \citenamefont {Griffiths}, \citenamefont {Jones},
  \citenamefont {Ritchie}, \citenamefont {Chen},\ and\ \citenamefont
  {Chen}}]{Luke2022}%
  \BibitemOpen
  \bibfield  {author} {\bibinfo {author} {\bibfnamefont {L.~W.}\ \bibnamefont
  {Smith}}, \bibinfo {author} {\bibfnamefont {H.-B.}\ \bibnamefont {Chen}},
  \bibinfo {author} {\bibfnamefont {C.-W.}\ \bibnamefont {Chang}}, \bibinfo
  {author} {\bibfnamefont {C.-W.}\ \bibnamefont {Wu}}, \bibinfo {author}
  {\bibfnamefont {S.-T.}\ \bibnamefont {Lo}}, \bibinfo {author} {\bibfnamefont
  {S.-H.}\ \bibnamefont {Chao}}, \bibinfo {author} {\bibfnamefont
  {I.}~\bibnamefont {Farrer}}, \bibinfo {author} {\bibfnamefont {H.~E.}\
  \bibnamefont {Beere}}, \bibinfo {author} {\bibfnamefont {J.~P.}\ \bibnamefont
  {Griffiths}}, \bibinfo {author} {\bibfnamefont {G.~A.~C.}\ \bibnamefont
  {Jones}}, \bibinfo {author} {\bibfnamefont {D.~A.}\ \bibnamefont {Ritchie}},
  \bibinfo {author} {\bibfnamefont {Y.-N.}\ \bibnamefont {Chen}},\ and\
  \bibinfo {author} {\bibfnamefont {T.-M.}\ \bibnamefont {Chen}},\ }\bibfield
  {title} {\bibinfo {title} {Electrically controllable {K}ondo correlation in
  spin-orbit-coupled quantum point contacts},\ }\href
  {https://doi.org/10.1103/PhysRevLett.128.027701} {\bibfield  {journal}
  {\bibinfo  {journal} {Phys. Rev. Lett.}\ }\textbf {\bibinfo {volume} {128}},\
  \bibinfo {pages} {027701} (\bibinfo {year} {2022})}\BibitemShut {NoStop}%
\bibitem [{\citenamefont {Gustafsson}\ \emph {et~al.}(2014)\citenamefont
  {Gustafsson}, \citenamefont {Aref}, \citenamefont {Kockum}, \citenamefont
  {Ekstr\"{o}m}, \citenamefont {Johansson},\ and\ \citenamefont
  {Delsing}}]{Gustafsson2014}%
  \BibitemOpen
  \bibfield  {author} {\bibinfo {author} {\bibfnamefont {M.~V.}\ \bibnamefont
  {Gustafsson}}, \bibinfo {author} {\bibfnamefont {T.}~\bibnamefont {Aref}},
  \bibinfo {author} {\bibfnamefont {A.~F.}\ \bibnamefont {Kockum}}, \bibinfo
  {author} {\bibfnamefont {M.~K.}\ \bibnamefont {Ekstr\"{o}m}}, \bibinfo
  {author} {\bibfnamefont {G.}~\bibnamefont {Johansson}},\ and\ \bibinfo
  {author} {\bibfnamefont {P.}~\bibnamefont {Delsing}},\ }\bibfield  {title}
  {\bibinfo {title} {Propagating phonons coupled to an artificial atom},\
  }\href {https://doi.org/10.1126/science.1257219} {\bibfield  {journal}
  {\bibinfo  {journal} {Science}\ }\textbf {\bibinfo {volume} {346}},\ \bibinfo
  {pages} {207} (\bibinfo {year} {2014})}\BibitemShut {NoStop}%
\bibitem [{\citenamefont {Manenti}\ \emph {et~al.}(2017)\citenamefont
  {Manenti}, \citenamefont {Kockum}, \citenamefont {Patterson}, \citenamefont
  {Behrle}, \citenamefont {Rahamim}, \citenamefont {Tancredi}, \citenamefont
  {Nori},\ and\ \citenamefont {Leek}}]{Manenti2017}%
  \BibitemOpen
  \bibfield  {author} {\bibinfo {author} {\bibfnamefont {R.}~\bibnamefont
  {Manenti}}, \bibinfo {author} {\bibfnamefont {A.~F.}\ \bibnamefont {Kockum}},
  \bibinfo {author} {\bibfnamefont {A.}~\bibnamefont {Patterson}}, \bibinfo
  {author} {\bibfnamefont {T.}~\bibnamefont {Behrle}}, \bibinfo {author}
  {\bibfnamefont {J.}~\bibnamefont {Rahamim}}, \bibinfo {author} {\bibfnamefont
  {G.}~\bibnamefont {Tancredi}}, \bibinfo {author} {\bibfnamefont
  {F.}~\bibnamefont {Nori}},\ and\ \bibinfo {author} {\bibfnamefont {P.~J.}\
  \bibnamefont {Leek}},\ }\bibfield  {title} {\bibinfo {title} {Circuit quantum
  acoustodynamics with surface acoustic waves},\ }\href
  {https://doi.org/10.1038/s41467-017-01063-9} {\bibfield  {journal} {\bibinfo
  {journal} {Nat. Commun.}\ }\textbf {\bibinfo {volume} {8}},\ \bibinfo {pages}
  {975} (\bibinfo {year} {2017})}\BibitemShut {NoStop}%
\bibitem [{\citenamefont {Iorsh}\ \emph {et~al.}(2020)\citenamefont {Iorsh},
  \citenamefont {Poshakinskiy},\ and\ \citenamefont {Poddubny}}]{Iorsh2020}%
  \BibitemOpen
  \bibfield  {author} {\bibinfo {author} {\bibfnamefont {I.}~\bibnamefont
  {Iorsh}}, \bibinfo {author} {\bibfnamefont {A.}~\bibnamefont
  {Poshakinskiy}},\ and\ \bibinfo {author} {\bibfnamefont {A.}~\bibnamefont
  {Poddubny}},\ }\bibfield  {title} {\bibinfo {title} {Waveguide quantum
  optomechanics: Parity-time phase transitions in ultrastrong coupling
  regime},\ }\href {https://doi.org/10.1103/PhysRevLett.125.183601} {\bibfield
  {journal} {\bibinfo  {journal} {Phys. Rev. Lett.}\ }\textbf {\bibinfo
  {volume} {125}},\ \bibinfo {pages} {183601} (\bibinfo {year}
  {2020})}\BibitemShut {NoStop}%
\bibitem [{\citenamefont {Benz}\ \emph {et~al.}(2016)\citenamefont {Benz},
  \citenamefont {Schmidt}, \citenamefont {Dreismann}, \citenamefont
  {Chikkaraddy}, \citenamefont {Zhang}, \citenamefont {Demetriadou},
  \citenamefont {Carnegie}, \citenamefont {Ohadi}, \citenamefont {de~Nijs},
  \citenamefont {Esteban}, \citenamefont {Aizpurua},\ and\ \citenamefont
  {Baumberg}}]{Benz2016}%
  \BibitemOpen
  \bibfield  {author} {\bibinfo {author} {\bibfnamefont {F.}~\bibnamefont
  {Benz}}, \bibinfo {author} {\bibfnamefont {M.~K.}\ \bibnamefont {Schmidt}},
  \bibinfo {author} {\bibfnamefont {A.}~\bibnamefont {Dreismann}}, \bibinfo
  {author} {\bibfnamefont {R.}~\bibnamefont {Chikkaraddy}}, \bibinfo {author}
  {\bibfnamefont {Y.}~\bibnamefont {Zhang}}, \bibinfo {author} {\bibfnamefont
  {A.}~\bibnamefont {Demetriadou}}, \bibinfo {author} {\bibfnamefont
  {C.}~\bibnamefont {Carnegie}}, \bibinfo {author} {\bibfnamefont
  {H.}~\bibnamefont {Ohadi}}, \bibinfo {author} {\bibfnamefont
  {B.}~\bibnamefont {de~Nijs}}, \bibinfo {author} {\bibfnamefont
  {R.}~\bibnamefont {Esteban}}, \bibinfo {author} {\bibfnamefont
  {J.}~\bibnamefont {Aizpurua}},\ and\ \bibinfo {author} {\bibfnamefont
  {J.~J.}\ \bibnamefont {Baumberg}},\ }\bibfield  {title} {\bibinfo {title}
  {Single-molecule optomechanics in
  {\textquotedblleft}picocavities{\textquotedblright}},\ }\href
  {https://doi.org/10.1126/science.aah5243} {\bibfield  {journal} {\bibinfo
  {journal} {Science}\ }\textbf {\bibinfo {volume} {354}},\ \bibinfo {pages}
  {726} (\bibinfo {year} {2016})}\BibitemShut {NoStop}%
\bibitem [{\citenamefont {Kuo}\ \emph {et~al.}(2020)\citenamefont {Kuo},
  \citenamefont {Lambert}, \citenamefont {Miranowicz}, \citenamefont {Chen},
  \citenamefont {Chen}, \citenamefont {Chen},\ and\ \citenamefont
  {Nori}}]{Po2020}%
  \BibitemOpen
  \bibfield  {author} {\bibinfo {author} {\bibfnamefont {P.~C.}\ \bibnamefont
  {Kuo}}, \bibinfo {author} {\bibfnamefont {N.}~\bibnamefont {Lambert}},
  \bibinfo {author} {\bibfnamefont {A.}~\bibnamefont {Miranowicz}}, \bibinfo
  {author} {\bibfnamefont {H.~B.}\ \bibnamefont {Chen}}, \bibinfo {author}
  {\bibfnamefont {G.~Y.}\ \bibnamefont {Chen}}, \bibinfo {author}
  {\bibfnamefont {Y.~N.}\ \bibnamefont {Chen}},\ and\ \bibinfo {author}
  {\bibfnamefont {F.}~\bibnamefont {Nori}},\ }\bibfield  {title} {\bibinfo
  {title} {Collectively induced exceptional points of quantum emitters coupled
  to nanoparticle surface plasmons},\ }\href
  {https://doi.org/10.1103/PhysRevA.101.013814} {\bibfield  {journal} {\bibinfo
   {journal} {Phys. Rev. A}\ }\textbf {\bibinfo {volume} {101}},\ \bibinfo
  {pages} {013814} (\bibinfo {year} {2020})}\BibitemShut {NoStop}%
\bibitem [{\citenamefont {Shi}\ \emph {et~al.}(2009)\citenamefont {Shi},
  \citenamefont {Chen}, \citenamefont {Nan}, \citenamefont {Xu},\ and\
  \citenamefont {Yan}}]{Shi2009}%
  \BibitemOpen
  \bibfield  {author} {\bibinfo {author} {\bibfnamefont {Q.}~\bibnamefont
  {Shi}}, \bibinfo {author} {\bibfnamefont {L.}~\bibnamefont {Chen}}, \bibinfo
  {author} {\bibfnamefont {G.}~\bibnamefont {Nan}}, \bibinfo {author}
  {\bibfnamefont {R.-X.}\ \bibnamefont {Xu}},\ and\ \bibinfo {author}
  {\bibfnamefont {Y.}~\bibnamefont {Yan}},\ }\bibfield  {title} {\bibinfo
  {title} {Efficient hierarchical liouville space propagator to quantum
  dissipative dynamics},\ }\href {https://doi.org/10.1063/1.3077918} {\bibfield
   {journal} {\bibinfo  {journal} {The Journal of Chemical Physics}\ }\textbf
  {\bibinfo {volume} {130}},\ \bibinfo {pages} {084105} (\bibinfo {year}
  {2009})}\BibitemShut {NoStop}%
\bibitem [{\citenamefont {Hu}\ \emph {et~al.}(2011)\citenamefont {Hu},
  \citenamefont {Luo}, \citenamefont {Jiang}, \citenamefont {Xu},\ and\
  \citenamefont {Yan}}]{Jie2011}%
  \BibitemOpen
  \bibfield  {author} {\bibinfo {author} {\bibfnamefont {J.}~\bibnamefont
  {Hu}}, \bibinfo {author} {\bibfnamefont {M.}~\bibnamefont {Luo}}, \bibinfo
  {author} {\bibfnamefont {F.}~\bibnamefont {Jiang}}, \bibinfo {author}
  {\bibfnamefont {R.-X.}\ \bibnamefont {Xu}},\ and\ \bibinfo {author}
  {\bibfnamefont {Y.}~\bibnamefont {Yan}},\ }\bibfield  {title} {\bibinfo
  {title} {Pad{\'e} spectrum decompositions of quantum distribution functions
  and optimal hierarchical equations of motion construction for quantum open
  systems},\ }\href {https://doi.org/10.1063/1.3602466} {\bibfield  {journal}
  {\bibinfo  {journal} {J. Chem. Phys.}\ }\textbf {\bibinfo {volume} {134}},\
  \bibinfo {pages} {244106} (\bibinfo {year} {2011})}\BibitemShut {NoStop}%
\bibitem [{\citenamefont {Garziano}\ \emph {et~al.}(2020)\citenamefont
  {Garziano}, \citenamefont {Settineri}, \citenamefont {Di~Stefano},
  \citenamefont {Savasta},\ and\ \citenamefont {Nori}}]{Garziano2020}%
  \BibitemOpen
  \bibfield  {author} {\bibinfo {author} {\bibfnamefont {L.}~\bibnamefont
  {Garziano}}, \bibinfo {author} {\bibfnamefont {A.}~\bibnamefont {Settineri}},
  \bibinfo {author} {\bibfnamefont {O.}~\bibnamefont {Di~Stefano}}, \bibinfo
  {author} {\bibfnamefont {S.}~\bibnamefont {Savasta}},\ and\ \bibinfo {author}
  {\bibfnamefont {F.}~\bibnamefont {Nori}},\ }\bibfield  {title} {\bibinfo
  {title} {Gauge invariance of the {D}icke and {H}opfield models},\ }\href
  {https://doi.org/10.1103/PhysRevA.102.023718} {\bibfield  {journal} {\bibinfo
   {journal} {Phys. Rev. A}\ }\textbf {\bibinfo {volume} {102}},\ \bibinfo
  {pages} {023718} (\bibinfo {year} {2020})}\BibitemShut {NoStop}%
\bibitem [{\citenamefont {Yoshihara}\ \emph {et~al.}(2022)\citenamefont
  {Yoshihara}, \citenamefont {Ashhab}, \citenamefont {Fuse}, \citenamefont
  {Bamba},\ and\ \citenamefont {Semba}}]{Yoshihara2022}%
  \BibitemOpen
  \bibfield  {author} {\bibinfo {author} {\bibfnamefont {F.}~\bibnamefont
  {Yoshihara}}, \bibinfo {author} {\bibfnamefont {S.}~\bibnamefont {Ashhab}},
  \bibinfo {author} {\bibfnamefont {T.}~\bibnamefont {Fuse}}, \bibinfo {author}
  {\bibfnamefont {M.}~\bibnamefont {Bamba}},\ and\ \bibinfo {author}
  {\bibfnamefont {K.}~\bibnamefont {Semba}},\ }\bibfield  {title} {\bibinfo
  {title} {Hamiltonian of a flux qubit-{LC} oscillator circuit in the
  deep{\textendash}strong-coupling regime},\ }\href
  {https://doi.org/10.1038/s41598-022-10203-1} {\bibfield  {journal} {\bibinfo
  {journal} {Sci. Rep.}\ }\textbf {\bibinfo {volume} {12}},\ \bibinfo {pages}
  {6764} (\bibinfo {year} {2022})}\BibitemShut {NoStop}%
\end{thebibliography}%


\end{document}